\documentclass[aps,twocolumn,groupedaddress,superscriptaddress]{revtex4-1}
\usepackage{graphicx,amsmath,amssymb}

\usepackage{lipsum}
\usepackage{color}

\newcommand\beq{\begin{equation}}
\newcommand\eeq{\end{equation}}
\newcommand\be{\begin{eqnarray}}
\newcommand\ee{\end{eqnarray}}

\begin{document}

\begin{titlepage}



\title{Annealing glasses by cyclic shear deformation}


\author{Pallabi Das}
\affiliation{Theoretical Sciences Unit, Jawaharlal Nehru Centre for Advanced Scientific Research,Jakkur Campus, Bengaluru, 560064,India}
\author{Anshul D. S. Parmar}
\altaffiliation[current address ]{Institute of Physical Chemistry, Westf{\"a}lische Wilhelms-Universit{\"a}t M{\"u}nster, Corrensstrae 28/30, 48149 M{\"u}nster, Germany}
%\affiliation{Theoretical Sciences Unit, Jawaharlal Nehru Centre for Advanced Scientific Research,Jakkur Campus, Bengaluru, 560064,India}
%\affiliation{Institute of Physical Chemistry, Westf{\"a}lische Wilhelms-Universit{\"a}t M{\"u}nster, Corrensstrae 28/30, 48149 M{\"u}nster, Germany}
\author{Srikanth Sastry}
\affiliation{Theoretical Sciences Unit, Jawaharlal Nehru Centre for Advanced Scientific Research,Jakkur Campus, Bengaluru, 560064,India}

\begin{abstract}

A major challenge in simulating glassy systems is the ability to generate configurations that may be found in equilibrium at sufficiently low temperatures, in order to probe static and dynamic behaviour close to the glass transition.  A variety of approaches have recently explored ways of surmounting this obstacle. Here, we explore the possibility of employing mechanical agitation, in the form of cyclic shear deformation, to generate low energy configurations in a model glass former. We perform shear deformation simulations over a range of temperatures, shear rates and strain amplitudes. We find that shear deformation induces faster relaxation towards low energy configurations, or {\it overaging}, in simulations at sufficiently low temperatures, consistently with previous results for athermal shear. However, for temperatures at which simulations can be run till a steady state is reached with or without shear deformation, we find that the inclusion of shear deformation does not result in any speed up of the relaxation towards low energy configurations. 
Although we find the configurations from shear simulations to have properties indistinguishable from an equilibrium ensemble, the cyclic shear procedure does not guarantee that we generate an equilibrium ensemble at a desired temperature. 
%We show, by performing cyclic shear deformation simulations over a range of temperatures, shear rates and strain amplitudes, that \pd{at sufficiently low temperatures where the system enters the aging regime, the incorporation of mechanical deformation enhances the aging caused by thermal fluctuation, causing faster aging or `overaging'} \sout{such an approach is capable of speeding up simulations by several orders of magnitude.} 
%Although \sout{low energy configurations inaccessible to normal simulations are generated, and }under suitable conditions we find the generated configurations to be indistinguishable from an equilibrium ensemble, the cyclic shear procedure does not guarantee that we generate an equilibrium ensemble at a desired temperature. 
In order to ensure equilibrium sampling, we develop a hybrid Monte Carlo algorithm that employs cyclic shear as a trial generation step, and has acceptance probabilities that depend not only on the change in internal energy but also on heat dissipated (equivalently, work done). 
We show that such an algorithm indeed generates an equilibrium ensemble. 
%\sout{We discuss possible extensions and applications to other systems with similar characteristics.}
%\pd{These generated low energy structures are found to be free from anisotropy, and their other properties also guarantee that generated structures are equilibrium-like. However, cyclic deformation does not lead to enhanced sampling at the temperature regime where the system can be equilibrated by the thermal fluctuations alone.

%\pd{We explore the possibility of generating low energy configurations in a model glass former under the application of oscillatory shear deformation with the optimal combination of temperature, the amplitude of deformation, and strain rate. We show that at sufficiently low temperatures where the system enters the aging regime, the incorporation of mechanical deformation enhances the aging caused by thermal fluctuation, causing faster aging or `overaging'. However, cyclic deformation does not lead to enhanced sampling at the temperature regime where the system can be equilibrated by the thermal fluctuations alone.   In that temperature regime, the employed hybrid Monte Carlo scheme that considers cyclic shear as a trial generation step shows that the configurations generated under suitable conditions construct an equilibrium ensemble. These generated low energy structures are found to be free from anisotropy, and their other properties also guarantee that generated structures are equilibrium-like.}
\end{abstract}

%Graphical abstract
%\includegraphics{graphicalabstract.pdf}

%Research highlights
%\item highlight1
%\item highlight2

%\begin{keyword}
%Fragility \sep Fragile-to-strong crossover \sep Mode coupling temperature \sep dynamical heterogeneity \sep Adam-Gibbs relation 
%\end{keyword}

\maketitle

\end{titlepage}


\section{Introduction}
\label{sec:level1}
The hallmark of glassy behaviour is the enormous slow down of dynamics upon decreasing temperature as the glass transition is approached. In studies of glass forming liquids through experiments and in computer simulations, such a slow down means that the observed glass transition is always a kinetic phenomenon whereby the liquid falls out of equilibrium in a protocol dependent fashion. Importantly, because glass formers (in particular in computer simulations) fall out of equilibrium  too far away from a putative ideal glass transition, definitive validation or refutation of proposed explanations for glassy behaviour becomes difficult \cite{kirkpatrick1989scaling,lubchenko2007theory,berthier2011theoretical,ediger2012perspective,karmakar2014growing}. A telling example is the growth of length scales that are considered to be associated with the approach to the glass transition. Whereas such length scales are expected to diverge at the glass transition, their growth in the observed range of temperatures is modest, varying by less than an order of magnitude \cite{karmakar2014growing}. Thus, extending the range of states that can be analysed is of great importance in developing a better understanding of the behaviour of glass forming systems. 
The reasons for the difficulty in accessing low temperature states is often expressed in terms of the complex energy landscape possessed by glassy systems \cite{goldstein1969viscous,debenedetti2001supercooled}, and going beyond glass formers, the ``rugged energy landscape'' problem is of relevance to a wide variety of physical systems and contexts. 

Recent years have witnessed encouraging progress in addressing the problem of preparing and simulating glass formers and glasses in well  annealed, low temperature (or high density) states. Seminal work by Ediger {\it et al.} opened new directions in experiments and simulations in generating deeper energy states in an efficient manner \cite{swallen2007organic,ediger2012perspective}. It has been shown experimentally that through physical vapour deposition (PVD) of particles on a substrate, maintained at an optimal temperature (15 \% below $T_{g}$), 
glasses that correspond to much lower temperatures compared to conventional methods can be prepared \cite{swallen2007organic,ediger2012perspective}.  Enthalpic measurements suggest that the PVD technique results in much lower enthalpy and higher density glasses, termed {\it ultrastable} glasses, compared to conventional approaches \cite{swallen2007organic,ediger2012perspective}.  In computer simulations, time scales accessible to conventional molecular dynamics and Monte Carlo simulations are many orders of magnitude shorter than in experiments. Some approaches using non-local moves and other methods have been attempted \cite{Krauth2000a,Kob2001b,Kranendonk_1989,gazzillo1989equation,grigera2001fast,Yan2004b,hukushima1996exchange} with varying degrees of success.  The experimental PVD method has motivated the corresponding method {\it in silico}  to generate extremely well annealed glass films \cite{singh2013ultrastable,lyubimov2013model,berthier2017origin}, by optimizing deposition rates and substrate temperature.  A limitation of this approach, however, is that it is restricted to the specific geometry required, namely that of a film, and  the glasses prepared {\it via} PVD are inhomogeneous, {\it i.e.}, the bulk density differs from that of the surface \cite{lyubimov2013model}. In simulations of polymeric glasses, the stability is correlated with the high degree of anisotropy, appearing from the layering of polymer along the normal direction to the substrate \cite{lin2014molecular}. More recently, the swap Monte Carlo method \cite{Kranendonk_1989,gazzillo1989equation,grigera2001fast}, in which non-local swaps of distinguishable particles are employed to achieve accelerated sampling of configuration space, have been employed with great success in simulating glass forming liquids \cite{Gutierrez_2015,berthier2016equilibrium,ninarello2017models}, and also promises to lead to other new simulation approaches \cite{brito2018theory}. A shortcoming currently of the swap Monte Carlo approach is that it relies on the presence of polydispersity in the simulated systems, although ways of circumventing this limitation are being explored \cite{AnshulMetGlass2020}. Even without doing so, the approach allows regimes previously unexplored in simulations to be explored \cite{Berthier2017d,ozawaPNAS2018,Bhaumik2019,Yeh2019}. These developments have greatly advanced the ability to simulate glass formers at low temperatures, and prompted the exploration of other approaches. 

%A combination of different approaches  \cite{Berthier2017d,berthier2017origin} is also another potential route to achieve the ability to simulate glass formers in regimes that have hitherto been inaccessible. 
 
%Here, we explore the ability of applying mechanical deformation to more efficiently explore the energy landscape of glass forming liquids.

Here, we explore whether the approach to low energy configurations in glass formers can be enhanced by the application of cyclic shear deformation, motivated by results for athermal cyclic shear. 
%attempt an \sout{efficient} exploration of the energy landscape of a model glassformer by applying mechanical deformation. 
%Aging and rejuvenation of glasses have been studied in the past experimentally and in simulations, owing to their obvious relevance in understanding the properties of glasses\cite{McKenna2003d,lacks2001energy,lacks2004energy}. 
The behaviour of glasses (or {\it inherent structures} (IS), local energy minima generated by energy minimization of liquid configurations) under (typically, but not restricted to) athermal quasistatic (AQS) shear deformation 
have recently been investigated in order to study the mechanical behaviour of glasses and related phenomena \cite{foffi2013,leishangthem2017yielding,priezjev2018molecular,Kawasaki2016a,Procaccia2017a,Regev2015a,Salerno2013c,Liu2016g,ozawaPNAS2018,PRIEZJEV2019,Bhaumik2019,Yeh2019,daspnas2020}. Under cyclic, or oscillatory, shear deformation, the energies of the glasses are found both to decrease from cycle to cycle, or increase, depending on relevant parameters \cite{lacks2004energy,foffi2013,leishangthem2017yielding}.  A detailed analysis of a model glass by Leishangthem {\it et al.} 
\cite{leishangthem2017yielding}, with the amplitude of shear deformation as the relevant variable, showed that below the yielding strain amplitude, progressively deeper energy minima are sampled, whereas above the yielding amplitude, energies become larger, accompanied by the formation of shear bands \cite{anshul-shearbanding}. The lowest energy, homogeneous, structures are attained at (but below) the yielding point. This observation (consistent with various theoretical investigations describing the yielding point as a limit of vanishing or low barriers to rearrangements \cite{lubchenko2003barrier,lubchenko2004theory,parisi2017shear,Jineaat6387}) suggests cyclic deformation at suitably chosen shear amplitudes as an approach to generate low energy configurations. In the present work, we investigate this possibility. In order to incorporate thermal relaxation and to explore the role of additional parameters, we study oscillatory shear deformation of liquids at finite temperature and shear rates.
\section{\label{model}Model and Methods}
We perform non-equilibrium molecular dynamics simulations (NEMD) to shear deform a model glass former at finite temperatures and strain rates. The trajectories are generated  {\emph via} the SLLOD algorithm \cite{PhysRevA.30.1528}, employing LAMMPS  \cite{plimpton1995fast} with a Nos\'e-Hoover thermostat, or (for simulations of equilibrium sampling) by an implementation of the Gaussian iso-kinetic thermostat as described in \cite{PanJCP2005}. We study the Kob-Andersen 80:20 mixture \cite{KAref} with a quadratic cutoff at $r_{c\alpha \beta} = 2.5 \sigma_{\alpha \beta}$, applying Lees-Edwards periodic boundary conditions \cite{lees1972computer}. The model parameters are 
  $\epsilon_{AB}/\epsilon_{AA}  = \epsilon_{BA}/\epsilon_{AA} = 1.5$, $\epsilon_{BB}/$ $\epsilon_{AA}  = 0.5 $, and
$\sigma_{AB}/\sigma_{AA} = \sigma_{BA}/\sigma_{AA} = 0.8 $, $\sigma_{BB}/\sigma_{AA} = 0.88 $. Energy values reported are energies per particle, in units of $\epsilon_{AA}$. Further details may be found in \cite{leishangthem2017yielding}.

The initial liquid configurations are generated {\it via} equilibrium molecular dynamics (MD) simulation (typically, at temperature $T=0.466$). Then, these configurations are subjected to oscillatory shear deformations for a range of temperatures, strain rates and amplitudes, solving the SLLOD equations: 
\begin{equation}
\dot{{\bf r}}_i = { {\bf p}_i \over m} + {\bf r}_i. {\bf \nabla} {\bf v} \nonumber
\end{equation} 
\beq
\dot{{\bf p}}_i = { {\bf F}_i \over m} - {\bf p}_i. {\bf \nabla} {\bf v} - \alpha(t) ~ {\bf p}_i
\eeq
where the strain rate tensor has only the $xy$ component being non-zero, and given by $\gamma_{xy}(t) = \gamma_{max} \sin(\omega t)$, where
$\omega$ is the frequency, and $\gamma_{max}$ is the amplitude of strain. The strain rates $\dot\gamma$ reported are the strain rate values at the initial time of each cycle, {\it i.e.}  $\dot\gamma = \gamma_{max} \omega$. The friction coefficient $\alpha$ depends on the thermostat used. The relaxation time from the equilibrium  MD simulation is denoted as $\tau$, whereas the time in the NEMD simulation is denoted by $t$. We consider $N = 4000$ particles at the reduced density $\rho = 1.2$ and perform simulations for a range of temperatures  ($T\in[0.1,0.4]$) across the Kauzmann temperature estimated in previous work to be ($T_{K}\approx0.3$),  shear rates ($\dot\gamma \in[10^{-6},10^{-3}]$), and strain amplitudes ($\gamma_{max}$) up to $0.06$.  We perform conjugate gradient minimization on the simulated (stroboscopic) configurations to obtain energy minimum configurations (inherent structures). We evaluate the potential energy of sheared liquid configurations and inherent structures,  stress anisotropy and two dimensional pair correlation functions to characterize their anisotropy, if present. The error bars on $\gamma_{max}$ indicate the grid spacing with which we sample $\gamma_{max}$.  The error bars on energies are the standard error of the mean of block averaged energies, with $20$ block considered within the measurement window (see Appendix for additional details).

%\sri{confirm the line below. You can remove the comment after you confirm/update, here and other small matters.}

%In addition to the cyclic shear simulations, we also perform constant temperature molecular dynamics simulations employing LAMMPS  \cite{plimpton1995fast} with a Nos\'e-Hoover thermostat.

In addition to the cyclic shear simulations, we also perform constant temperature MD simulations using the Gaussian isokinetic thermostat and employing LAMMPS  \cite{plimpton1995fast} with a Nos\'e-Hoover thermostat.

\section{\label{results}Results}

We carry out a series of simulations  for different shear amplitudes, for each of a set of shear rates and temperatures. The grid of values is limited by the significant computational effort for each data set. 
In order to monitor the extent of structural relaxation, we compute the average IS energies as a function of cycles, and map the IS energies to corresponding temperatures, as explained below.

\subsection{\label{sec:level2}Cyclic shear simulations}

\begin{figure}[htp]
\centering{}
\includegraphics[scale=0.25]{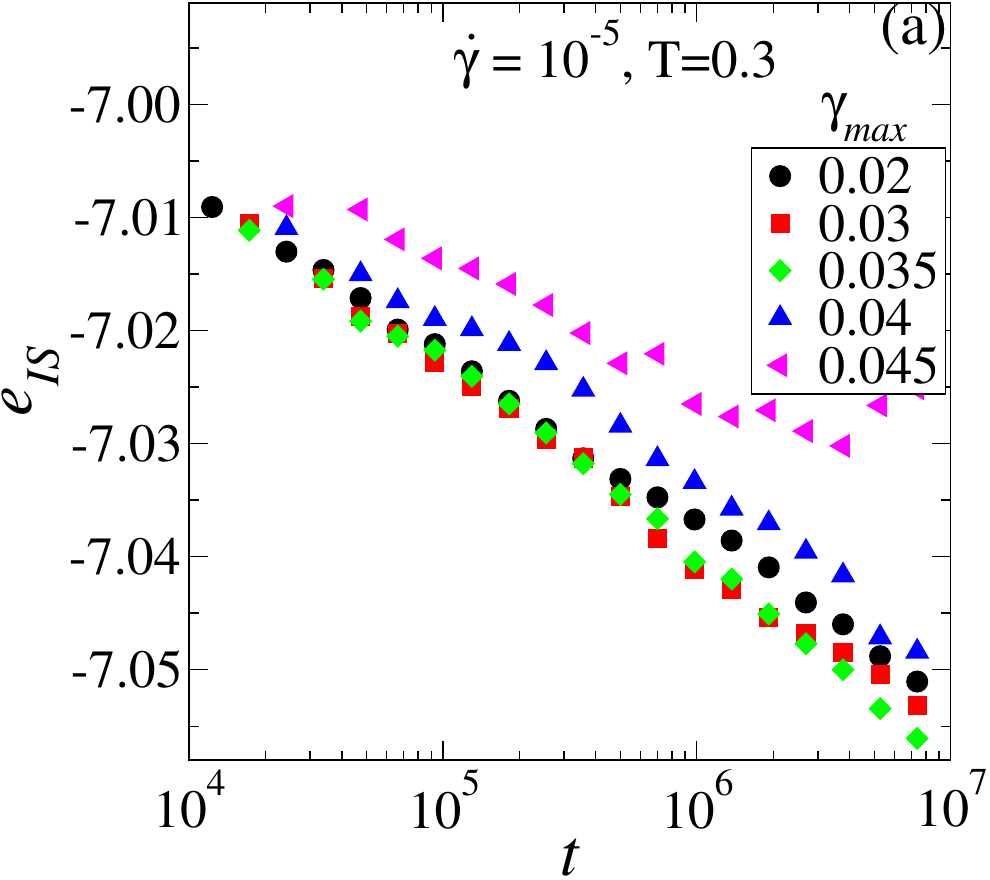} 
\includegraphics[scale=0.25]{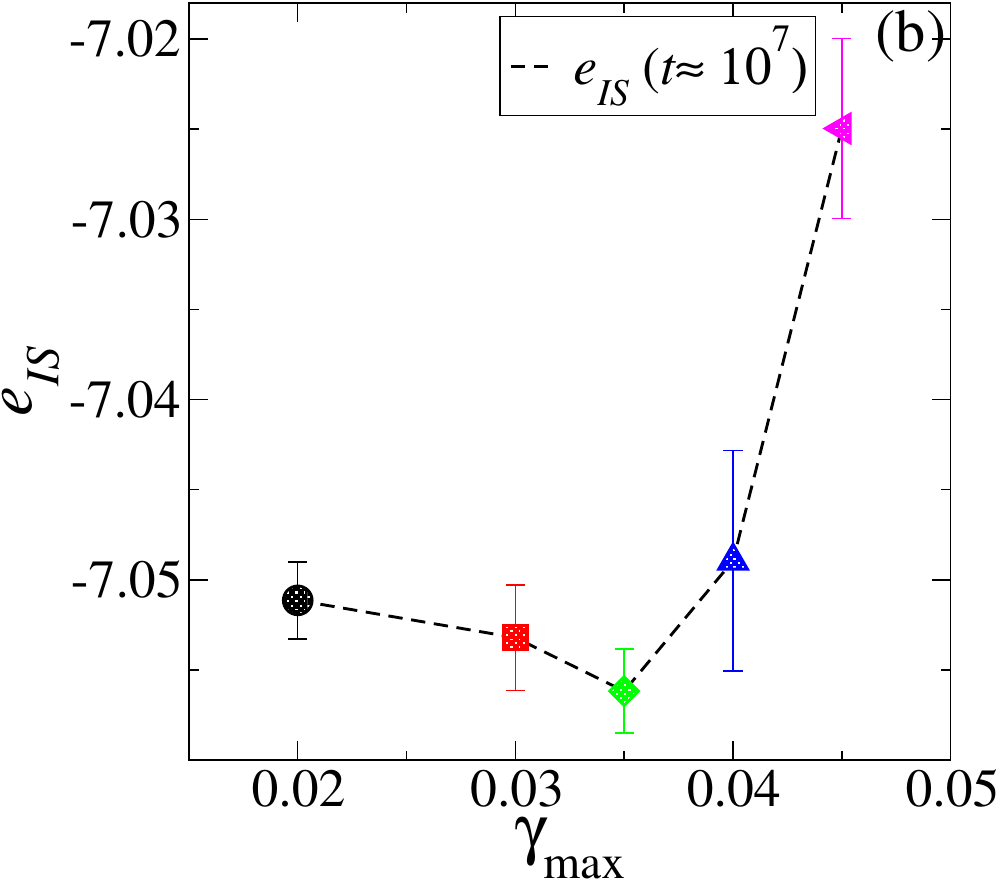} 
\includegraphics[scale=0.26]{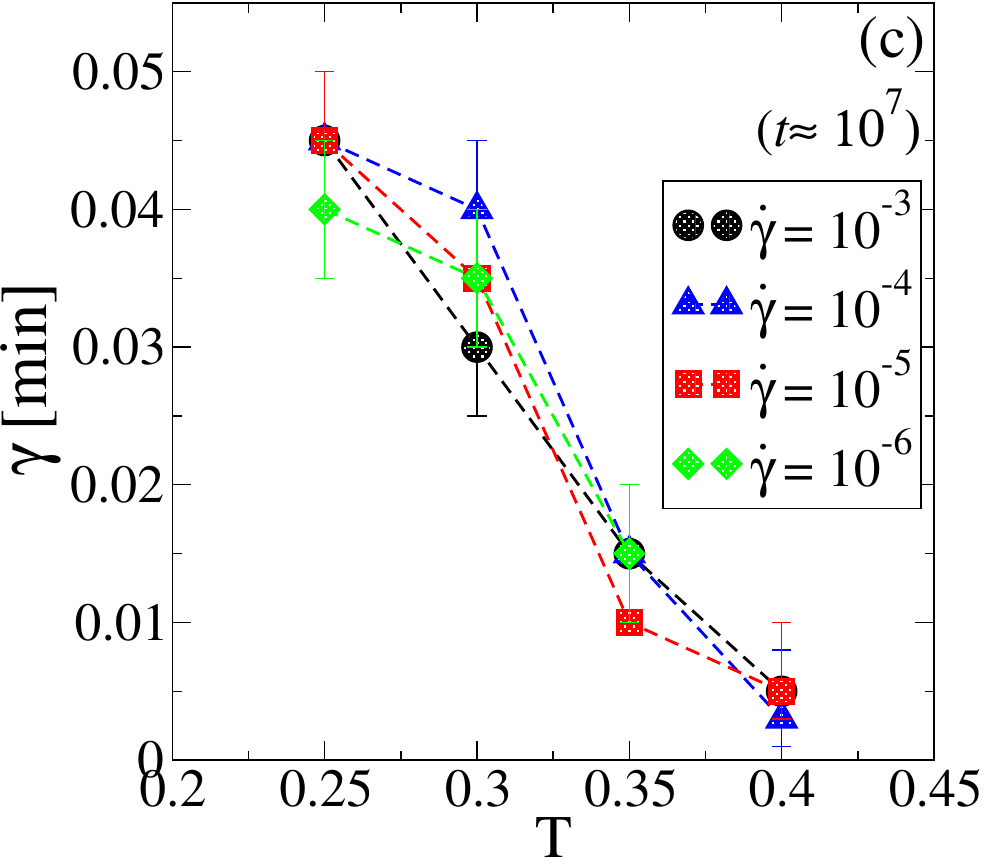}\includegraphics[scale=0.26]{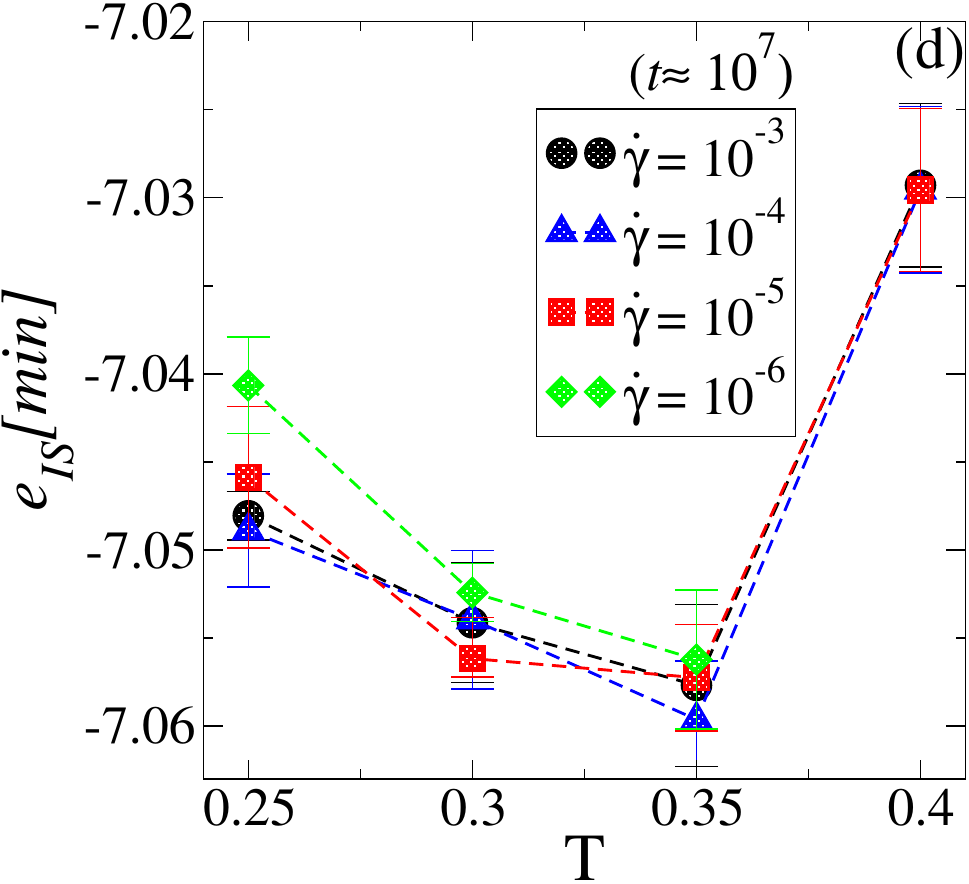}
\caption{(\emph{a}) Time evolution of inherent structure (IS) energy at a fixed temperature ($T=0.3$) and shear rate ($\dot\gamma=10^{-5})$. (\emph{b}) The long time values of IS energies {\it vs.} strain amplitude, obtained as an average over the interval from  $t = 8 \times 10^{6}$ to $t = 10^{7}$. The strain amplitude for which the energy is lowest is identified as the optimal amplitude for annealing. 
(\emph{c}) The strain amplitude for minimum $IS$ energy is a decreasing function of temperature. 
(\emph{d}) The minimum inherent structure energy attained
{\it vs.} temperature, shown for three different shear rates. The results indicate that the optimum temperature for annealing is $T  \approx 0.35$. } 
\label{fig:amp}
\end{figure}

To find out the amplitude dependence of the $IS$ energy (which we report stroboscopically, {\it i. e.}, at the end of each cycle, unless otherwise noted),  we show in Fig. \ref{fig:amp} the $IS$ energies ($e_{IS}$), for temperature $T=0.3$ and strain rate $\dot\gamma=10^{-5}$. Fig. \ref{fig:amp} (a) shows that with the increase in strain amplitude the system descends towards lower energy minima, up to a particular amplitude, whereas for larger amplitudes the energies begin to increase. Fig. \ref{fig:amp} (b) shows the values of the $IS$ energies at a large time of $t \sim 10^{7}$, as a function of strain amplitude $\gamma_{max}$, with  $\gamma_{max}=0.035$ generating the lowest energy configurations. As shown in the Appendix, Fig. A1, $\gamma_{max}=0.035$ displays characteristics of the yield strain amplitude observed earlier in AQS simulations\cite{leishangthem2017yielding} and we henceforth refer to the $\gamma_{max}$ value displaying the lowest energies as the yield amplitude. Corresponding data for other cases are presented in the Appendix. We note that the decrease of energies is generally logarithmic, a feature observed in aging systems, granular compaction, {\it etc.} \cite{McKenna2003d,knight1995density,PhysRevLett.82.916,Bandi_2018}. 

We next consider the  strain rate and temperature dependence of the optimal strain amplitude considering a range of strain rates ($\dot\gamma \in [10^{-3},10^{-6}]$) and temperatures ($T\in[0.25,0.4]$). Although with limited data, shown in the Appendix, we see that with increasing strain rates, the yield strain shifts towards higher values \textcolor{black}{(see Appendix, Fig. A2-A5)}, consistently with previous results \cite{varnik2004study,shrivastav2016yielding}. Likewise, as  the temperature is lowered the yield strain amplitude  shifts towards higher strain amplitudes, as shown in Fig. \ref{fig:amp} (c). For each strain rate, we consider the IS energies obtained at the optimal strain amplitude $\gamma_{max}$ and plot it as a function of temperature in  Fig. \ref{fig:amp} (d) \textcolor{black}{(see appendix, Fig. A2-A5)}.  Similar to observations for PVD \cite{lyubimov2013model}, we find the maximum extent of annealing for $T = 0.35$, near the estimated Kauzmann temperature ($T_{K}\approx 0.3$).

%\sout{, and we perform further analysis at that temperature.} 

\begin{figure*}[]
\centering{}
\includegraphics[scale=0.35]{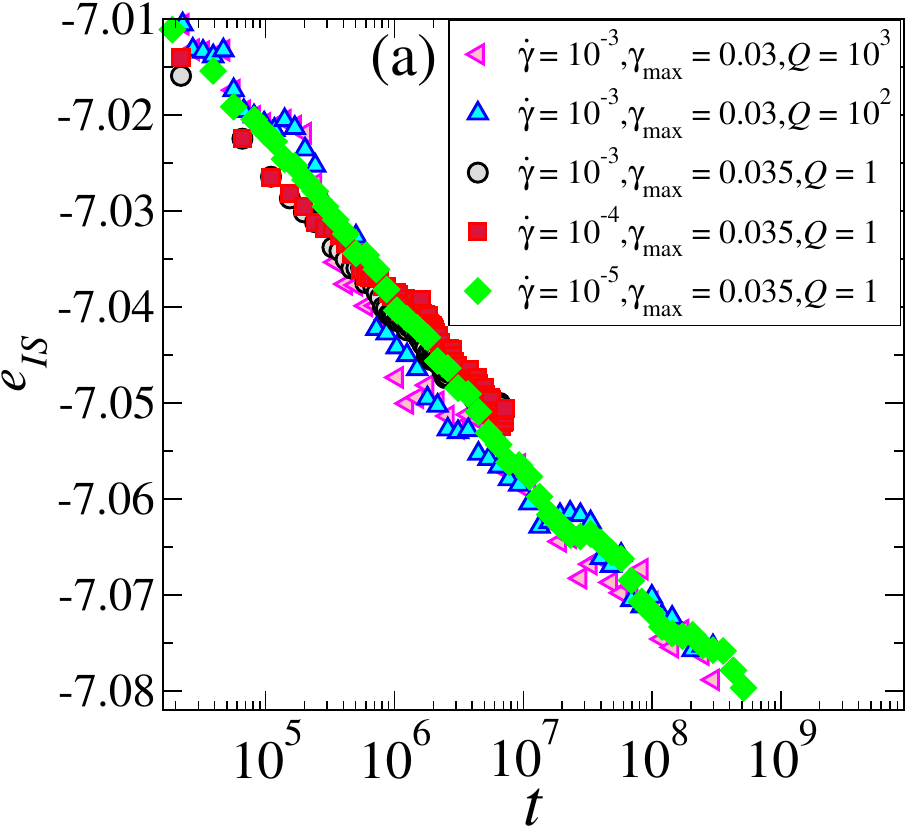} 
\includegraphics[scale=0.35]{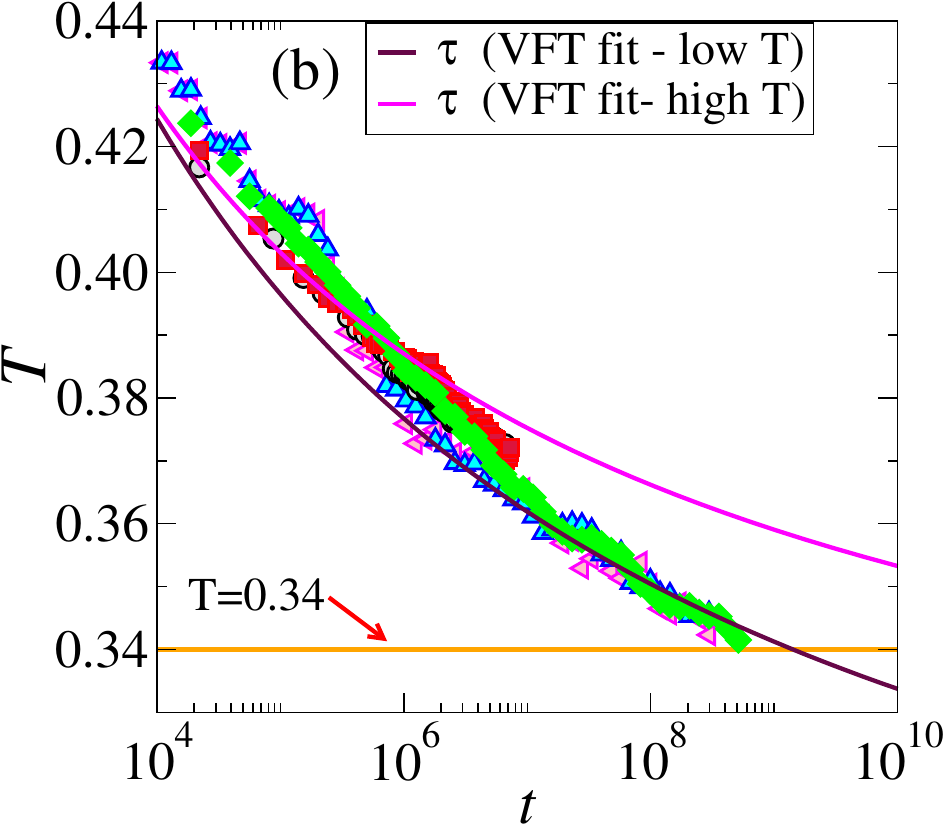} 
\includegraphics[scale=0.35]{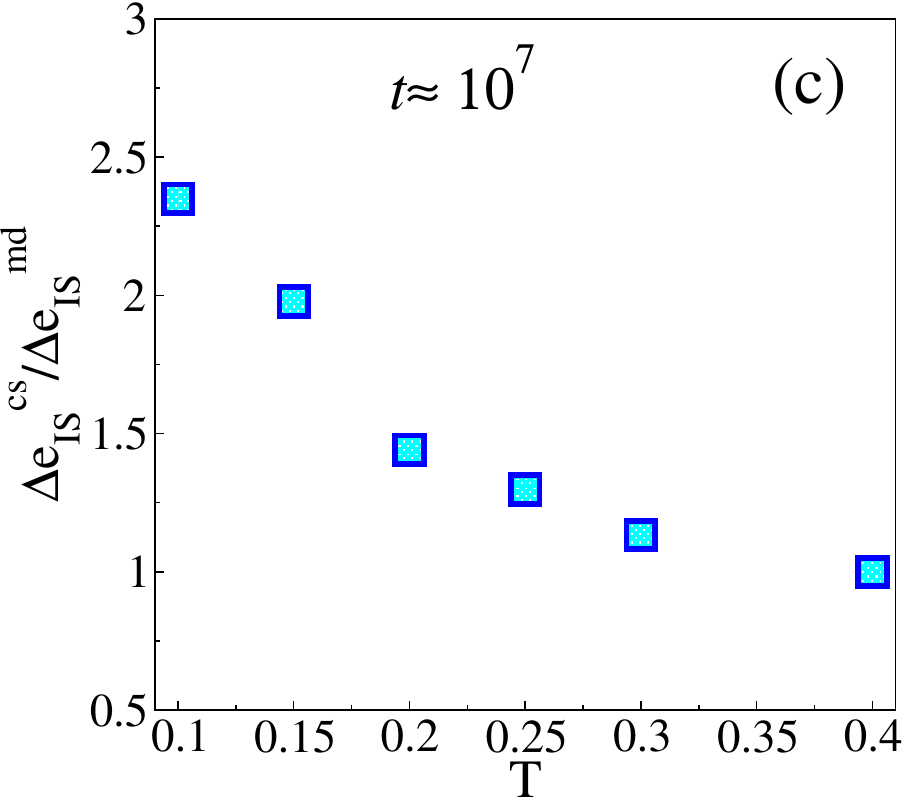}  

\caption{Inherent structure energies $e_{IS}$ {\it vs.} time for different shear rates and different damping parameters ($Q$), at the simulation temperature $T=0.3$. (\emph{b}) The energies $e_{IS}$ is  transformed to temperatures corresponding to those energies for equilibrated samples. Also shown are VFT fits to relaxation times obtained in MD simulations, (i) above $T = 0.435$ (magenta line), and (ii) for temperatures extending to $T = 0.365$ (maroon line). (\emph{c}) The ratio of the difference between the energy reached at $(t\sim 10^7)$ and the initial value, for MD and cyclic shear ($\Delta e_{IS}^{md}$ and $\Delta e_{IS}^{cs}$ respectively), as a function of the simulation temperature.  %\sri{Fix the figure}
%\sout{Also shown is the VFT relationship between the temperatures and relaxation times for comparison}
%\pd{Two extrapolated VFT curves are shown as they are fitted to two different temperature regimes (up to $T_{MCT}$ and up to substantially low temperatures below $T_{MCT}$)}. 
%(\emph{c})  \sout{The ratio of relaxation times $\tau$ corresponding to the state at a given time $t$ plotted against $t$ indicates a gain of roughly  $5$ orders of magnitude, and a strong growth in the gain vs simulation time}
%\pd{The ratio of $T_{md}$ and $T_{cs}$ at an equal time indicates the acceleration achieved to reach lower energy levels when cyclically sheared compared to the energies reached by evolving through thermal fluctuations only. (alternately: Temperature accessed through cyclic driving ($T_{cs}$) and equilibrium thermal fluctuation  ($T_{md}$) at long time $(t\sim 10^7)$ has been shown as a function of temperature $T$ at which the simulation is performed. At a higher temperature regime, the system reaches equilibrium within $(t\sim 10^7)$, but at lower temperatures, it progressively reaches higher temperatures. The difference between $T_{md}$ and $T_{cs}$ increases.)}
}

\label{Fig:com}
\end{figure*}

Though the optimal strain amplitude depends on the temperature and rate, we focus below on the amplitude $\gamma_{max}= 0.035$, the optimal value for strain rate $10^{-5}$ at $T = 0.3$. The evolution of the IS energy for different shear rates (Fig. \ref{Fig:com} (a)) shows that for higher shear rates the energies in the initial times are lower, but the rate of decrease is marginally larger for lower shear rates. In order to estimate the efficiency of the approach, we need a mapping of the inherent structure energies and temperatures. For this, we use the observation\cite{Sastry2001} that the inherent structure energies at low temperatures display the behaviour
\begin{equation} 
e_{IS}(T) = E_{\infty}-{\mathcal A}/{T},
\label{eq:eisvst}
\end{equation}
where $E_{\infty}$ is the extrapolated infinite temperature IS energy and $\mathcal A$ represent the slope of $e_{IS}$ {\it vs.} $1/T$ (see Appendix). We estimate the  equilibrium temperature corresponding to a given inherent structure energy using this relationship. As an extrapolation, however, the meaning of the temperature so estimated should be treated with due caution (see Appendix for further details). Next, we compute relaxation times corresponding to a given temperature using the Vogel Fulcher Tamman relation (VFT) expression $ \tau = \tau_0 \exp \left [{(K_{VFT}({T}/{T_{VFT}}-1))^{-1}}\right]$
where parameters $\tau_0$, $K_{VFT}$ and $T_{VFT}$ are obtained from fits to MD simulations\cite{DAS2022100098} 
%\sout{for temperatures above $T = 0.435$} 
(see Appendix). A comparison of the estimated temperature reached for a given simulation time with the relaxation time at that temperature provides a way of judging the extent to which cyclic shear may accelerate the accessing of low energy configurations. Fig. \ref{Fig:com} (b) shows the estimated temperatures {\it vs.}  simulation time, along with reference curves that indicates the dependence of the relaxation times on temperature.  A long run of duration \textcolor{black}{$t \sim 6\times10^8$} at $\dot\gamma = 10^{-5}$ shows the lowest temperature accessed is approximately \textcolor{black}{$T = 0.34$ ($e_{IS}\approx-7.08$}, to be compared with the lowest estimated value of $-7.15$; see Appendix). 
%\sout{Comparison with VFT times indicate a speed-up of close to \textcolor{black}{$5$} orders of magnitude. We next compute, as a function of time $t$ the corresponding relaxation time $\tau$ reached (through the estimated temperature), and show the ratio  $\tau/t$ {\it vs.} $t$ in Fig. \ref{Fig:com} (c), which indicates that the speed up is faster than linear, but slower than exponential (the form $\tau/t \sim exp(a~t^{1/4}$) describes the data reasonably). A number of factors may influence the observed speed-up and as an illustration, we shown in  Fig. \ref{Fig:com} data obtained from thermostatted SLLOD simulations but with smaller damping coefficients, for a higher shear rate of $\dot\gamma_{xy} = 10^{-3}$.}
%\sout{We observe that smaller damping coefficients lead to faster annealing, leading to a speed-up, for the simulations performed, of close to $5$ orders of magnitude.}
We show extrapolated VFT curves for two cases in Fig.\ref{Fig:com} (b). The first curve (magenta) is the VFT fit to relaxation times for temperatures above $T_{MCT} = 0.435$, which is extrapolated for lower temperatures. A comparison with this VFT fit leads to the conclusion that the relaxation to low energy configurations is significantly accelerated by the application of cyclic shear. To make a direct comparison, MD runs are performed at low temperatures (reported in \cite{DAS2022100098}), with the lowest temperature being $T=0.365$. These simulations reveal that the temperature dependence of relaxation times undergoes a crossover, with low temperature relaxation times being significantly smaller than the VFT extrapolation from temperatures above $T = 0.435$.  The relaxation times over this extended temperature range are also fitted to the VFT form, which is also shown in  Fig.\ref{Fig:com} (b) (maroon curve). A comparison of the cyclic shear results with results from low temperature MD simulations indicates that, contrary to the earlier conclusion, no significant acceleration is obtained through the application of cyclic shear. 

Although the application of cyclic shear deformation does not therefore appear promising as an approach to performing accelerated sampling at low temperatures, it does lead to overaging when simulations are performed at temperatures lower than $T = 0.3$ considered above. We thus consider a series of temperatures $T = 0.1, 0.15, 0.2, 0.25, 0.3, 0.4$, and perform cyclic shear at each temperature for a range of strain amplitudes $\gamma_{max}$ and shear rate $\dot{\gamma} = 10^{-5}$.  For each temperature, we consider the IS energies reached for a simulation time $t \approx 10^7$, and compare with the corresponding results for MD simulations. Results shown in the Appendix (Fig.s A4 - A6) %\sri{Confirm} 
reveal that the gap between the energies reached in MD simulations and cyclic shear simulations increases upon lowering the temperature. To quantify this, we compute the difference between the energy 
reached at  $t \approx 10^7$ and the initial value, $\Delta e_{IS}$, in MD and cyclic shear simulations respectively. The ratio $\Delta e_{IS}^{cs}/ \Delta e_{IS}^{md}$, shown in Fig. \ref{Fig:com} (c), indeed grows as the temperature decreases, indicating a greater degree of overaging induced by the cyclic shear deformation, the lower the temperature. 

\subsection{\label{sec:level2}Properties of sheared configurations}

%\sout{We next examine possible anisotropies and inhomogeneities in the structures we generate, since we shear the liquid in a particular shear plane, as well as possible crystallinity (see below). 
%From results shown in the Appendix, we conclude that the samples we generate are isotropic, homogeneous and free of significant crystallinity, and can be considered to be of similar nature to configurations  generated through normal molecular dynamics.} \\

Apart from the question of whether cyclic shear induces acceleration of relaxation, addressed in Sec. \ref{sec:level2} above, it is of importance to ascertain whether the properties of the sheared liquids are comparable to the equilibrium liquid. We perform a comparison of properties of the sheared liquid and that simulated with conventional MD. Results presented here and in the Appendix show that the configurations generated by cyclic shear are isotropic, homogeneous, and have properties that are indistinguishable from the equilibrated liquid configurations. We have chosen $T=0.4$ (and higher temperatures) for a comparison of the structures generated by conventional constant temperature MD and cyclic shear, at which the liquid can easily be equilibrated with moderate effort in a constant temperature molecular dynamics simulation and (as described later) cyclic shear simulations reach steady states within the simulated time window. For cyclic shear, we have kept our shear rate fixed at $\dot \gamma = 10 ^{-5}$ and based on short runs across different strain amplitudes, identify $\gamma_{max} = 0.005$ as being a reasonable choice based on data shown in Fig. \ref{fig:S5}, for comparison with MD simulations.

\begin{figure}[h]
\centering
\includegraphics[scale=0.38]{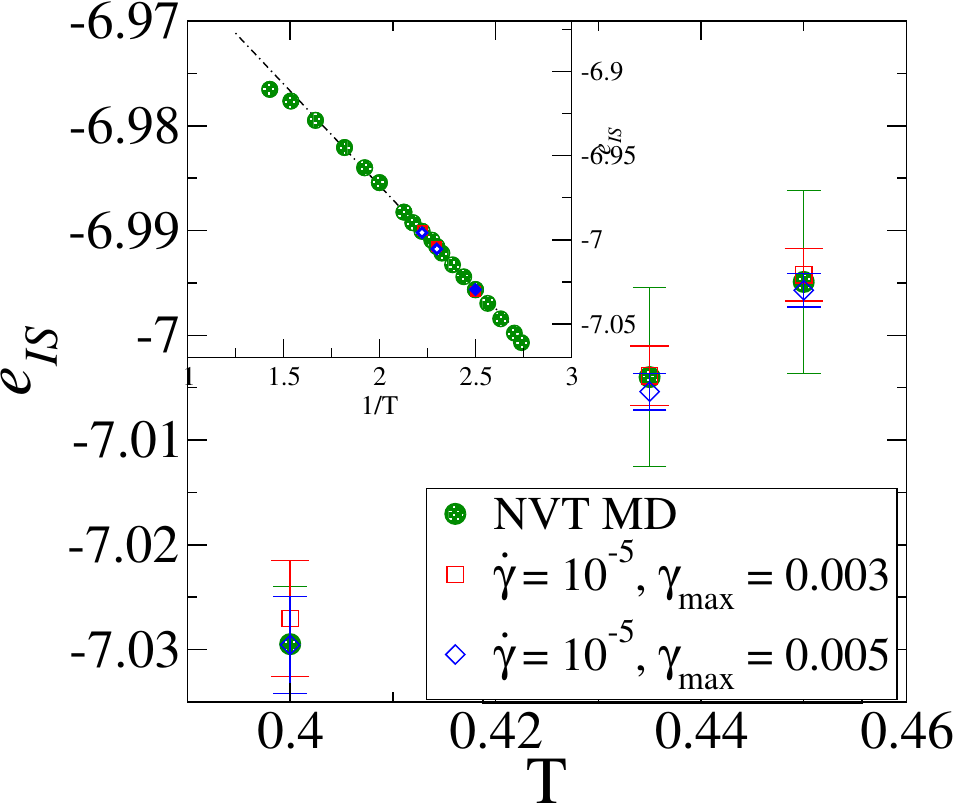}
\caption{The $e_{IS}$ energies obtained from the equilibrium MD is compared to the $e_{IS}$ energy obtained from sheared configurations at amplitudes  of $\gamma_{max} = 0.003,0.005$.  The energies show good quantitative agreement. The temperature varies in the range  $[0.4-0.45]$, where the system can be equilibrated easily by normal molecular dynamics. The shear rate has been kept fixed at $\dot\gamma = 10^{-5}$. The inset shows the inherent structure energies against inverse temperature, indicating that the energy obeys a $1/T$ dependence on $T$. }
\label{fig:eis_dist}
\end{figure}

In Fig. \ref{fig:eis_dist} we compare the average inherent structure energy obtained from molecular dynamics simulations and finite temperature, finite shear rate, cyclic deformation, as a function of temperature, and show that the energies in these cases are comparable to each other and follow $1/T$ behaviour in the low temperature range.

\begin{figure}[htp]
\centering
\includegraphics[scale=0.28]{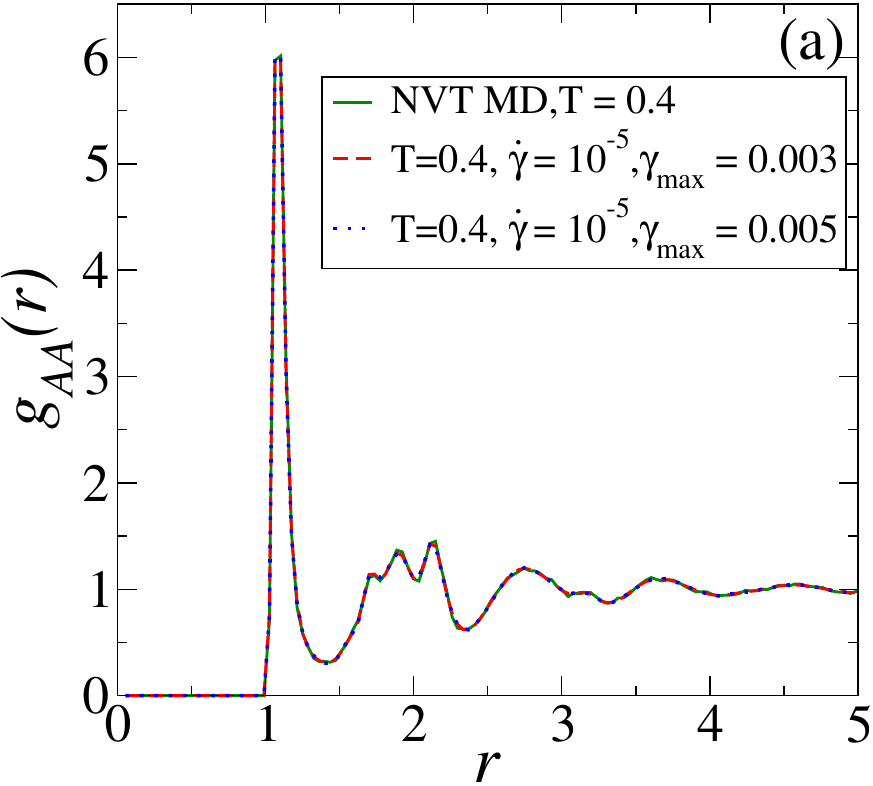}
\includegraphics[scale=0.28]{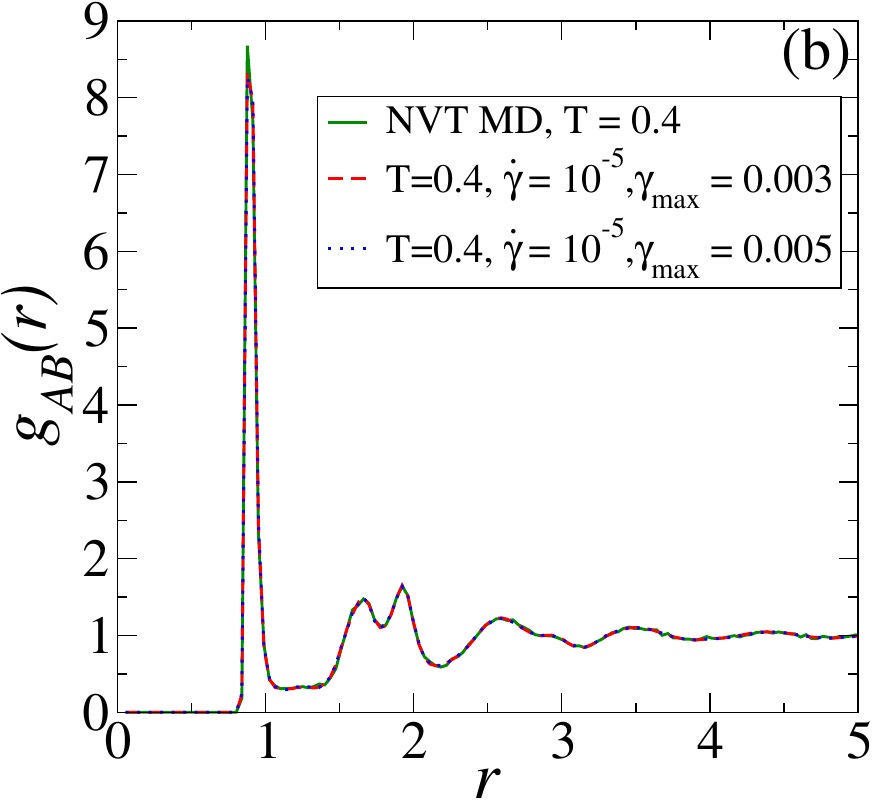}
\includegraphics[scale=0.28]{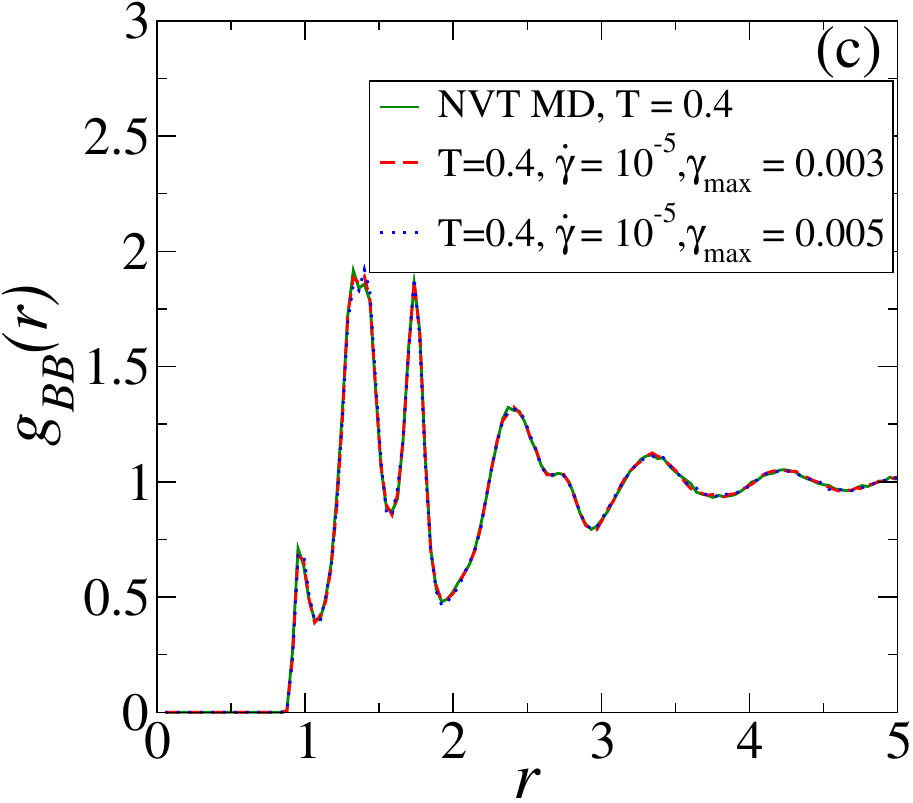}
\caption{ \textcolor{black}{The partial radial distribution functions of the IS configurations obtained through equilibrium MD and cyclic shear at a high enough temperature, ($T=0.4$) where the system can be equilibrated easily through conventional MD. The data  show that there is no significant structural difference between the configurations generated from the two approaches.}}
\label{fig:eqMDcomp_rdf}
\end{figure}

Fig. \ref{fig:eqMDcomp_rdf} shows that the partial radial distribution functions of the inherent structures obtained by NVT MD and cyclic shear simulations agree with each other quantitatively, which implies that the structures generated by cyclic shear are the same as those generated by MD.

\begin{figure}[h]
\centering
\includegraphics[scale=0.29]{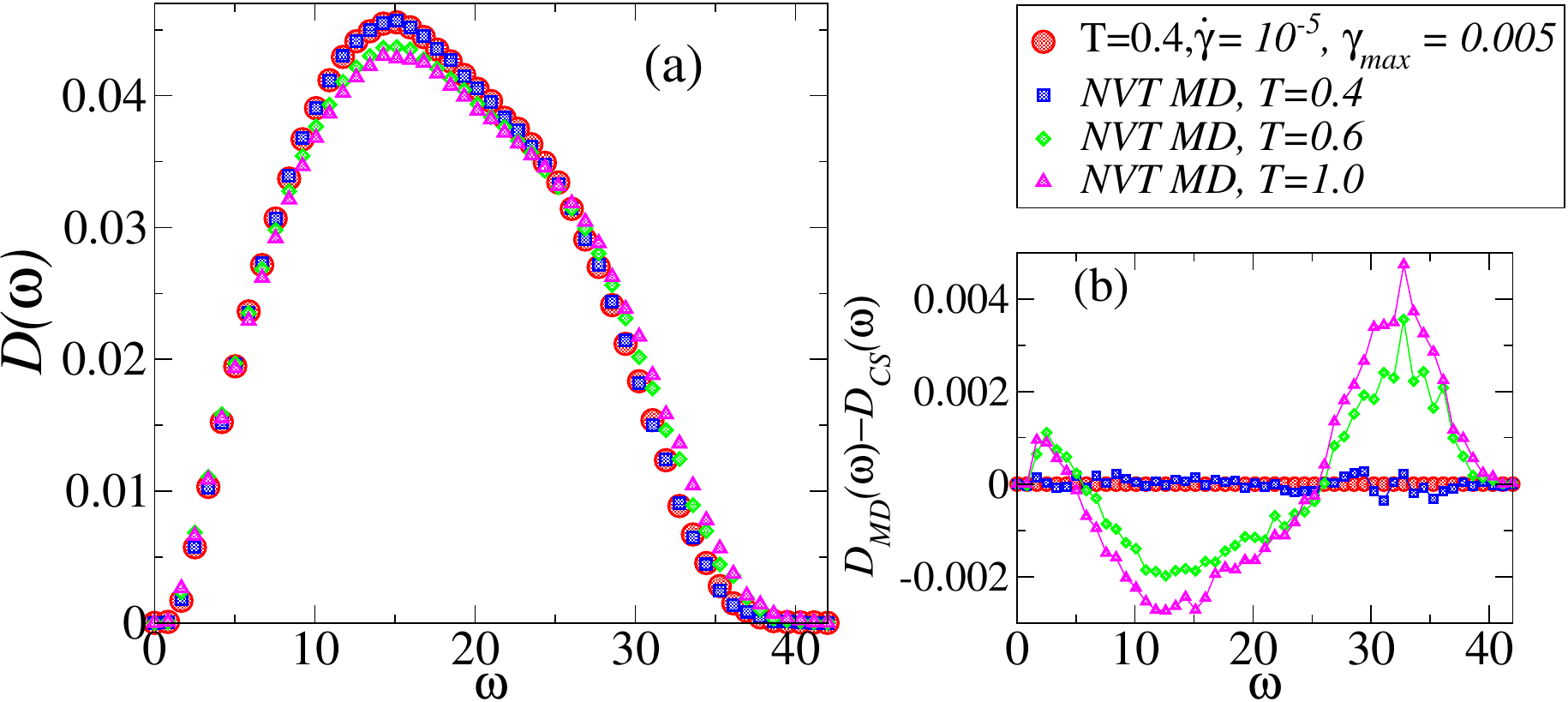} 
\caption{ \emph{(a)} Vibrational density of states (DOS) for cyclically sheared configurations at $T = 0.4$, and configurations generated by molecular dynamics at $T = 0.4, 0.6, 1.0$. The DOS at $T = 0.4$ are indistinguishable for molecular dynamics and cyclic shear, whereas they are clearly different from those at $T = 0.6, 1.0$ obtained from MD simulations. (b) The difference of the DOS at different temperatures from the cyclically sheared configurations at $T = 0.4$, indicating that MD and cyclic shear results at $T = 0.4$ are indistinguishable.}
\label{fig:DOS}
\end{figure}

We  also show, in Fig. \ref{fig:DOS}, that the vibrational density of states of configurations obtained by NVT MD and cyclic shear at $T = 0.4$ are the same, whereas they are clearly different from those of configurations generated by MD at higher temperatures.

\begin{figure}[]
\centering
\includegraphics[scale=0.25]{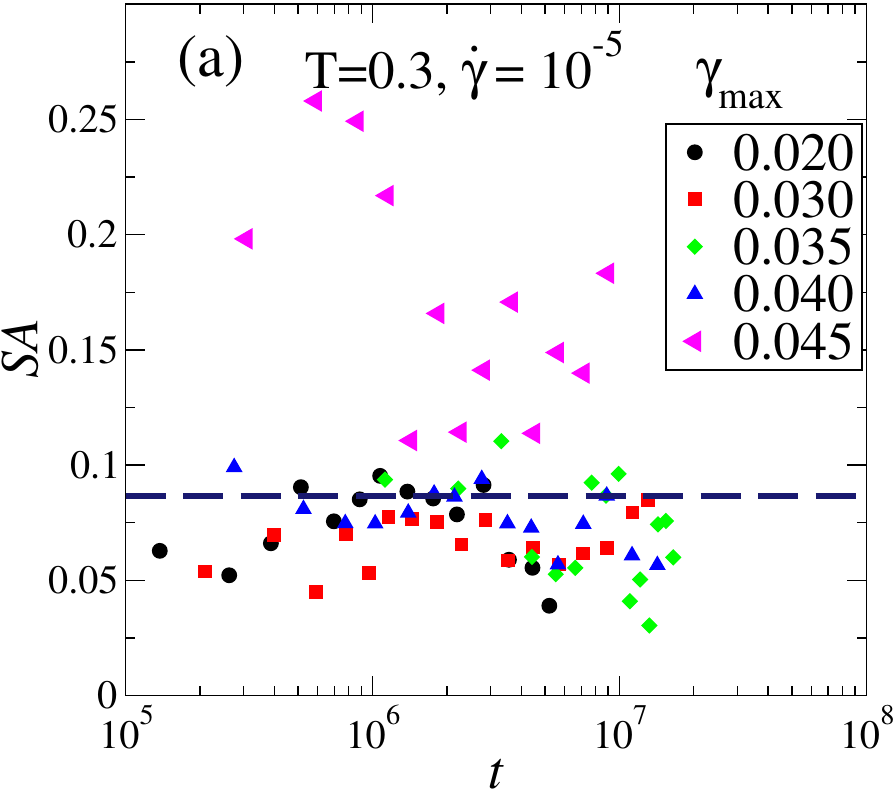}
\includegraphics[scale=0.25]{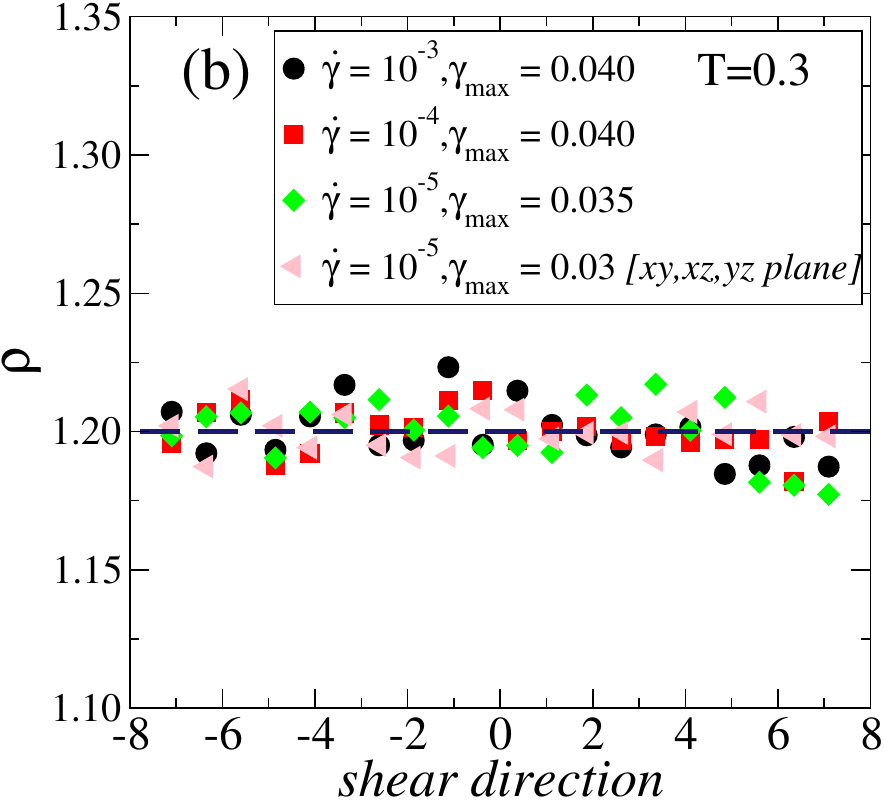}
\caption{ 
(\emph{a}) For strain amplitudes lower than or close to the yielding amplitude, stress anisotropies for strain rate $10^{-5}$ compare with those of isotropic inherent structures (indicated by the horizontal line). 
(\emph{b}) Variation of density in the shear direction indicating that the system is homogeneous in all cases.}
 \label{Fig:aniso}
\end{figure}

We next examine possible anisotropies in the structures we generate.  The ultrastable glasses produced by physical vapor deposition (PVD) method have shown features that can be connected to more anisotropic packing, compared to its ordinary glass counterpart formed by cooling the liquid. Since we perform shear in a given plane, anisotropies are also possible in our case. 
%\sou{We also consider the effect of shearing in alternating shear planes ($xy$, $xz$, and $yz$, repeated after 3 cycles).  As shown in Fig. \ref{Fig:aniso} (a) the generated energies are essentially the same (however, it has been reported that for larger systems at lower temperatures and shear amplitudes, alternating shear planes leads to better annealing \cite{PRIEZJEV2019}.} 
To characterise anisotropy and inhomonegeity, we first consider the stress anisotropy, which is defined as  $SA = (S_1-S_3)/(S_1+S_2+S_3)$, where $S_{i}$ ($S_1>S_2>S_3$) are the eigenvalues of the stress tensor. Excepting for very large strain amplitudes, we find the stress anisotropies to be small, and comparable to those of inherent structures quenched directly from liquid configurations, as shown in Fig. \ref{Fig:aniso} (a). We also test for the possibility of shear localisation accompanied by inhomogeneities in the local density. Density values obtained for slabs in the shear direction show no evidence of density inhomogeneities (as they do beyond the yield strain in AQS simulations \cite{anshul-shearbanding}, although for larger samples), as shown in Fig, \ref{Fig:aniso} (b).
%The measured stress anisotropies, density profiles and pair correlation functions, as shown in the SI Appendix, reveal no anisotropy or inhomogeneity in the configurations we generate as compared with configurations obtained from normal molecular dynamics simulations.  We also examine the degree of crystallinity (see below) and ensure that the samples we analyse are free of significant crystallinity.

Additional comparisons presented in the Appendix (A.5 and A.6) show that the IS configurations obtained by cyclic shear are isotropic, and that they exhibit the same response to shear (in the form of the distribution of strain required to undergo a plastic rearrangement) as those obtained from conventional MD. Finally, in Appendix A.7 we show results from applying shear in alternating shear planes ($xy$, $xz$, and $yz$) which does not lead to significant change in relaxation of the energy, or the stress anisotropy.

%These two curves show significant differences as it has been reported\cite{DAS2022100098} there is a non-Arrhenius to Arrhenius-like crossover across $T_{MCT}$. VFT prediction overestimates relaxation times at lower temperatures. Hence VFT fits at two different temperature regimes show a significant difference when extrapolated at much lower temperatures. The evolution of the temperature for the case when the lowest temperature has been accessed through cyclic shear follows the second VFT form closely at the low temperature regime. }

%The data shown clearly indicates that the gain in time is faster than linear. Attempts to fit the functional form indicate that $\tau$ values vary faster than a quadratic function of $t$ for the lowest shear rate (but always vary faster than linear) and slower than an exponential function (as is clear from the Fig. \ref{Fig:com}(c)) though the exact form is difficult to ascertain owing to the quality of the data. In any event, the data shown clearly indicates that the gain in time grows faster than linear with the simulation times. In the simulations we perform, it reaches three orders of magnitude. 

\subsection{\label{sec:level3}Equilibrium sampling}

%To summarise results thus far, we have explored the possibility of generating low energy configurations by the application of oscillatory shear deformation, and have identified optimal conditions for doing so, namely temperatures close to the Kauzmann temperature ($T \approx T_K$), strain amplitudes close to yield strain ($\gamma_{max} \approx \gamma_y$) and small strain rates. And we have shown that the resulting configurations are isotropic and homogeneous. 

The results described above illustrate that configurations generated by cyclic shear, under the right conditions, display properties that are indistinguishable from those obtained from MD simulations, an observation that is extremely useful (see also \cite{BPBhowmik2020}). Nevertheless, it does not guarantee that an equilibrium ensemble is always generated, which is desirable. %We provide an affirmative answer to this question, in two ways. First, we show empirically that the configurations we generate are indistinguishable from those obtained through a normal molecular dynamics (MD) simulation. Second, 
In order to achieve this, we develop an equilibrium Monte Carlo sampling scheme which guarantees  equilibrium sampling, and demonstrate that it can be implemented as an efficient method and generates equilibrium ensembles of configurations.  

%We generate inherent structures with energies smaller than any that have been generated by conventional simulations, reaching within $15\%$ of $T_K$, with an estimated simulation time gain of $3$ orders of magnitude. Whether such gains may be enhanced further to make the studied approach competitive with other methods currently probed remains to be seen, and will depend on a better understanding of factors governing the speed-up. But our results clearly demonstrate that cyclic deformation is among the attractive options available for accelerated sampling, and at the very least, it may be interesting to develop approaches for using it in tandem with other acceleration schemes. The presented method has the advantage that it does not depend on a particular choice of system, and is capable of generating bulk samples. A potential concern may be that such a procedure will generate anisotropic structures, but we  have shown that such is not the case. Future investigations, we hope, will shed light on the degree to which this method can be fine tuned to maximize performance, and on the reasons why cyclic deformation leads to an accelerated approach to low energy configurations. Further, the procedure followed here can easily be modified, in the spirit of hybrid Monte Carlo \cite{Duane1987}, employing  a generalization applicable to non-equilibrium systems \cite{Nilmeier2011}, to obtain equilibrium sampling at a desired temperature.
% investigation of which is under progress.

\begin{figure*}[]
\centering{}
\includegraphics[scale=0.37]{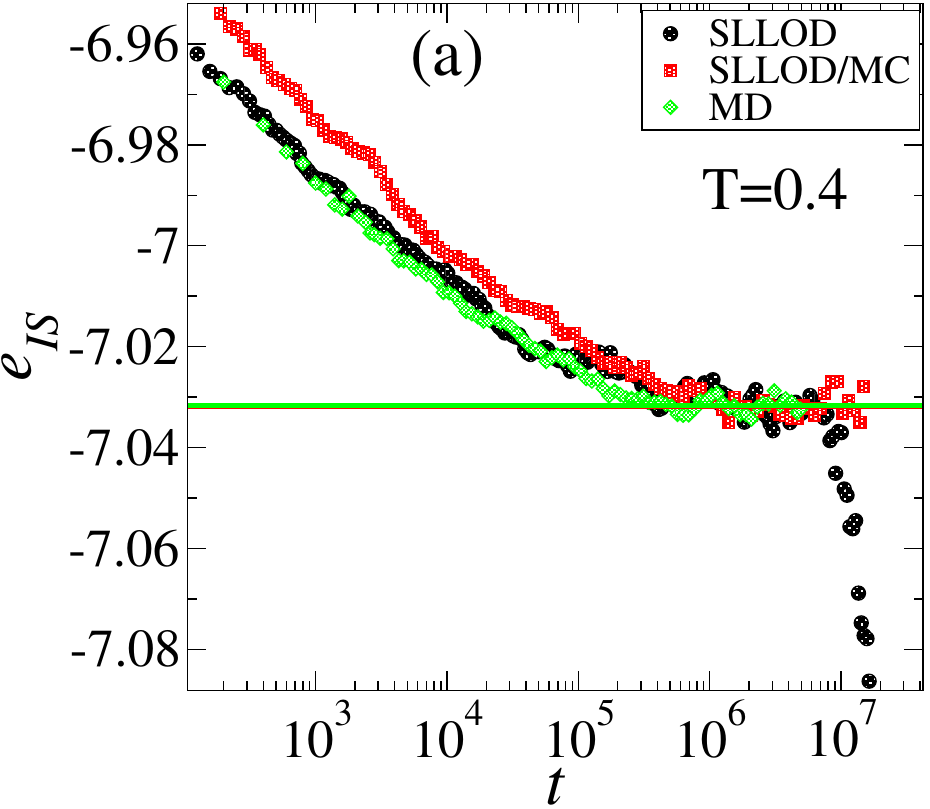} 
\includegraphics[scale=0.37]{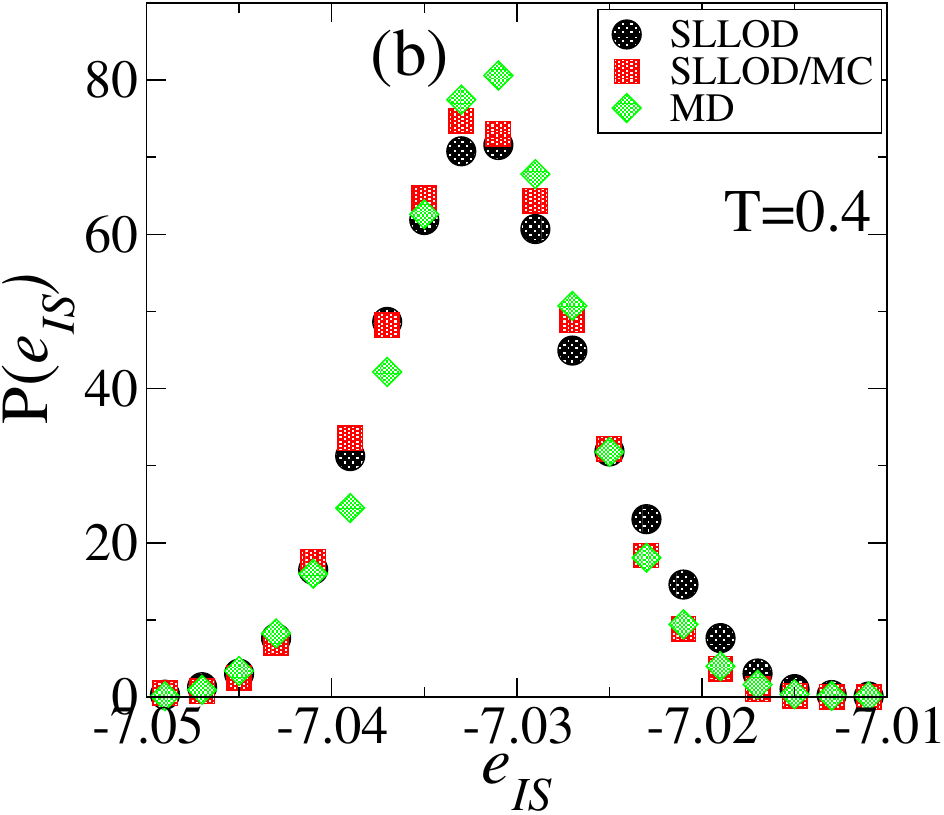}
\includegraphics[scale=0.37]{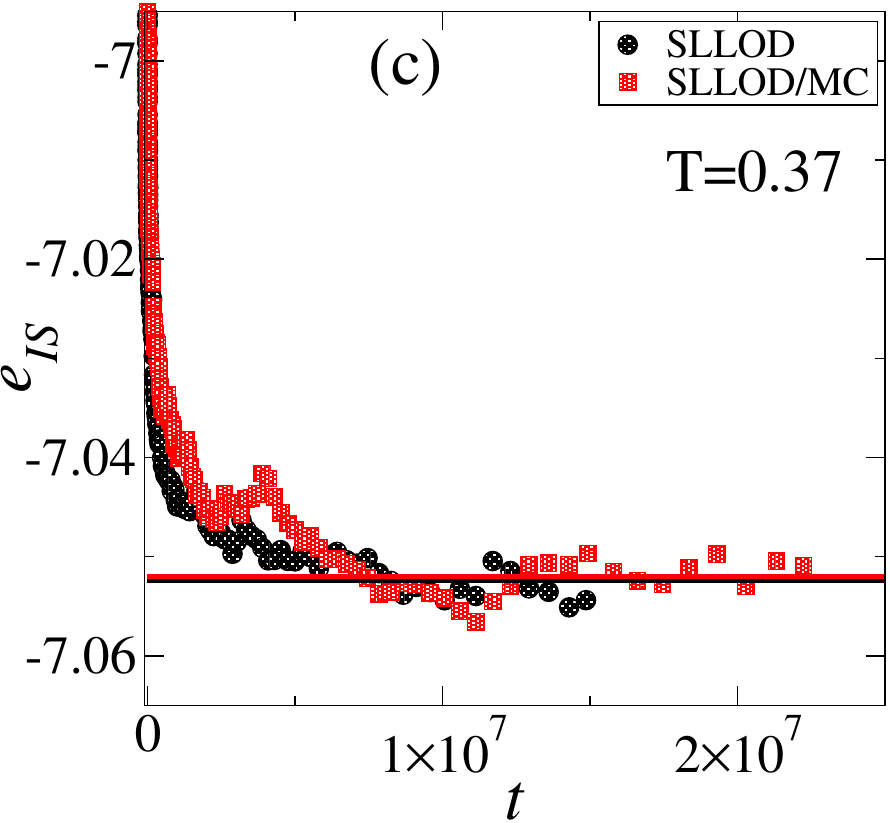}
\caption{ {\emph{(a)} Time evolution of the inherent structure energies as the system, at temperature $T=0.4$ (at which the system can be easily equilibrated using normal molecular dynamics), is sheared at a fixed amplitude $\gamma_{max} = 0.005$ and shear rate $\dot{\gamma} = 10^{-3}$, without, and with Monte Carlo sampling. Note that the sheared system without Monte Carlo sampling crystallizes at long times. {\emph{(b)}} The inherent structure energy distribution is shown for the MD and sheared configurations at $T=0.4$ after the system reaches equilibrium  or steady state respectively. The distribution of IS energies from the sheared simulations and MD simulations are indistinguishable. {\emph{(c)}} Time evolution of the inherent structure energies for sheared systems without and with Monte Carlo sampling at $T = 0.37$. Horizontal lines in (a) and (c) indicate average values evaluated in a steady state window of time.}}
\label{fig:MCcomp}
\end{figure*}

In order to compare with equilibrium properties, we consider $T = 0.4$, at which it is relatively easy to equilibrate using a conventional MD (NVT) simulation. Based on results shown in the Appendix, we choose a strain amplitude $\gamma_{max} = 0.005$ and perform cyclic shear at $\dot{\gamma}_{xy} = 10^{-3}$ and compare with molecular dynamics simulations, both using the Gaussian iso-kinetic thermostat.   Fig. \ref{fig:MCcomp} (a) shows the evolution of the IS energy for sheared simulations as well as for molecular dynamics, averaging results over $5$ independent simulations.  
%Although at this high temperature cyclic shear does not offer any improvement in speed, 
We note that sheared simulations reach steady state values that correspond to the expected equilibrium value.  
As shown in Sec. \ref{sec:level2}, this is true for a range of temperatures, and further, structural and vibrational properties of sheared and MD configurations are indistinguishable. 
In one of the shear simulations, the energies drop to significantly lower values at longer times, as also shown in Fig. \ref{fig:MCcomp} (a). This results from crystallisation\cite{PhysRevXCryst2019}, as we confirm from analysis presented in the Appendix (A.9). We thus check all simulation results to ensure that the analysed trajectories are free of a significant degree of crystallisation. Fig. \ref{fig:MCcomp} (b) shows that the distribution of energies in the steady state for the cyclic shear simulations is identical to the MD trajectory. 

Next we describe an equilibrium sampling algorithm employing cyclic shear, which is in the spirit of hybrid Monte Carlo (HMC) \cite{Duane1987}. In hybrid Monte Carlo, a molecular dynamics trajectory is employed as a 'generation step' for a trial move. At the end of each such generation step, a Metropolis Monte Carlo acceptance of the resulting final state is attempted. As with most Monte Carlo algorithms, one must obey detailed balance. In the original hybrid Monte Carlo algorithm, detailed balance is guaranteed by the use of Hamiltonian dynamics, which is time reversible, and phase space volume preserving. However, nonequilibrium simulations, such as ours do not obey these properties. Although SLLOD equations preserve phase space volume and are reversible,  thermostatting leads to a compression of phase space volume \cite{ToddDaivisBook} (see Appendix for details). We modify the HMC acceptance probability by compensating for the phase space volume contraction. Representing a point in phase space by $\Gamma$, and with the {\it phase space compression factor}  $\Lambda(\Gamma) \equiv  {\partial \over \partial \Gamma} . \dot{\Gamma}$, the change in phase space volume is given by 
\begin{equation}
C(t) =  \exp\left( \int^{t} \Lambda(s) ds \right) = e^{\Delta Q \over k_B T}
\label{eq:compression}
\end{equation}
where $\Delta Q$ is the change in heat (heat dissipation $Q_d = -\Delta Q$). Thus, the acceptance probability for a trial phase space point $\Gamma^{'}= (x^{'},p^{'})$ starting with $\Gamma = (x,p)$, that ensures detailed balance, is 

 \begin{equation} 
 p_A (\Gamma \rightarrow \Gamma^{'}) = min\{1,  exp(-\beta_s \Delta U) exp(\beta_s \Delta Q)  \}
 \label{eq:accept}
 \end{equation}
 \begin{equation}\nonumber
 ~ = min\{1,  exp(-\beta_s \Delta W)   \}
\end{equation}
\begin{equation}\nonumber 
~ = min\{ 1, \exp{(-\beta_s \Delta \Phi )  \exp(-\beta_t \Delta K)  \exp(\beta_t \Delta Q)   } \}
\end{equation}

\noindent where $U$ is the internal energy, $\Phi$ the potential energy, $K$ the kinetic energy, $\Delta W$ the work done, $T_s = 1/k_B \beta_{s}$ the sampling temperature. The final, general, expression holds even when the thermostat temperature $T_t = 1/ k_B \beta_t$ is not the same as the sampling temperature $\beta_s$, a possibility permitted and exploited in HMC schemes. The same acceptance rule can be arrived at, for systems that obey microscopic reversibility, using Crooks theorem \cite{Crooks1998}, as discussed in \cite{Nilmeier2011}. The resulting algorithm in the present case involves the following steps each cycle: (1) Generate velocities from a Maxwell distribution corresponding to the temperature at which we perform the SLLOD cyclic deformation ($\beta_t$, $=\beta_s$ in this work), for each cycle regardless of the acceptance of the previous trial configuration. (2) Perform a cycle of shear deformation with the thermostatted SLLOD equations of motion. (3) Compute the change in total energy, heat or equivalently, the work done during the cycle. (4) Accept or reject the final configuration according to the acceptance probability in Eq. \ref{eq:accept}. Note that the heat can be computed by integrating $\Lambda$ as in Eq. \ref{eq:compression}, which in turn is given in terms of the friction coefficient $\alpha$ used in thermostatting the dynamics, as $\Lambda = -3N\alpha$ where $N$ is the number of particles. The work done can be computed by integrating the stress $\sigma$ over the cycle: $\Delta W = \int^t  V \mathbf{\sigma}_{xy}^{T}(s)(s) \dot{\gamma}_{xy}(s) ds$ where $V$ is the volume, $\dot{\gamma}_{xy}$ is the shear rate.

We next describe results that demonstrate that the algorithm described generates an equilibrium ensemble effectively. We first consider the case $T = 0.4$, with $\gamma_{max} = 0.005$,  $\dot\gamma_{xy} = 10^{-3}$, considered earlier. Fig. \ref{fig:MCcomp} shows IS energies, averaged over five independent runs, with Monte Carlo sampling, in addition to the MD and shear results  described above. We find that indeed, the equilibrium sampling results agree with the molecular dynamics and cyclic shear simulations without equilibrium sampling. We next consider shear simulations without and with equilibrium sampling at $T = 0.37$, at which molecular dynamics simulations are harder, since the relaxation time is $\tau \sim 2.3 \times 10^6$. %\sri{Put the correct value that you now know}. 
Results shown in  Fig. \ref{fig:MCcomp} (c), indicate that indeed, the shear simulations reach steady state values of the IS energy close to $-7.05$, which is the expected value based on Eq. \ref{eq:eisvst} and obtained from MD simulations reported in \cite{DAS2022100098}. 
%\sri{Correct? Confirm}. confirmed
%The steady state values are reached around $t \sim 5 \times 10^6$, indicating a speed up of roughly a decade at this temperature. 
Although for results presented thus far, cyclic shear without Monte Carlo sampling also generates results indistinguishable from cyclic shear with Monte Carlo sampling, this need not be the case. In the Appendix, we show that energies obtained under large amplitude shear deviate from the expected equilibrium values, whereas with Monte Carlo sampling, values close to expected equilibrium values are generated. 

We note from Eq. \ref{eq:accept} that the acceptance probability depends on the (extensive) work done during a cycle and not the change in energy. This poses a potential problem for large systems. Indeed, as shown in the Appendix, we find that for the case considered so far,  $N = 4000$, the acceptance rate is negligible with $T = 0.3$, $\gamma_{max} = 0.03$,  $\dot\gamma_{xy} = 10^{-3}$. For $N = 500$, the work done fluctuates around zero (with a small positive mean), and the acceptance rate is sufficiently large. 

\begin{figure}[htp]
\centering{}
\includegraphics[scale=0.26]{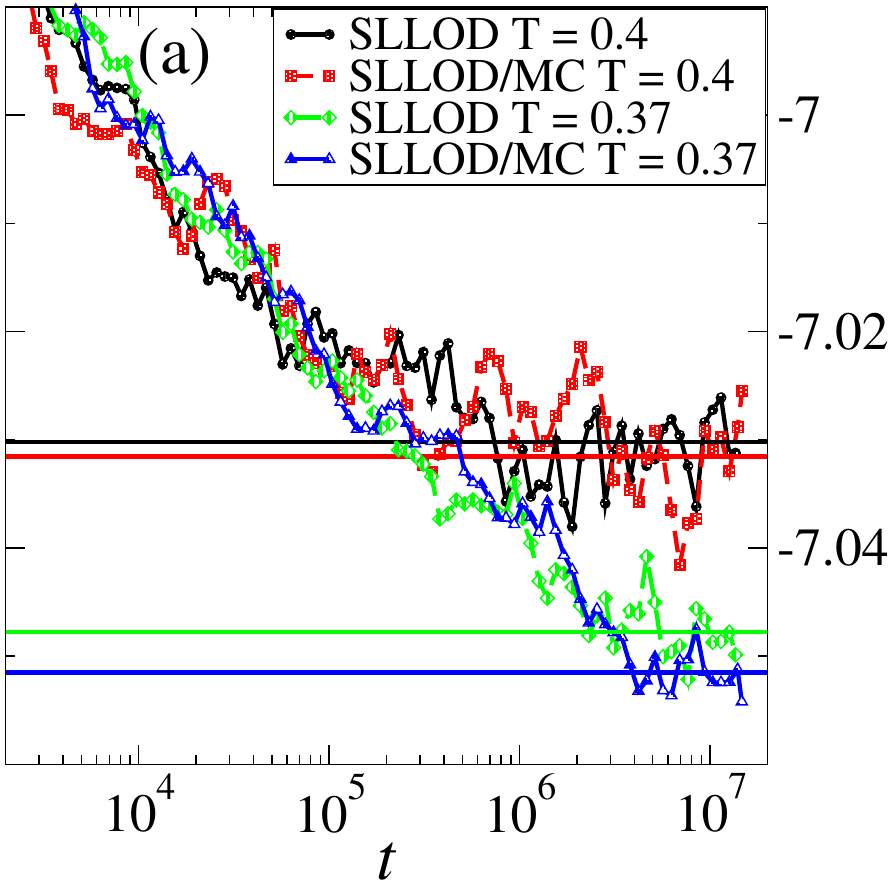}
\includegraphics[scale=0.26]{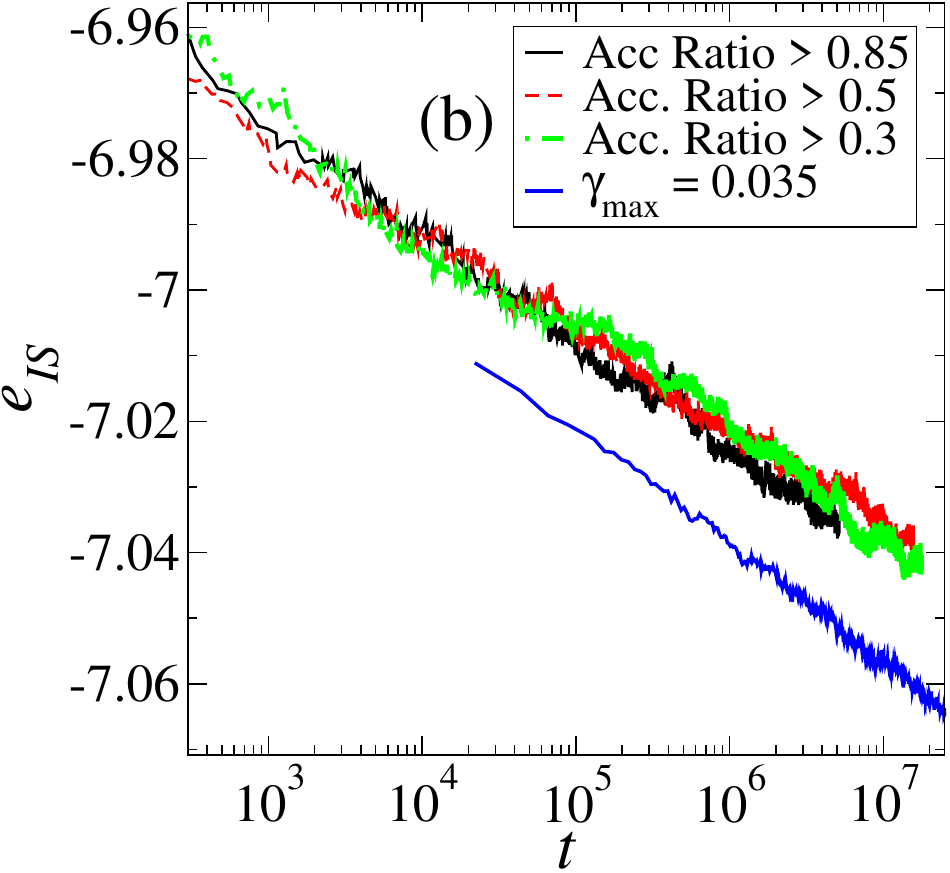}
\caption{Equilibrium sampling for a system of $N = 4000$ particles by performing cyclic shear (a) for a randomly chosen subvolume and (b) employing adaptive modification of shear amplitude.}
\label{fig:N4000sa}
\end{figure}

We discuss two strategies that lead to feasible acceptance rates for larger systems and demonstrate them for $N = 4000$. 
The first strategy is to perform shear for a randomly chosen subvolume of a size that leads to feasible acceptance rates. 
The second strategy is to adaptively modify the strain amplitude to achieve a desired acceptance rate. In implementing the first strategy, for each cycle, we randomly pick the centre of a cubic subvolume of pre-specified size. During the shear cycle, the rest of the system is frozen and strained affinely, whereas the subvolume is subjected to SLLOD dynamics at the chosen temperature and shear rate. At the end of the cycle, the entire system is simulated with normal unsheared molecular dynamics for a specified duration. We choose the subvolume to be of linear dimension equal to half the box size $L$ ($1/8$ of the total system) and the duration of molecular dynamics in the second step to be $1/10$ of the cycle period. As before, we perform cyclic shear simulations for $T = 0.4$ and $T = 0.37$,
with $\gamma_{max} = 0.005$ and  $\dot\gamma_{xy} = 10^{-3}$. A single trajectory is simulated in each case. Fig. \ref{fig:N4000sa} (a) shows that indeed this method works efficiently to equilibrate the system in both cases to the correct value of the inherent structure energy. Though the following observations need to be verified by better averaging, we note that cyclic shear without sampling results in higher energies for $T = 0.37$. With sampling, the asymptotic range of energies is reached by $t \sim 2.5 \times 10^6$, indicating a factor of $10$ speed up. As earlier results demonstrate, however, the speed-up increases dramatically with a lowering of the target temperature. The goal here, however, has been to present evidence of effective equilibrium sampling.

In implementing the strategy of adaptive strain amplitude, we choose a cutoff for the acceptance ratio, which is evaluated over $10-20$ cycles. If the acceptance ratio is below the cutoff, the strain amplitude $\gamma_{max}$ is reduced, whereas if it is above, it is increased. We test this approach for $T = 0.3$, choosing three different cutoff values of $0.3$, $0.5$ and $0.85$. The  $\gamma_{max}$ fluctuate around values $0.015$, $0.01$ and $0.001$ in these cases. As shown in Fig. \ref{fig:N4000sa} (b), in each case, one obtains a logarithmic relaxation of the energy as in the earlier case of $\gamma_{max} = 0.035$ but less efficiently by roughly a factor of $10$. An adaptive strategy that targets the degree of annealing, rather than acceptance rate, may thus be more desirable, and needs to be explored.

%Thus, we first show results for $N = 500$. We then discuss two strategies that lead to feasible acceptance rates for larger systems and demonstrate them for $N = 4000$. The first strategy is to adaptively modify the strain amplitude to achieve a desired acceptance rate. The second strategy is to perform shear for a randomly chosen subvolume of a size that leads to feasible acceptance rates. 

%We first perform simulation for a small system with $N = 500$, for which we find that the work done is always small enough to ensure reasonable acceptance rates. We consider two temperatures, $T = 0.4$ and $T = 0.37$, at both of which long enough MD simulations can be run to compare ($\tau(T=0.37) \sim 2.5 \times 10^7$). We thus generate distributions of IS energies (expected average $E_{IS}(T=0.4) = -7.028$ and $E_{IS}(T=0.37) = -7.052$) through different protocols and show, in Fig. \ref{fig:N500}, that (a) the equilibrium sampling algorithm generates the equilibrium distribution of energies, regardless of the strain amplitude employed, and (b) cyclic shear without sampling also generates the correct distribution for small strain amplitudes but not large amplitude shear. 

\bigskip
% Will need rewriting based on final graphs and data. 
\subsection{\label{sec:level1}Discussion and Conclusions}

To summarise, we have explored the possibility of generating low energy configurations by the application of cyclic shear deformation, more efficiently than conventional MD simulations. In the range of temperatures ($T \ge 0.365$) where equilibrium MD simulations can be performed for the model glass former studied \cite{DAS2022100098}, our results indicate that cyclic shear does not lead to an accelerated approach to low energy configurations. Nevertheless, our results show that cyclic deformation at low temperatures (below the estimated ideal glass transition) leads to accelerated relaxation to low energy structures, the degree of which increases with decreasing temperature. These results may help rationalise conflicting results regarding overaging of glasses 
\cite{lequeux,ediger_nooveraging} and are of relevance for understanding mechanically induced aging of glasses in general. Our results demonstrate that, at the temperatures investigated, cyclic shear simulations generate configurations that are indistinguishable from those generated by conventional MD simulations. 
%and have identified optimal conditions for doing so, namely temperatures close to the Kauzmann temperature ($T \approx T_K$), strain amplitudes close to yield strain ($\gamma_{max} \approx \gamma_y$) and small strain rates. We generate inherent structures with energies smaller than any that have been generated by conventional simulations for the model system we study, reaching within $10\%$ of $T_K$, with an estimated simulation time gain of close to \textcolor{black}{$5$ orders of magnitude}. 
%Our results clearly demonstrate that cyclic deformation is an attractive option for accelerated sampling. The presented method has the advantage that it does not depend on a particular choice of system, and is capable of generating bulk samples. A potential concern may be that such a procedure will generate anisotropic structures, but we show that such is not the case. For the lowest energies obtained in this work, the speed-up is competitive in comparison with other generally applicable methods ({\it e.g.} \cite{AnshulMetGlass2020}) but ways of optimising performance based on the observations made in the present work need to be further pursued. Future investigations, we hope, will also shed light why cyclic deformation leads to an accelerated approach to low energy configurations. 
We have further developed an equilibrium sampling method based on cyclic shear as a trial generating step, and demonstrated that it works effectively to generate equilibrium ensembles of configurations. We have presented strategies that make it generally feasible to employ the algorithm, and demonstrated their effectiveness.  This algorithm also suggests a general methodology for employing driving of various kinds for accelerating dynamics, with generalised cyclic deformation, including bulk strain, being an obvious example. Under what conditions such driving may result in accelerated sampling remains an open question to be understood.

%It will also be of interest to apply these methodologies to other systems, such as biomolecular assemblies, where overcoming hierarchies of barriers poses challenges. The observation that for moderate strain amplitudes, cyclic shear generates an equilibrium ensemble even without Monte Carlo sampling, implies that the method can be more widely used with a judicious choice of protocol, including experimentally. A significant outstanding problem is the estimation of dynamical quantities from such accelerated sampling methods\cite{Tiwary2013}. 

% \section*{Acknowledgments}

\noindent{\bf Acknowledgements:} We gratefully acknowledge Arunkumar Bupathy, Rajneesh Kumar and Himangsu Bhaumik for help with some parts of the computations, Daan Frenkel, H A Vinutha, Itamar Procaccia, Francis Starr and Francesco Sciortino for discussions and TUE-CMS, SSL, JNCASR, and the National Supercomputing Mission,  (Param Yukti) at the Jawaharlal Nehru Centre for Advanced Scientific Research (JNCASR), for computational resources. SS acknowledges support through the JC Bose Fellowship (JBR/2020/000015) from the Science and Engineering Research Board, Department of Science and Technology, India.

%\showacknow{} % Display the acknowledgments section

% Will need rewriting based on final graphs and data. 
\part*{}   %%%% SI begins
\setcounter{equation}{0}
\setcounter{figure}{0}
\renewcommand{\thefigure}{A\arabic{figure}}%

\appendix
\section*{Appendix}
\subsection{Characterization of the optimal strain amplitude}
\begin{figure*}[]
 \centering
\includegraphics[scale=0.35]{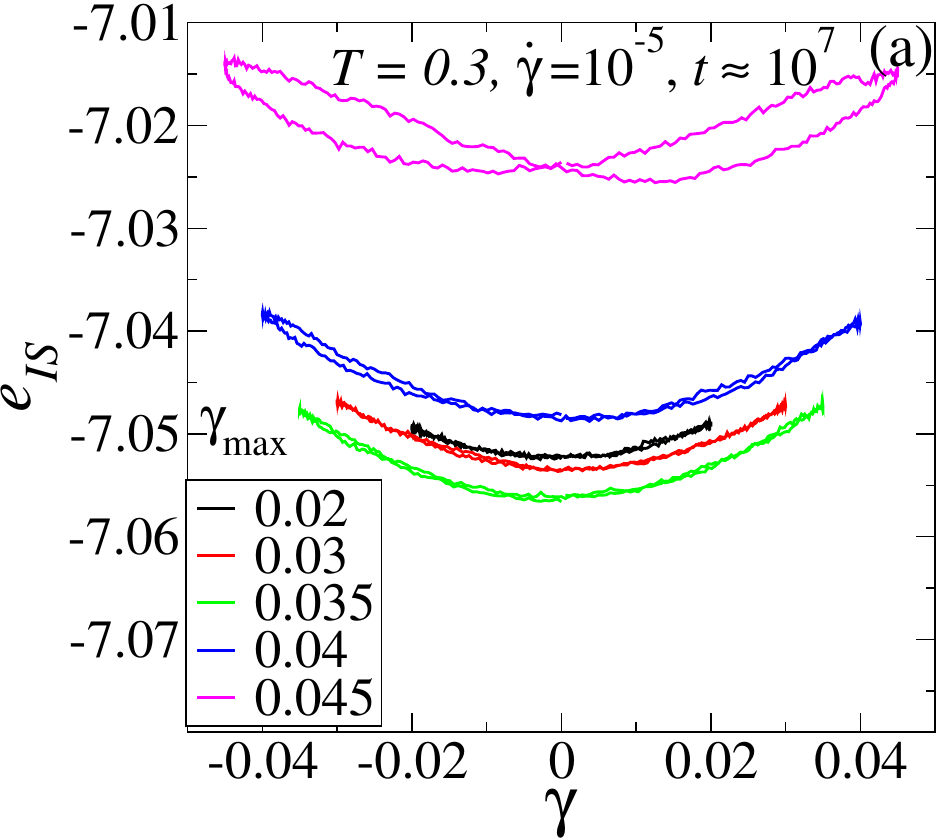} 
\includegraphics[scale=0.35]{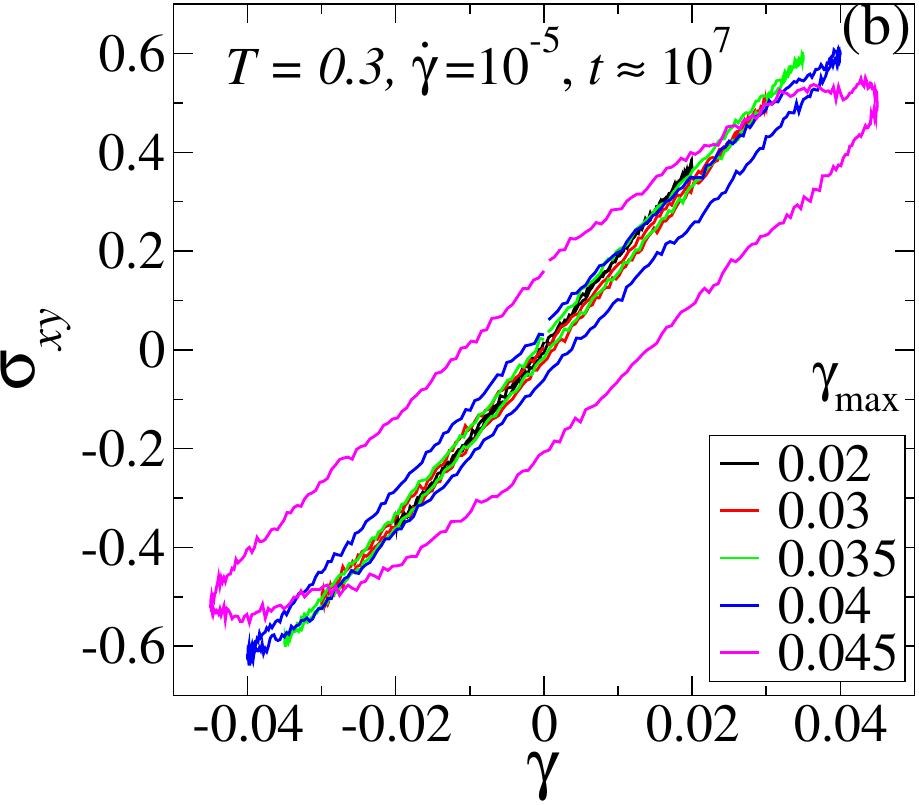} 
\includegraphics[scale=0.35]{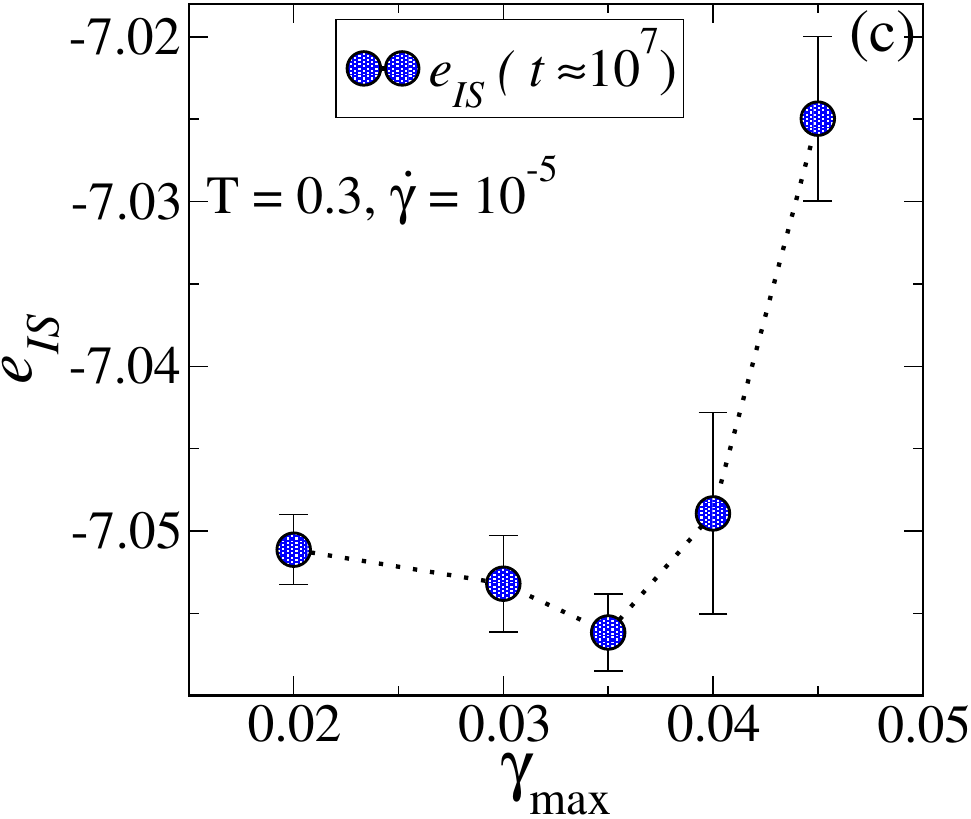} 
\caption{ (\emph{a}) Inherent structure energy variation within a cycle for different amplitudes at { a} fixed rate and  { a} fixed temperature. The energy minimum at zero strain { shifts} to finite strain values above a certain strain amplitude.  (\emph{b}) Variation of stress $\sigma_{xy}$ of inherent structures over a cycle.  After a certain amplitude of strain, 
the stress-strain curves begin to enclose a finite area. (\emph{c}) The optimal strain identified in the manuscript is the location of the minimum in the energy at zero strain is consistent with criteria for the yield strain in earlier work. }
\label{fig:S1}
\end{figure*}

Onset of yielding in athermally sheared glasses has been studied by considering the energy and stress within a strain cycle and as function of cycles. At the yield strain amplitude ($\gamma_{y}$), the system accesses the minimum energy states based on past work\cite{leishangthem2017yielding}. Beyond the yielding point, the location of the minimum  of energy 
shifts from zero strain to finite strain values, and the area enclosed by the stress-strain curve becomes finite, indicating the onset of plasticity in the system. We show here that the optimal strain amplitude which we identify as the strain at which the inherent structure energy is minimum (stroboscopically) also displays the 
characteristic features mentioned above. Note that in the steady state, the evolution of the inherent structure energy within a cycle is expected to be symmetric with respect to strain. In Fig. \ref{fig:S1}, we show the variation of energy at some intermediate time at which a steady state has not necessarily been reached. The average $e_{IS}$ we report as at time $t\approx 10^7$ corresponds to the average energy calculated within a time window of $t \approx 8\times 10^6 - t\approx 10^7$. The Error bars shown are the standard error of the mean of block averaged energies obtained for $20$ blocks within this time window. At the highest temperature, (T=0.4) as the system equilibrates faster, we have chosen the time window to be
$t \approx 2 \times 10^5 - t \approx 6\times 10^5$ and indicate the average within this block to be the long time average.

\subsection{Dependence on shear amplitude and rate}
In this section, we show results of the inherent structure energy variation with time for a range of temperatures and strain rates. The deformation amplitude corresponding to the minimum $e_{IS}$ in a long time window is identified as the optimal amplitude. As noted earlier, such an optimal strain amplitude shares characteristics with the yielding strain amplitude previously studied in athermally sheared glasses. The optimal point shifts to higher values of strain amplitude as the strain rate is decreased and it shifts to a lower value as the temperature is increased. At the highest temperature, even though we have analysed amplitudes below $\gamma_{max} = 0.005$, the behavior for such amplitudes are not significantly distinguishable from that of  $\gamma_{max} = 0.005$ and no clear minimum in the evaluated energies is present. However, above $\gamma_{max} = 0.005$ the energies increase to higher than the equilibrium energy at T=0.4, and hence we identify for this temperature $\gamma_{max} = 0.005$ as the optimal amplitude. 

\begin{figure*}[]
 \centering
\includegraphics[scale=0.32]{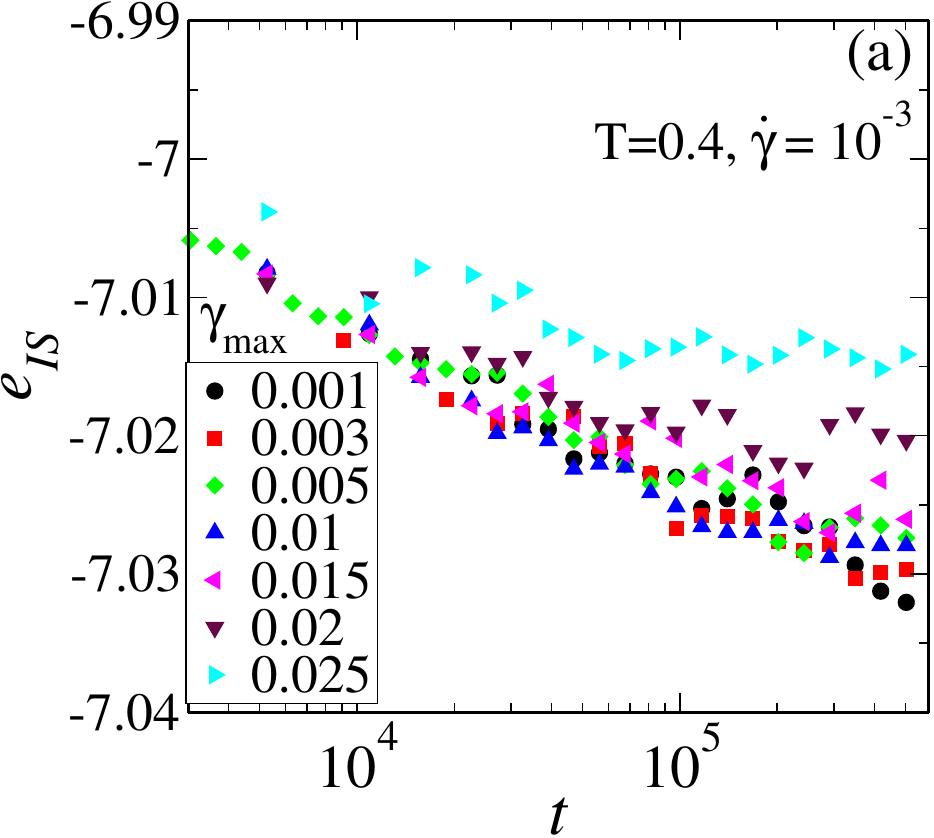} 
\includegraphics[scale=0.32]{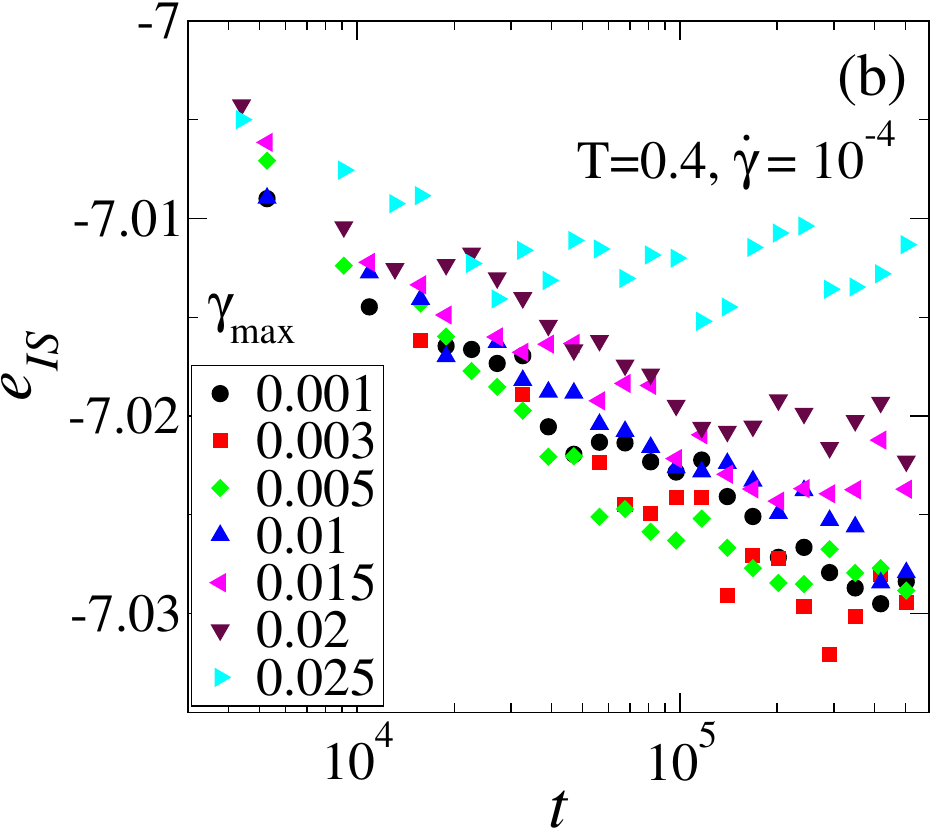} \includegraphics[scale=0.32]{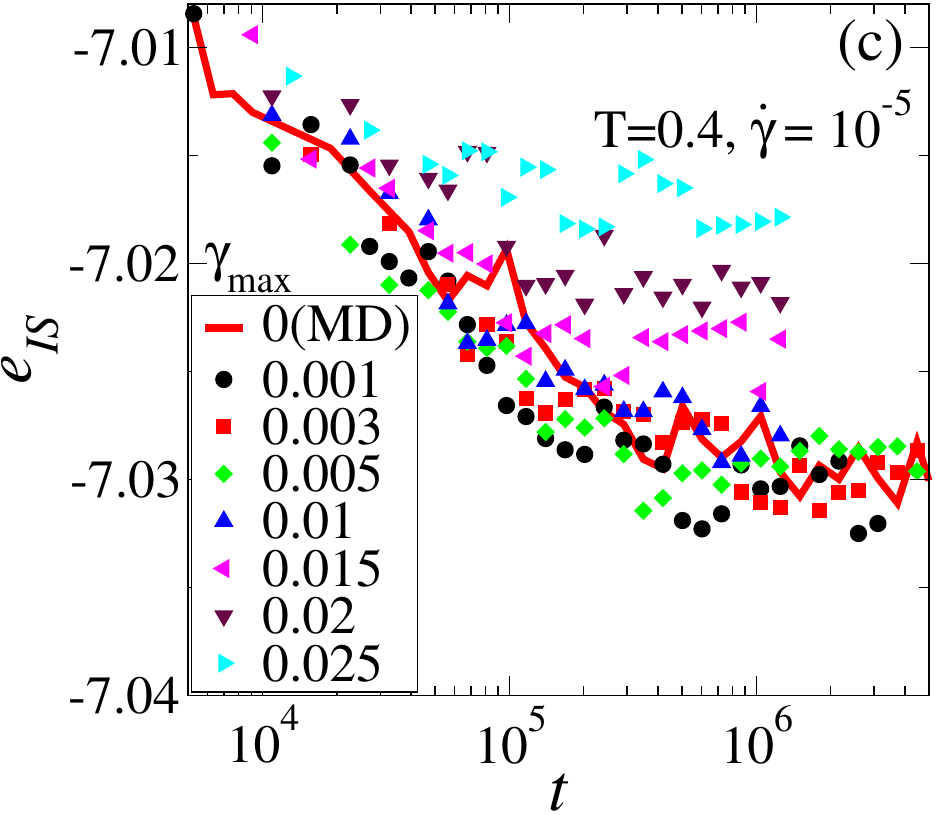}
\includegraphics[scale=0.32]{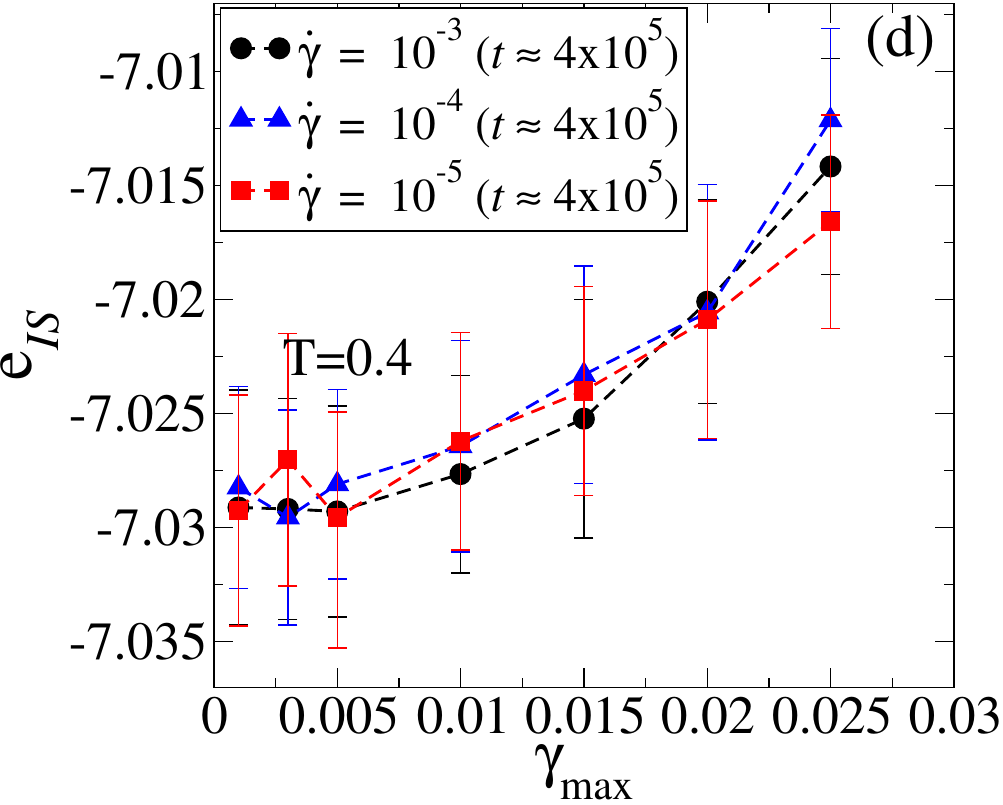}
\caption{\textcolor{black}{(\emph{a})-({\emph{c}}) The evolution of IS energy for different shear rates has been shown for $T = 0.4$. The amplitude at which the long time energy value { reaches} a minimum is identified as the optimal amplitude $\gamma_y$.  (\emph{d})  The long time values of IS energies {\it vs.} strain amplitude, obtained as an average within a time window from $t = 2 \times 10^5$ to $ 6 \times ~ 10^5$.} The evolution of IS energy for molecular dynamics at $T=0.4$ is shown in (c) for comparison (red line). %\sri{Fix figure accordingly} 
}
\label{fig:S5}
\end{figure*}
\begin{figure*}[]
 \centering
\includegraphics[scale=0.32]{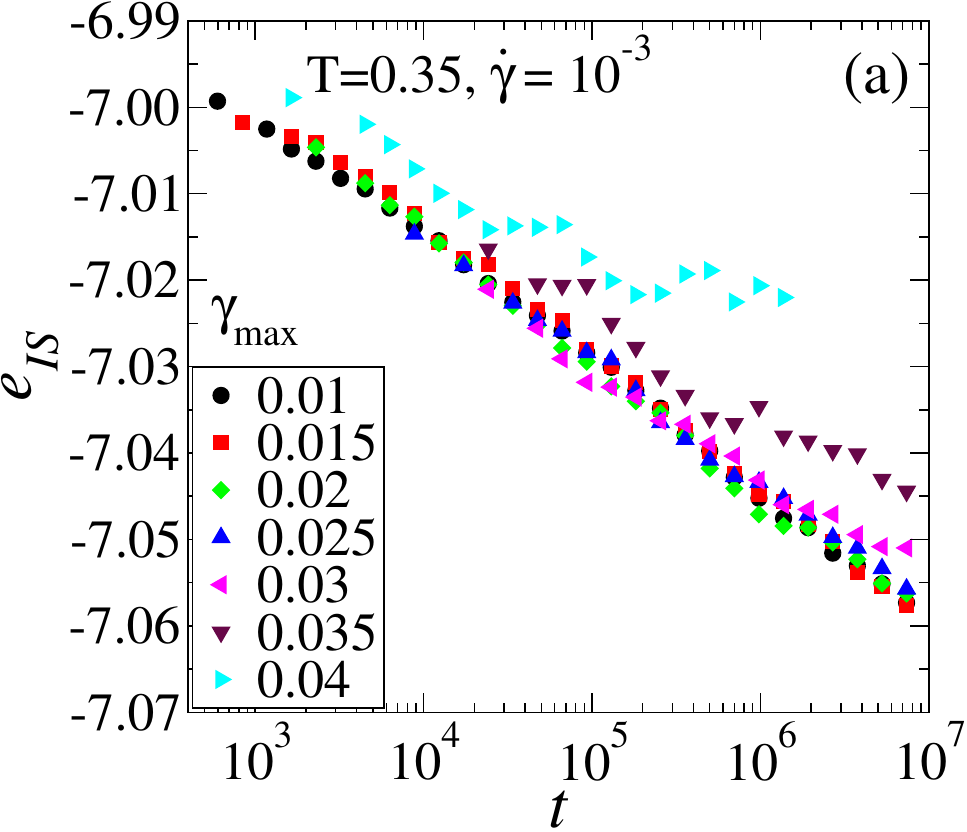} 
\includegraphics[scale=0.32]{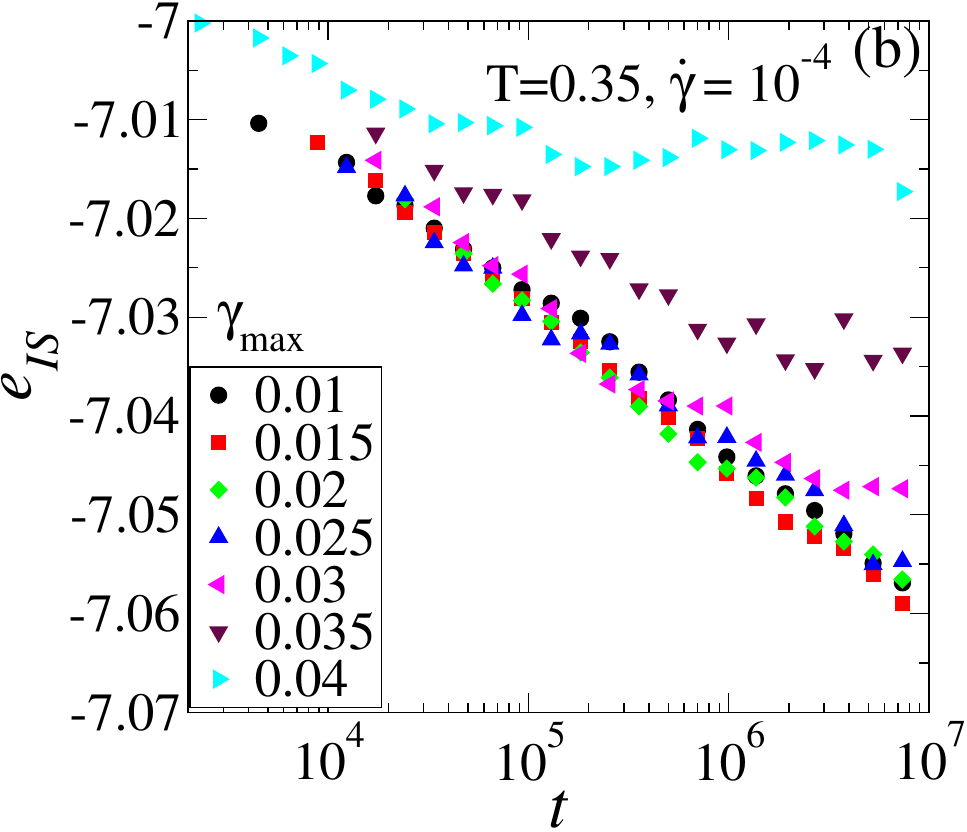} 
\includegraphics[scale=0.32]{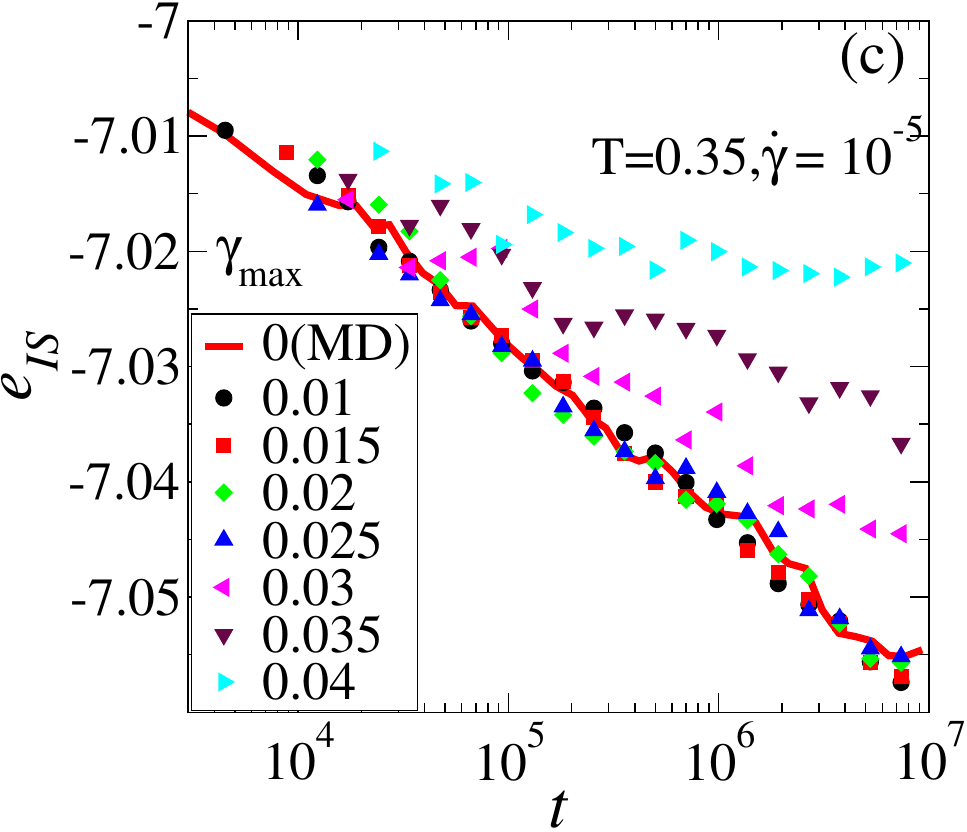} 
\includegraphics[scale=0.32]{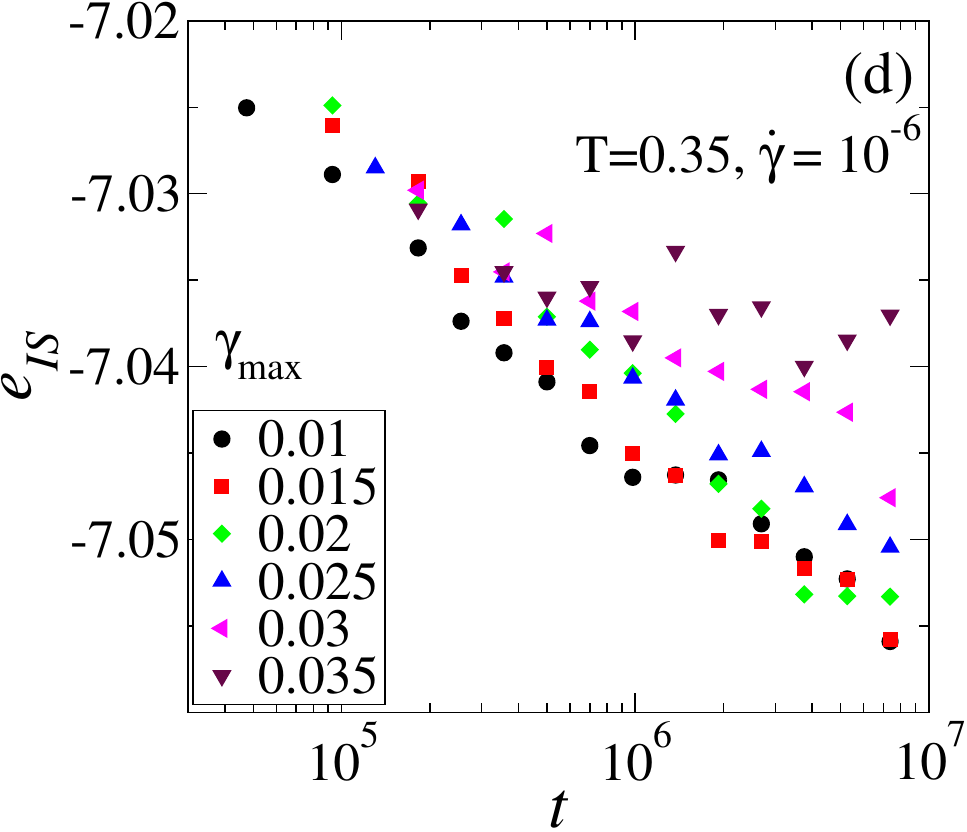}
\includegraphics[scale=0.32]{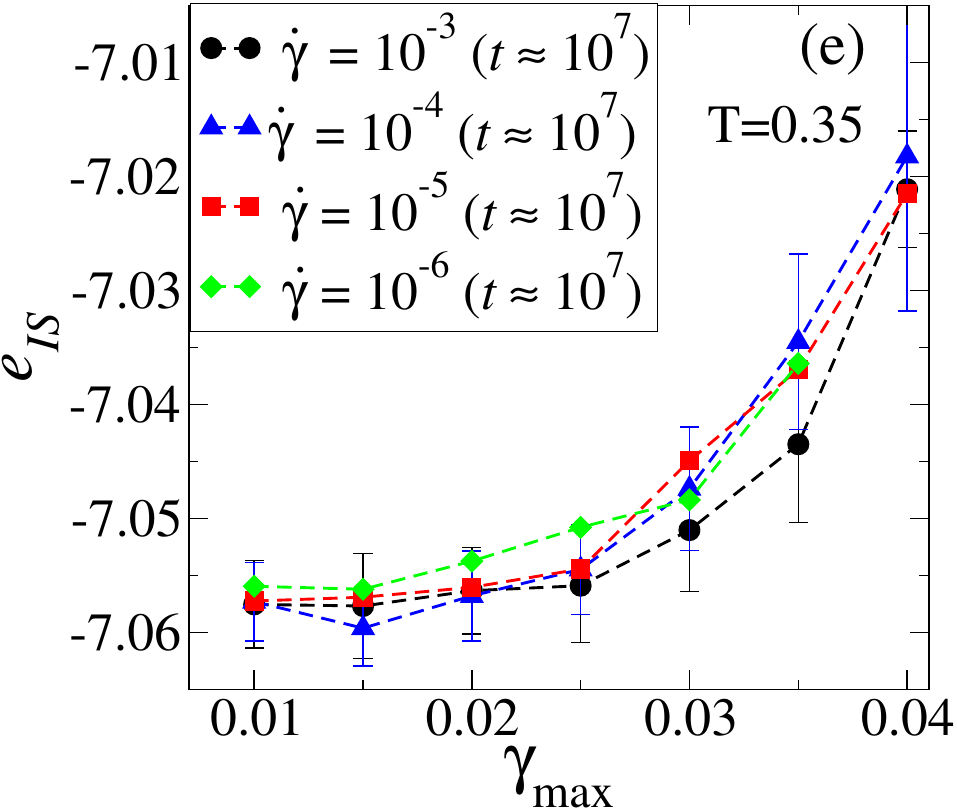}
\caption{\textcolor{black}{(\emph{a})-({\emph{d}}) The evolution of IS energy for different shear rates has been shown for $T = 0.35$. The amplitude at which the long time energy value reaches a minimum is identified as the optimal amplitude $\gamma_y$. (\emph{e}) The long time values of IS energies {\it vs.} strain amplitude, obtained as an average for  $t = 8 \times 10^{6}$ to $t = 10^{7}$.  The evolution of IS energy for molecular dynamics at $T=0.4$ is shown in (c) for comparison (red line). %\sri{Fix figure accordingly} 
}}
%\pd{ \emph{f}shows the evolution of IS energy for molecular dynamics at $T=0.35$}.}
\label{fig:S4}
\end{figure*}
%\clearpage
\begin{figure*}[h!]
 \centering
\includegraphics[scale=0.32]{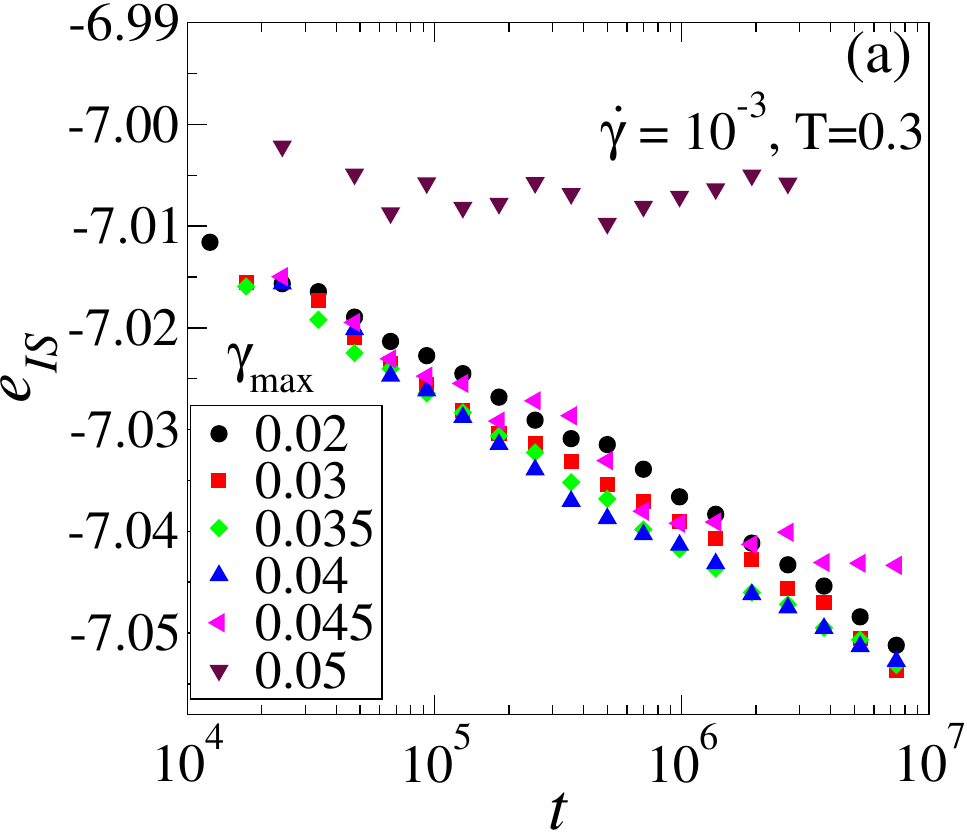} 
\includegraphics[scale=0.32]{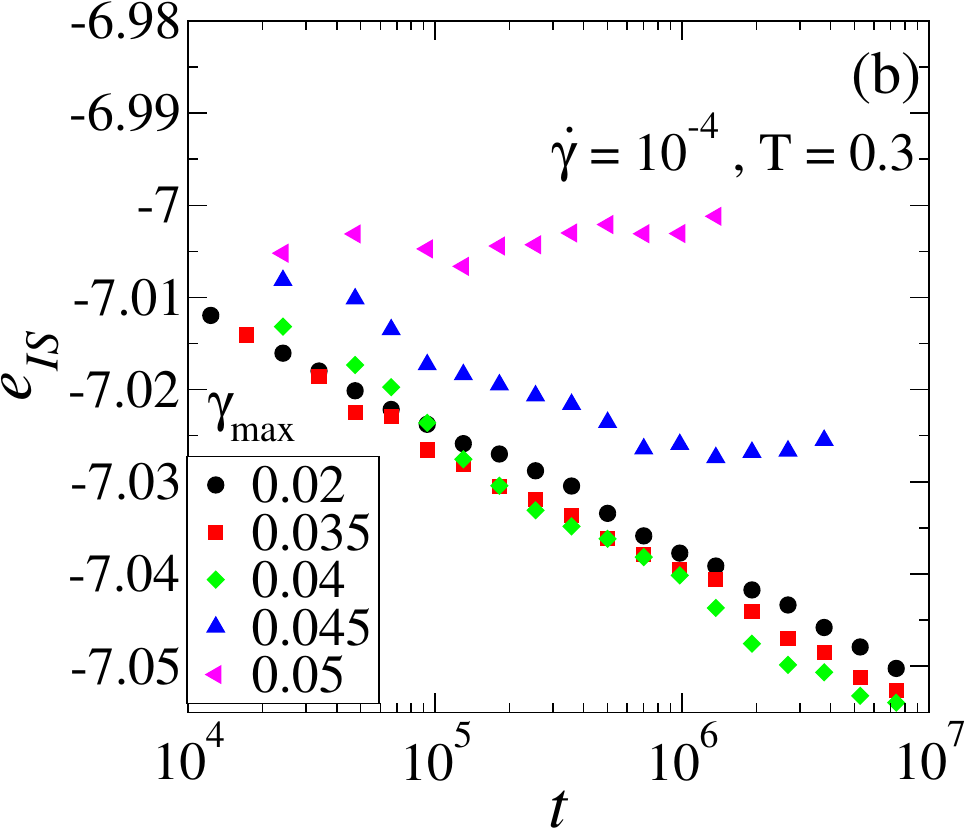} 
\includegraphics[scale=0.32]{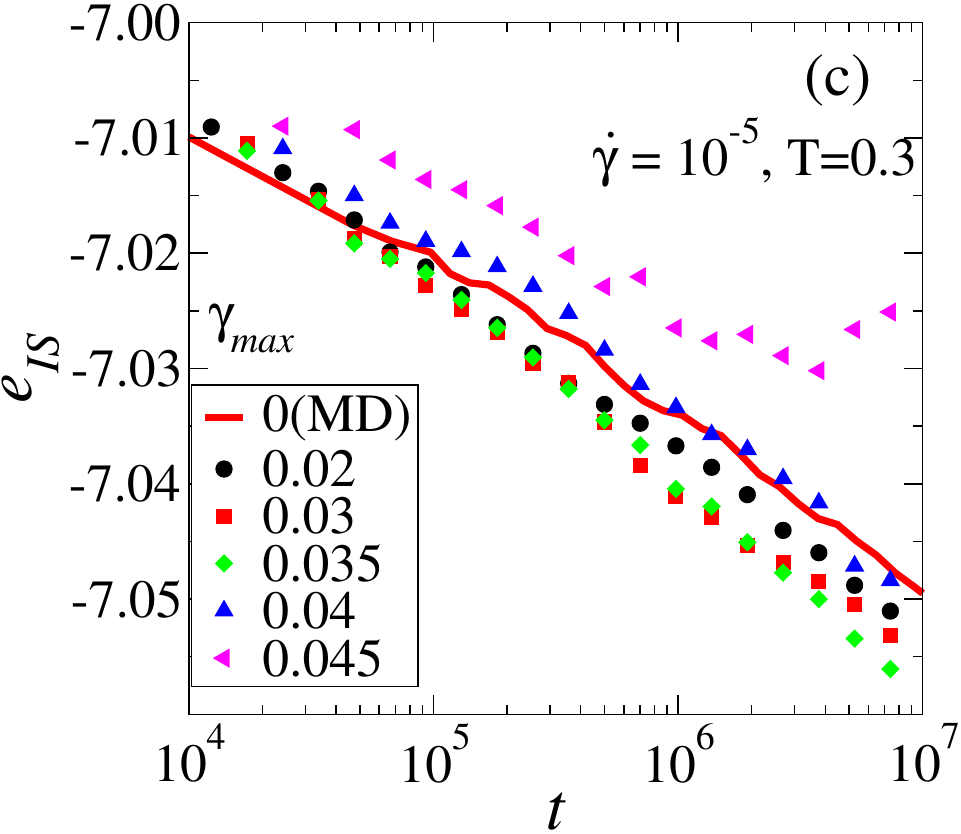}
\includegraphics[scale=0.32]{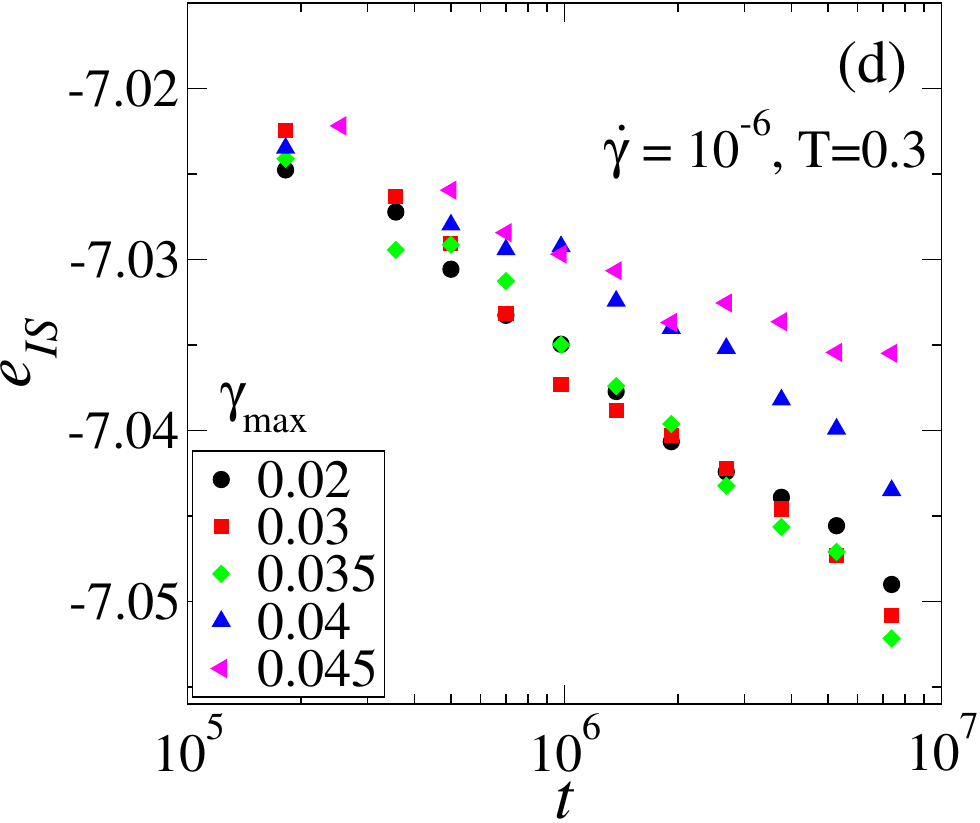}
\includegraphics[scale=0.32]{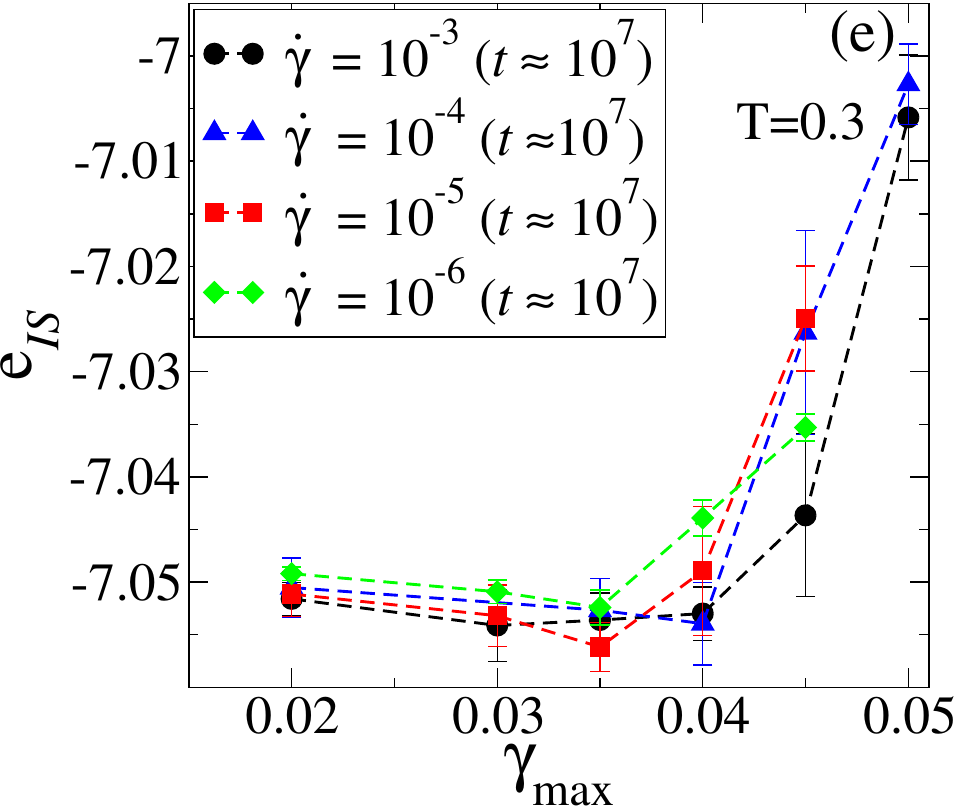}
\caption{\textcolor{black}{(\emph{a})-({\emph{d}}) The evolution of IS energy for different shear rates has been shown for $T = 0.3$. The amplitude at which the long time energy value reaches a minimum is identified as the optimal amplitude $\gamma_y$. (\emph{e}) The long time values of IS energies {\it vs.} strain amplitude, obtained as an average for  $t = 8 \times 10^{6}$ to $t = 10^{7}$.  The evolution of IS energy for molecular dynamics at $T=0.4$ is shown in (c) for comparison (red line). %\sri{Fix figure accordingly}
}}
%\pd{ \emph{f}shows the evolution of IS energy for molecular dynamics at $T=0.3$}.}
\label{fig:S3}
\end{figure*}
\begin{figure*}[h]
 \centering
\includegraphics[scale=0.32]{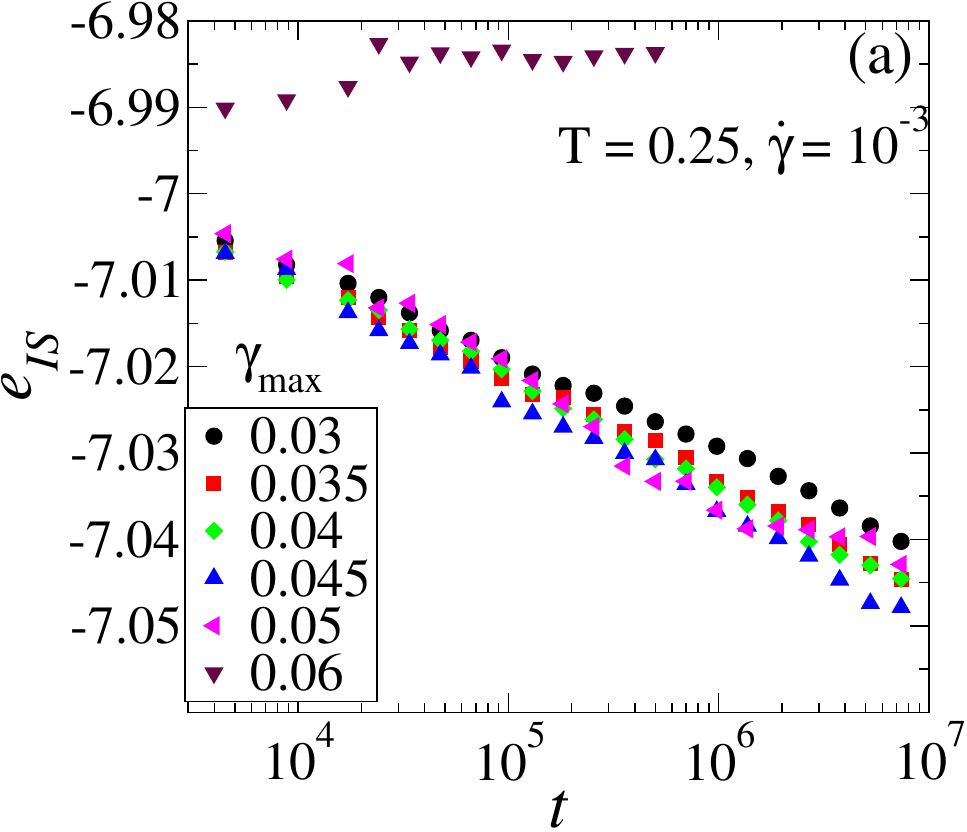} 
\includegraphics[scale=0.32]{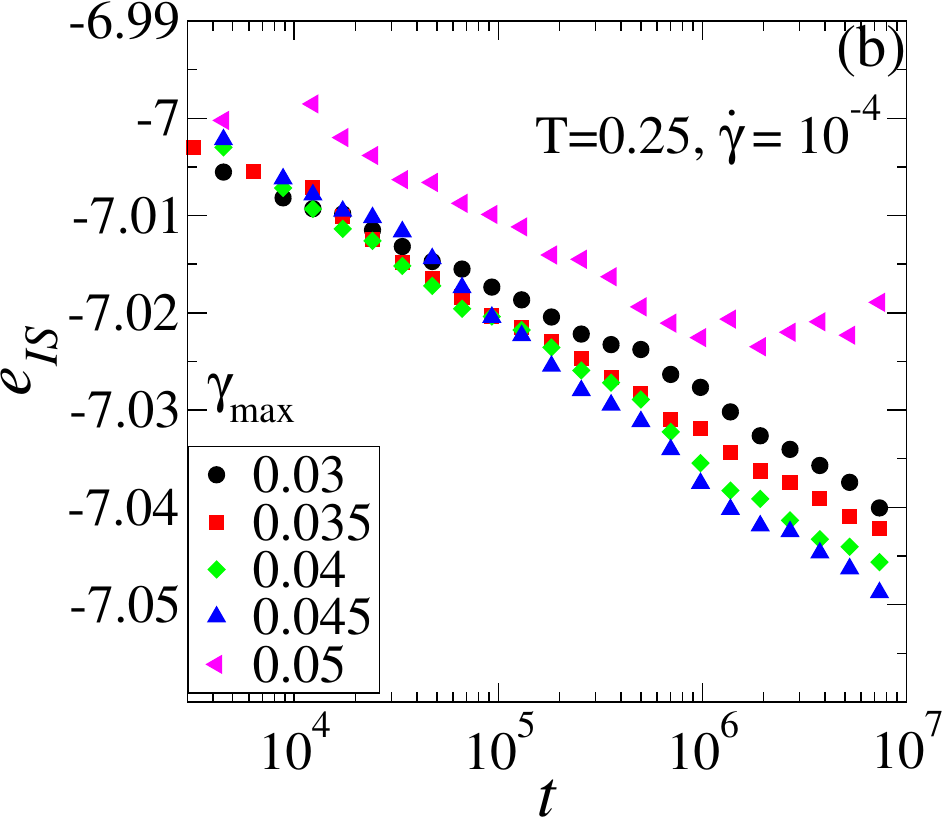} 
\includegraphics[scale=0.32]{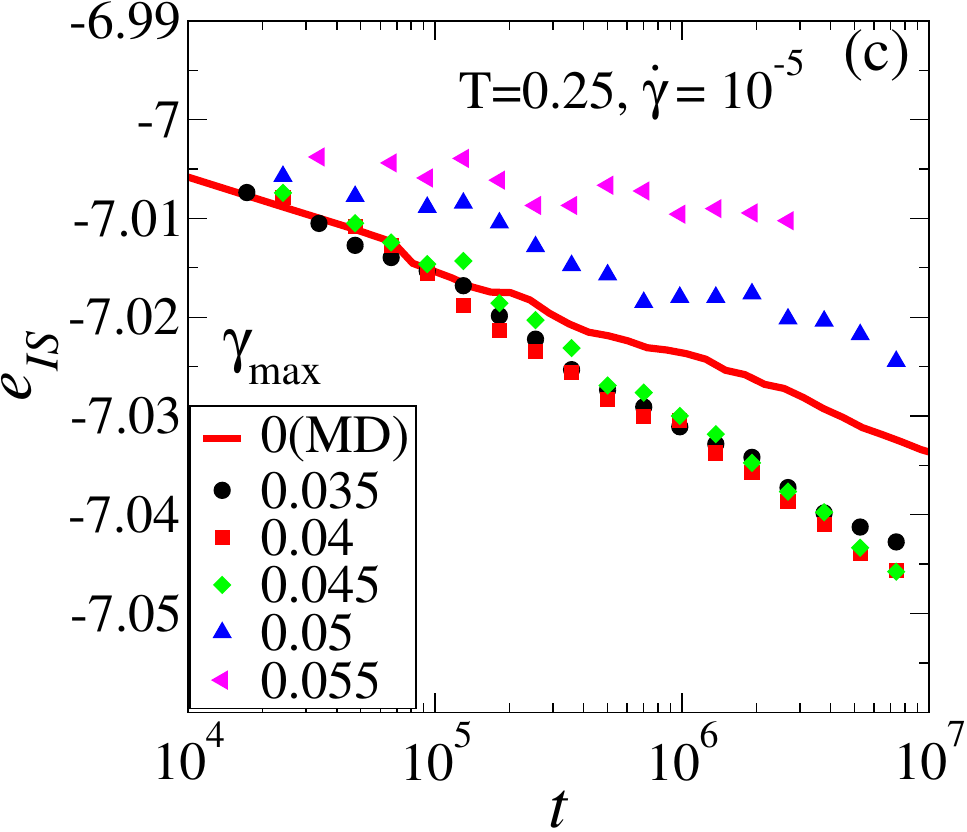} 
\includegraphics[scale=0.32]{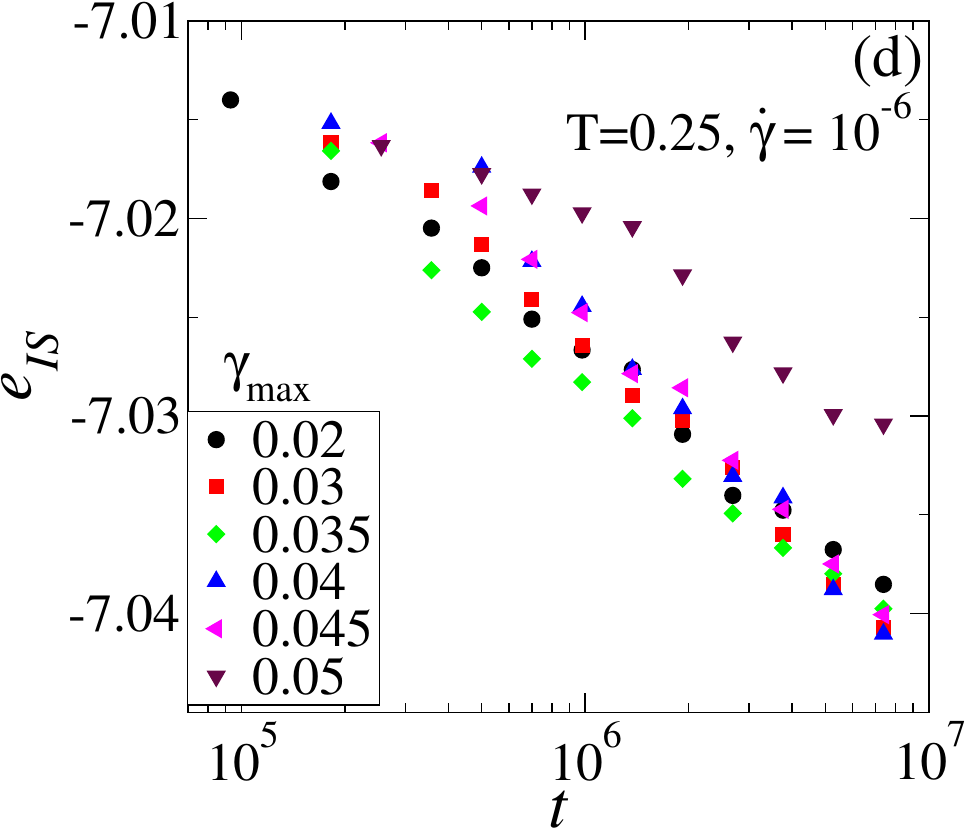}
\includegraphics[scale=0.32]{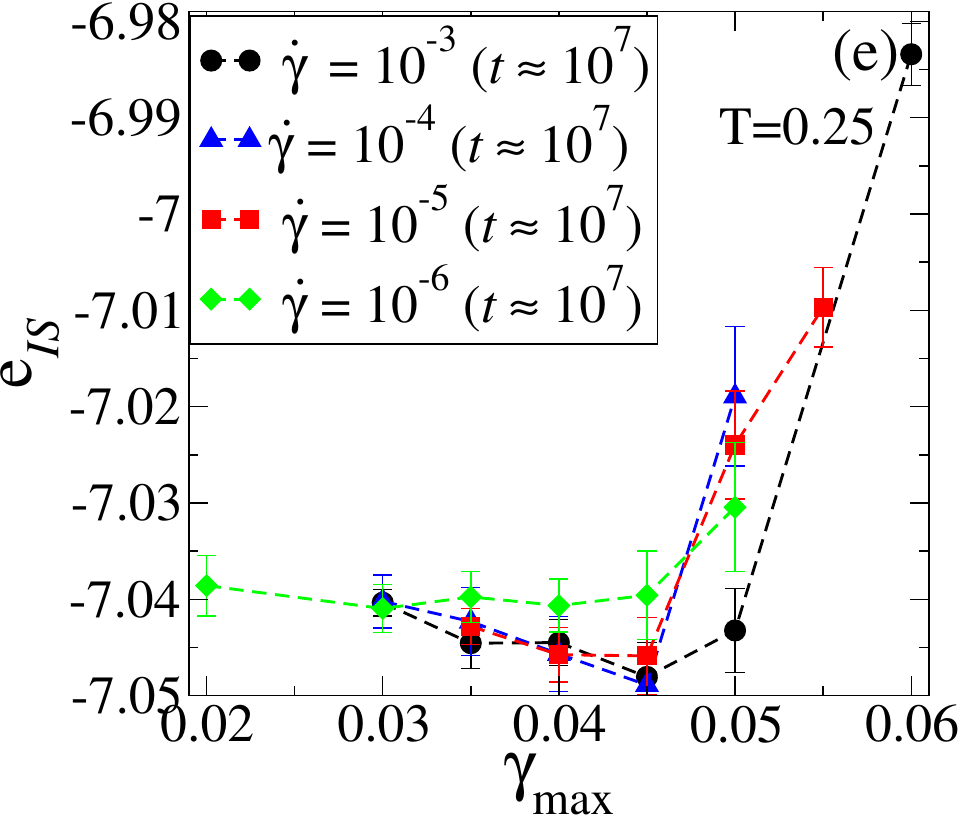}
\caption{\textcolor{black}{(\emph{a})-({\emph{d}}) The evolution of IS energy for different shear rates has been shown for $T = 0.25$. The amplitude at which the long time energy value  reaches a minimum is identified as the optimal amplitude $\gamma_y$ . (\emph{e}) The long time values of IS energies {\it vs.} strain amplitude, obtained as an average for  $t = 8 \times 10^{6}$ to $t = 10^{7}$.  The evolution of IS energy for molecular dynamics at $T=0.4$ is shown in (c) for comparison (red line). % \sri{Fix figure accordingly}
}}
%\pd{ \emph{f}shows the evolution of IS energy for molecular dynamics at $T=0.25$}.}
\label{fig:S2}
\end{figure*}
\begin{figure*}
 \centering
\includegraphics[scale=0.35]{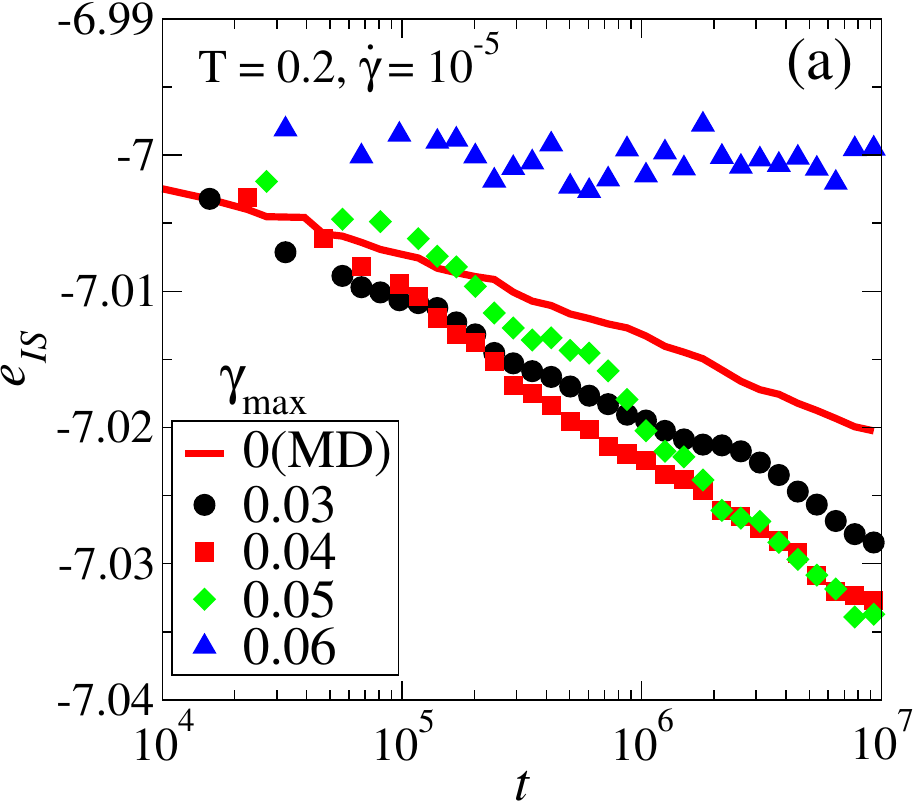} 
\includegraphics[scale=0.35]{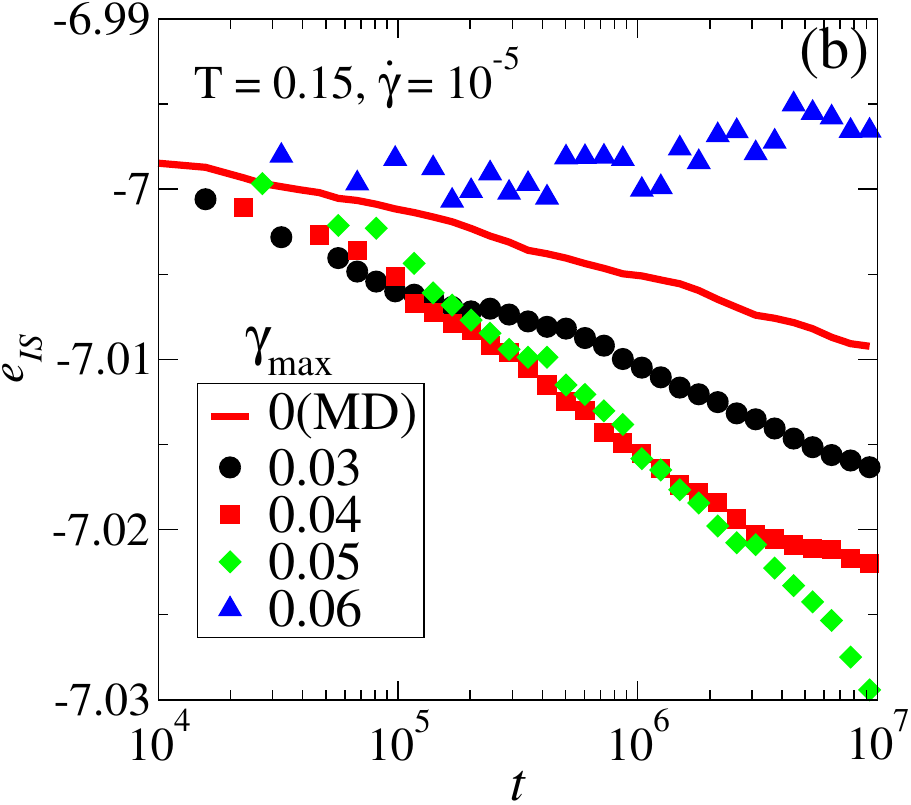} 
\includegraphics[scale=0.35]{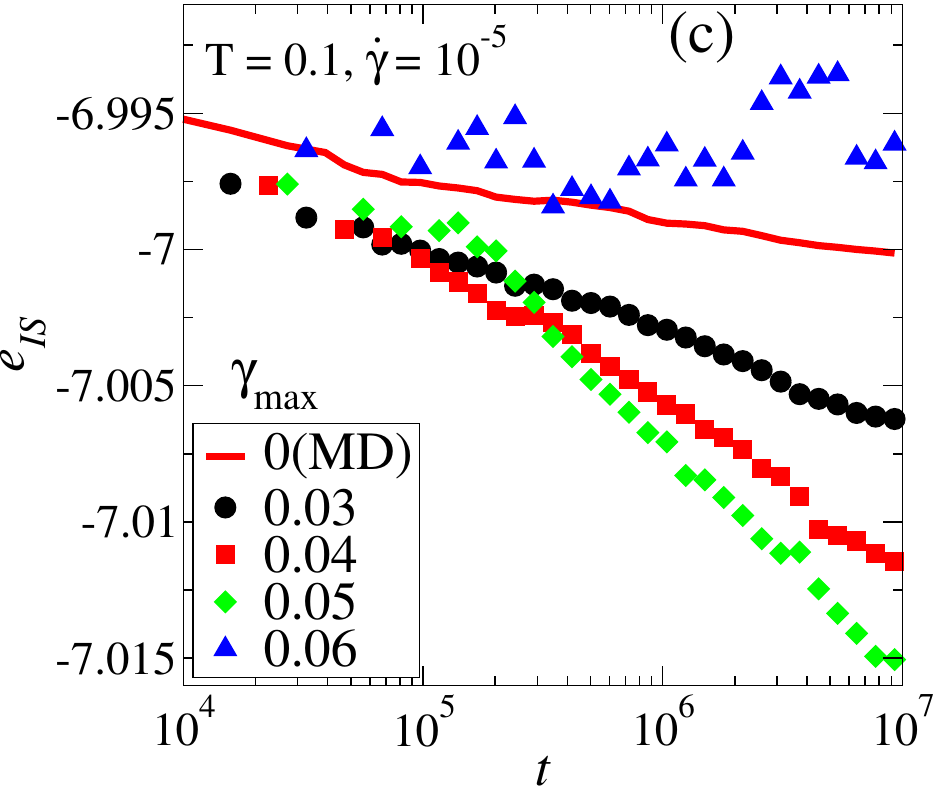} 
\caption{(\emph{a})-( \emph{c} ) The evolution of IS energy for the three lowest temperatures ($T=0.1,0.15,0.2$) is shown. The amplitude at which the long time energy value reaches a minimum is identified as the optimal amplitude $\gamma_y$. The evolution of IS energy for molecular dynamics at $T=0.4$ is shown for comparison (red lines). %
%\sri{Fix figure accordingly} 
}
%
%(\emph{a})-({\emph{c}}) shows the evolution of IS energy for molecular dynamics at $T=0.1,0.15,0.2$ }.}
%\includegraphics[scale=0.32]{PDF/T3_rate000001_allamp.pdf}\caption{\textcolor{black}{(\emph{a})-({\emph{c}}) The evolution of IS energy for different shear rates has been shown for $T = 0.4$. The amplitude at which the long time energy value { reaches} a minimum is identified as the optimal amplitude $\gamma_y$.  (\emph{d})  The long time values of IS energies {\it vs.} strain amplitude, obtained as an average within a time window from $t = 2 \times 10^5$ to $ 6 \times ~ 10^5$.}}
\label{lowTnew}
\end{figure*}

\clearpage
\subsection{Equilibrium Molecular Dynamics: Relaxation time and configurational entropy}

\begin{figure}[h]
\centering{}
\includegraphics[scale=0.4]{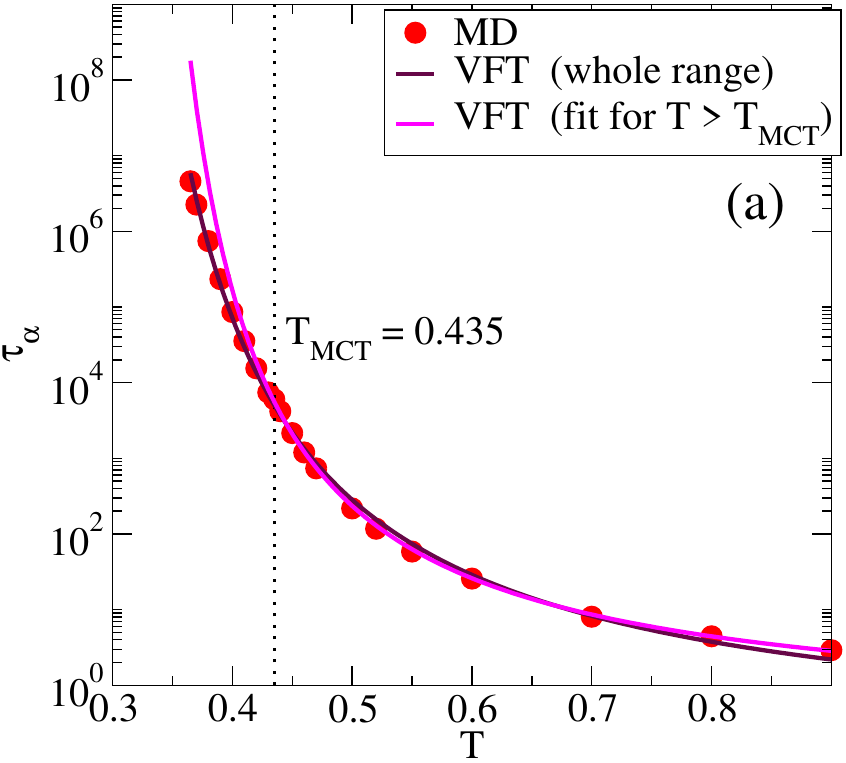}
\includegraphics[scale=0.4]{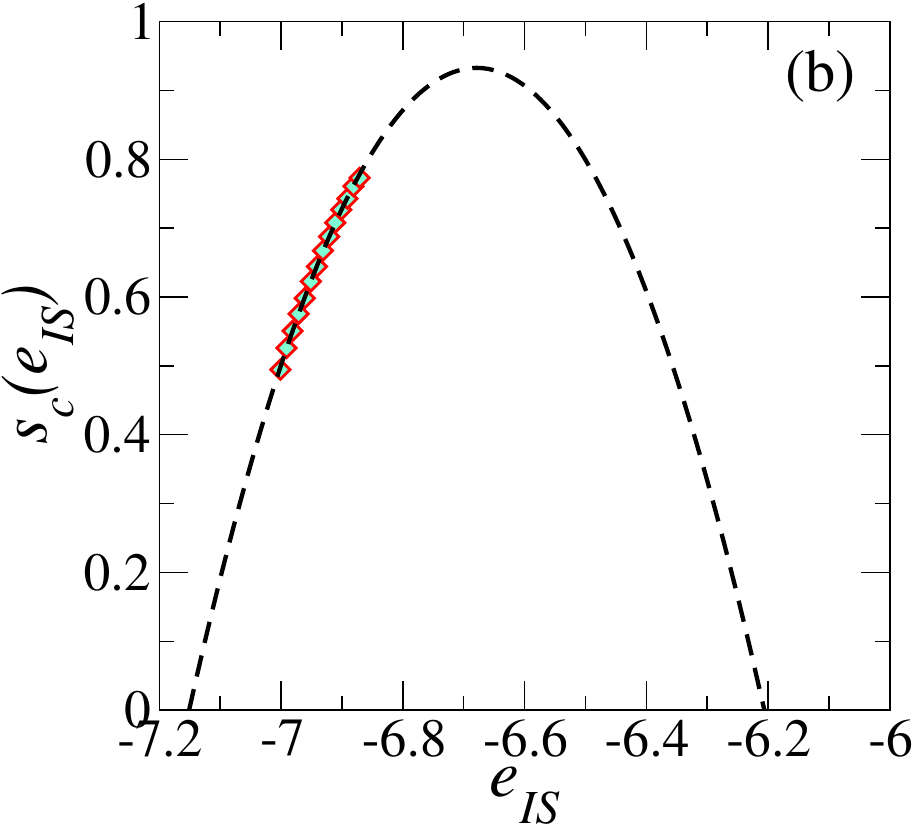}
\caption{(\emph{a}) Relaxation times from molecular dynamics simulations {\it vs.} temperature, and the corresponding fit to the VFT form. 
(\emph{b}) The configurational entropy density as  a function of the inherent structure energy, and a quadratic fit.  
The extrapolated value of IS energy at which configurational entropy vanishes is ($\approx-7.15$).}
\label{VFT}
\end{figure}

A comparsion of the time scales accessed by the non-equilibrium MD (NVT-SLLOD, NEMD) simulations with the equilibrium NVT-MD simulation results is an important aspect of our analysis. In Fig. \ref{VFT} (a), we show VFT fits to the relaxation times, first by considering temperatures only for $T > T_{MCT}$ $(\tau_0 = 0.3101$, $K_{VFT} = 0.2243$, $T_{VFT} = 0.2989$). The VFT fits to the high temperature data clearly overestimate the relaxation times for $T < T_{MCT}$. On the other hand, for the VFT fit to the full range ($\tau_0 = 0.1175$, $K_{VFT} = 0.1383$, $T_{VFT} = 0.2592$.), the VFT form does not provide a good description of the data at lower temperatures\cite{DAS2022100098}.
%\sout{The relaxation time over a large temperature range is fitted to the VFT functional form in Fig. \ref{VFT}(a) (data from the Ph. D. thesis of S. Sengupta, JNCASR \cite{ShiladityaThesis2013}. With the VFT fit, $\tau = \tau_0 \exp \left[{(K_{VFT}({T}/{T_{VFT}}-1))^{-1}}\right]$ the fit parameters are $\tau_0 = 0.31051$, $K_{VFT} = 0.2243$ and $T_{VFT} = 0.2983$. }
The fit values are used in our analysis in the manuscript. Fig. \ref{VFT} (b) shows data (from \cite{Sastry2001}) for the configurational entropy density  {\it vs.} inherent structure energy and a quadratic fit (logarithm of a Gaussian density of states). From the shown data and the fit, the configurational entropy vanishes at the Kauzmann energy of $e_{IS} = -7.15$. 

\subsection{Correspondence between inherent structure energies and temperatures}

\begin{figure}[h!]
\centering
\includegraphics[scale=0.35]{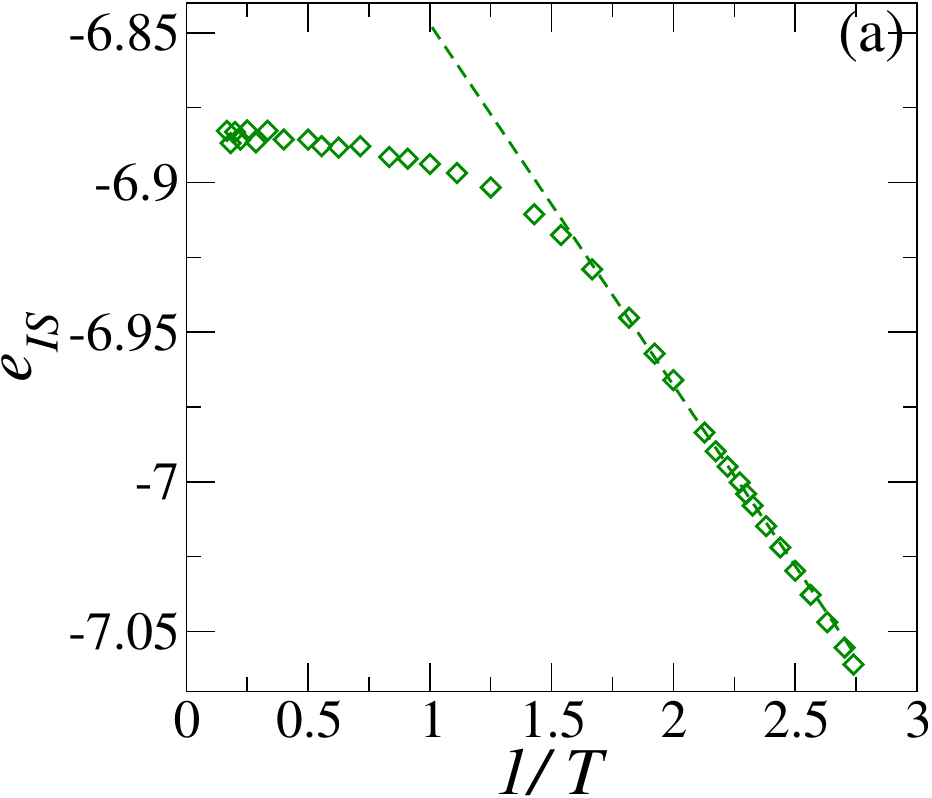}
\includegraphics[scale=0.35]{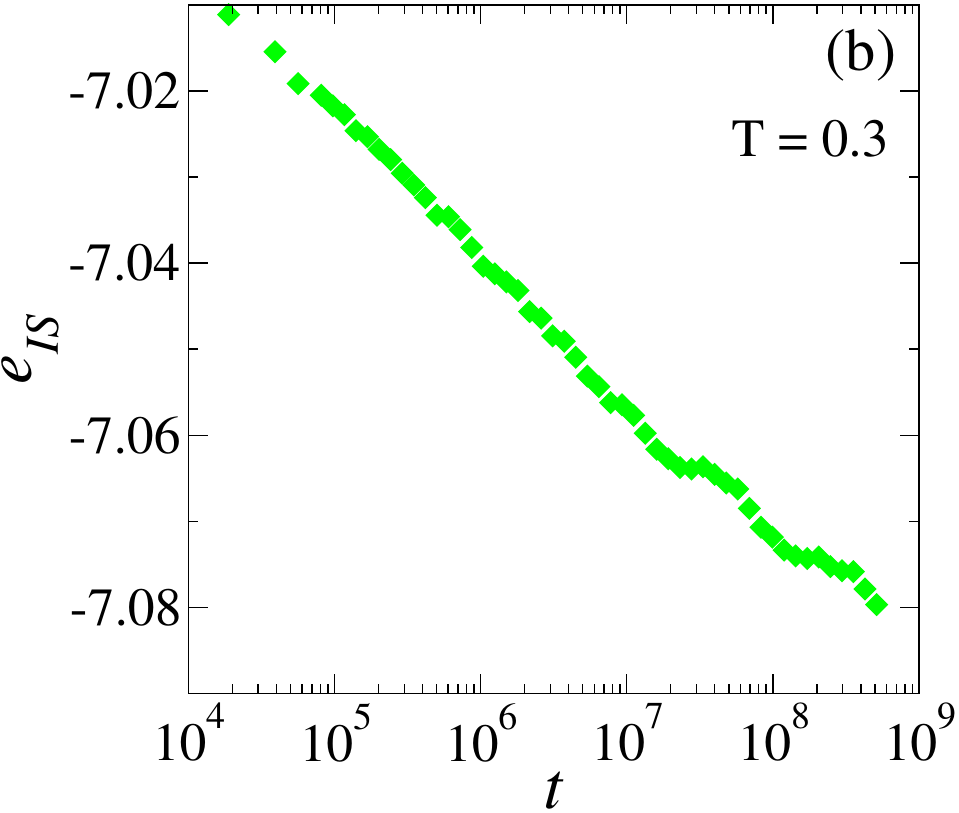}
\includegraphics[scale=0.35]{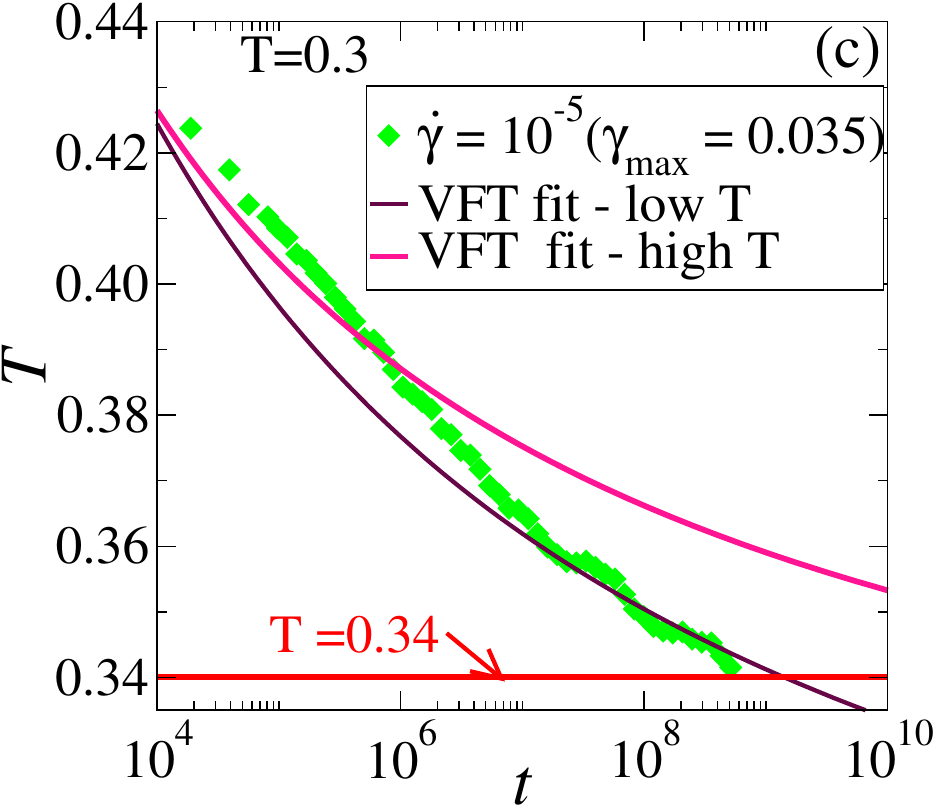}
\caption{(\emph{a}) Temperature dependence of  the inherent structure (IS) energy for a liquid equilibrated in a molecular dynamics simulation. 
The equilibrium molecular dynamics simulation data has been fitted below  temperature $T = 0.7$ to obtain a mapping between the IS energy and temperature, which is used to map the IS energy {\it vs.} time data in panel (\emph{b}) to { the} temperature values in panel (\emph{c}). Panel (\emph{c}) also shows the VFT relationship between temperature $T$ and relaxation times $\tau$. }
\label{ISref}
\end{figure}

In order to obtain an estimate of the temperature to which the inherent structures we generate correspond, we use the observation \cite{Sastry2001} that inherent structure energies at low temperatures vary as $1/T$. Fitting the energies to the form $e_{IS}(T) = E_{\infty}-{\mathcal A}/{T}$, we are able to map IS energies and temperatures, as illustrated in Fig. \ref{ISref}. The fit parameters are $E_{\infty} = -6.7264$ and ${\mathcal A} =  0.12065$.

%\clearpage

\subsection{Structural analysis}

\begin{figure}[]
\centering
\includegraphics[scale=0.4]{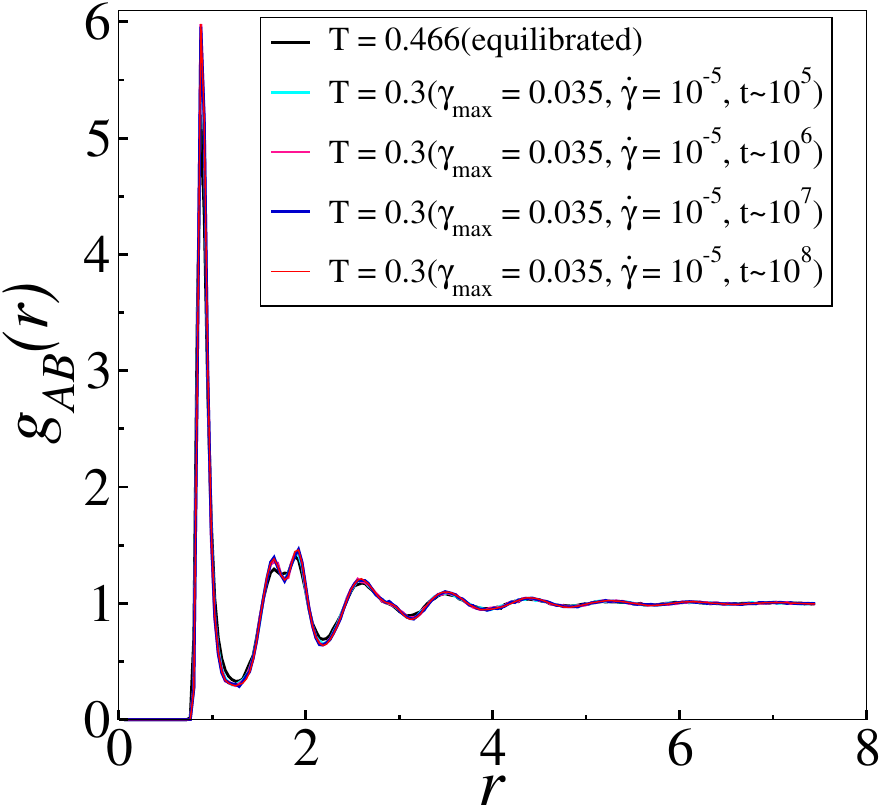}
\includegraphics[scale=0.4]{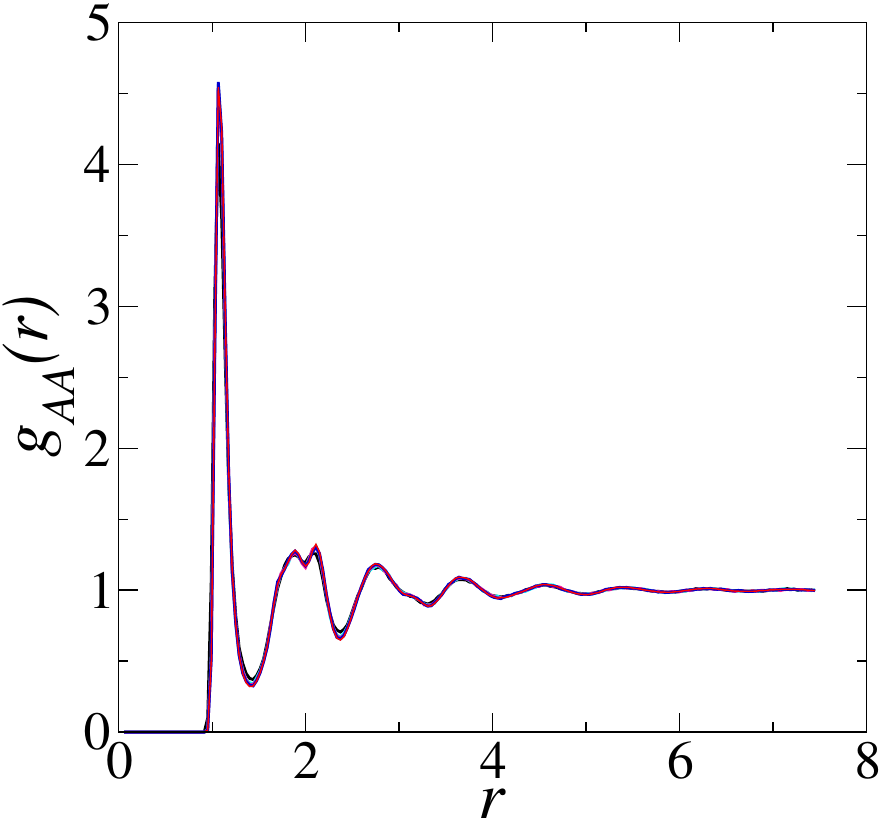}
\includegraphics[scale=0.4]{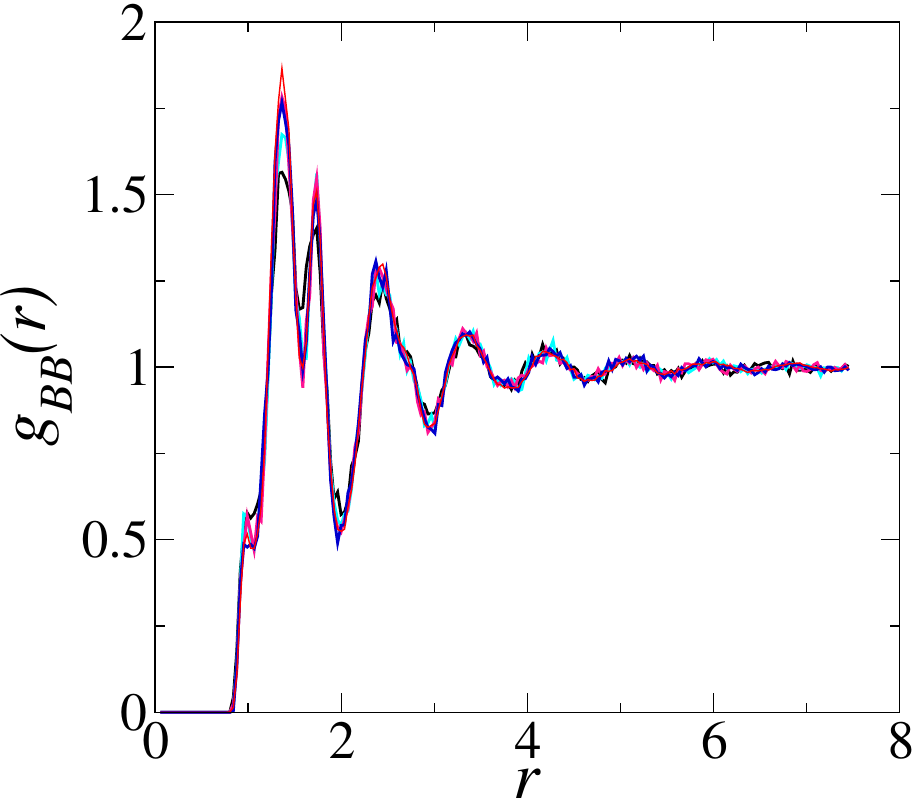}
\caption{The partial pair correlation functions of the liquid configurations at the end of different cycles compared with the initial configuration.}
\label{goftdir}
\end{figure}

We compute the radially averaged and two dimensional (partial) pair correlation functions, for stroboscopic configurations, in order to analyse the structure. 
For temperature T = 0.3 $\dot\gamma_{xy} = 10^{-5}$, $\gamma_{max} = 0.035$ has been identified as the yielding strain amplitude. 
For this amplitude, the radially averaged pair correlation function  is calculated after $0$ (newly quenched from the liquid), $5$, $50$ and $500$ cycles.
The Increase in the number of shear cycles corresponds to a decrease in energy and the temperature. The pair correlation functions show very small amounts of change, comparable to what is observed in the equilibrium liquid at different temperatures (Fig.\ref{goftdir}). 
%The results after a large number of cycles are shown in Fig. 4. \sri{Confirm and state more precisely if possible}. 

We also calculate the two dimensional radial distribution function ($g(x,y)$), in the $xy-$ (shear) plane, defined as

%\begin{widetext}
%\begin{equation}
%	g(x,y) = \frac{1}{2Na\rho}\times \left < \sum_{i =1}^{N-1} \sum_{j\neq i}^{N} \delta(x-(x_{i}-x_{j})) \delta(y-(y_{i}-y_{j})) \theta(a-|z_{i}-z_{j}|) \right >,
%\end{equation}
%\end{widetext}

% \begin{widetext}
\begin{align}
	g(x,y) = \frac{1}{2Na\rho}\times \Biggl < \sum_{i =1}^{N-1} \sum_{j\neq i}^{N} \delta(x-(x_{i}-x_{j}))\Biggr. \nonumber \\
	\Biggl.\delta(y-(y_{i}-y_{j})) \theta(a-|z_{i}-z_{j}|) \Biggr >,
\end{align}
% \end{widetext}

where ``$\langle\rangle$" represents the averaging over independent samples. $x_{i},~y_{i}$, and $z_{i}$ are coordinates of particles. A pair of particles is considered to be in  the same plane if the separation between { them} does not exceed a threshold value $a = 0.2\sigma_{AA}$, which is enforced by the Heaviside function above. 

\begin{figure*}[h!]
\centering
\includegraphics[scale=0.25]{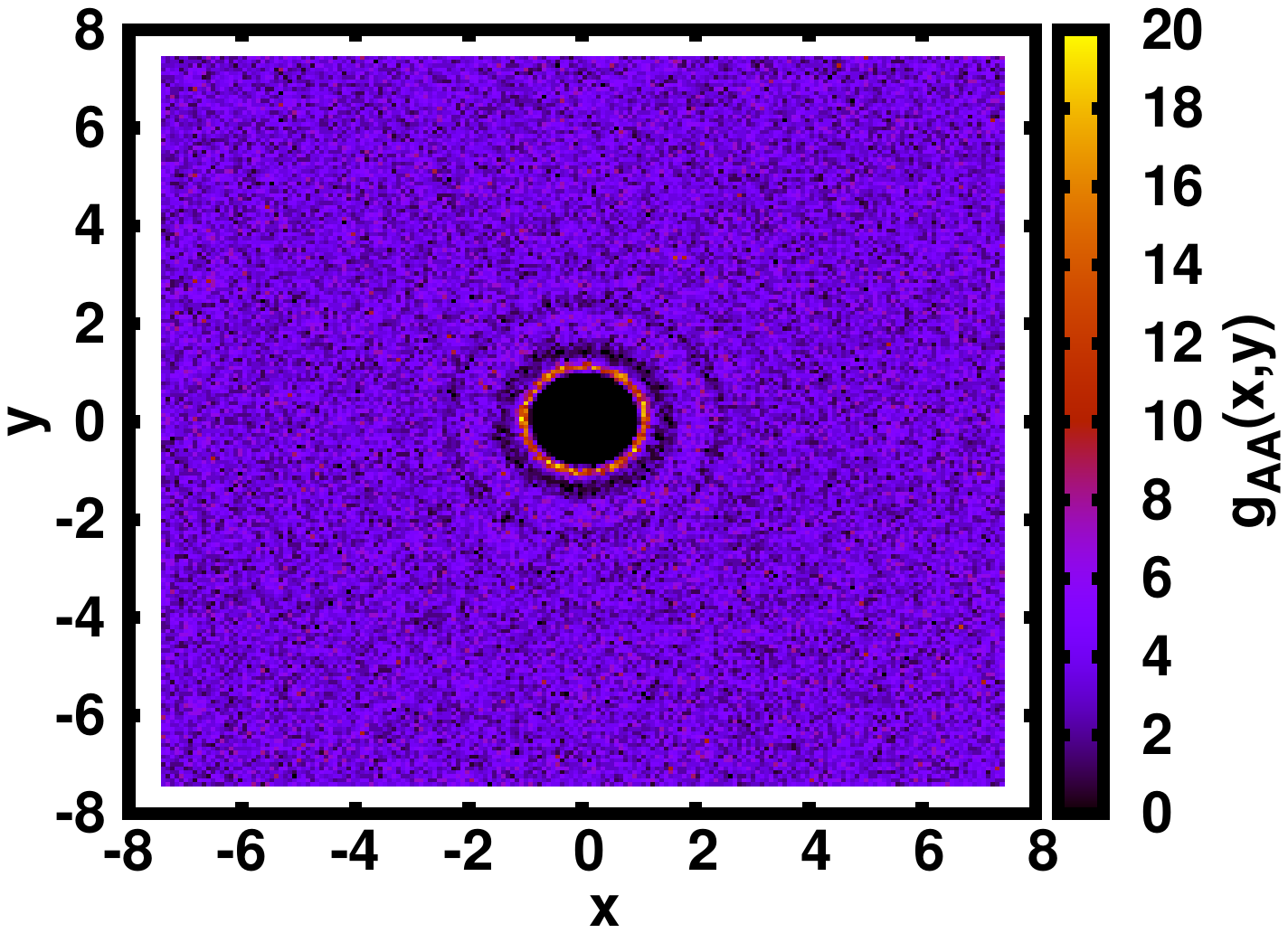}
\includegraphics[scale=0.25]{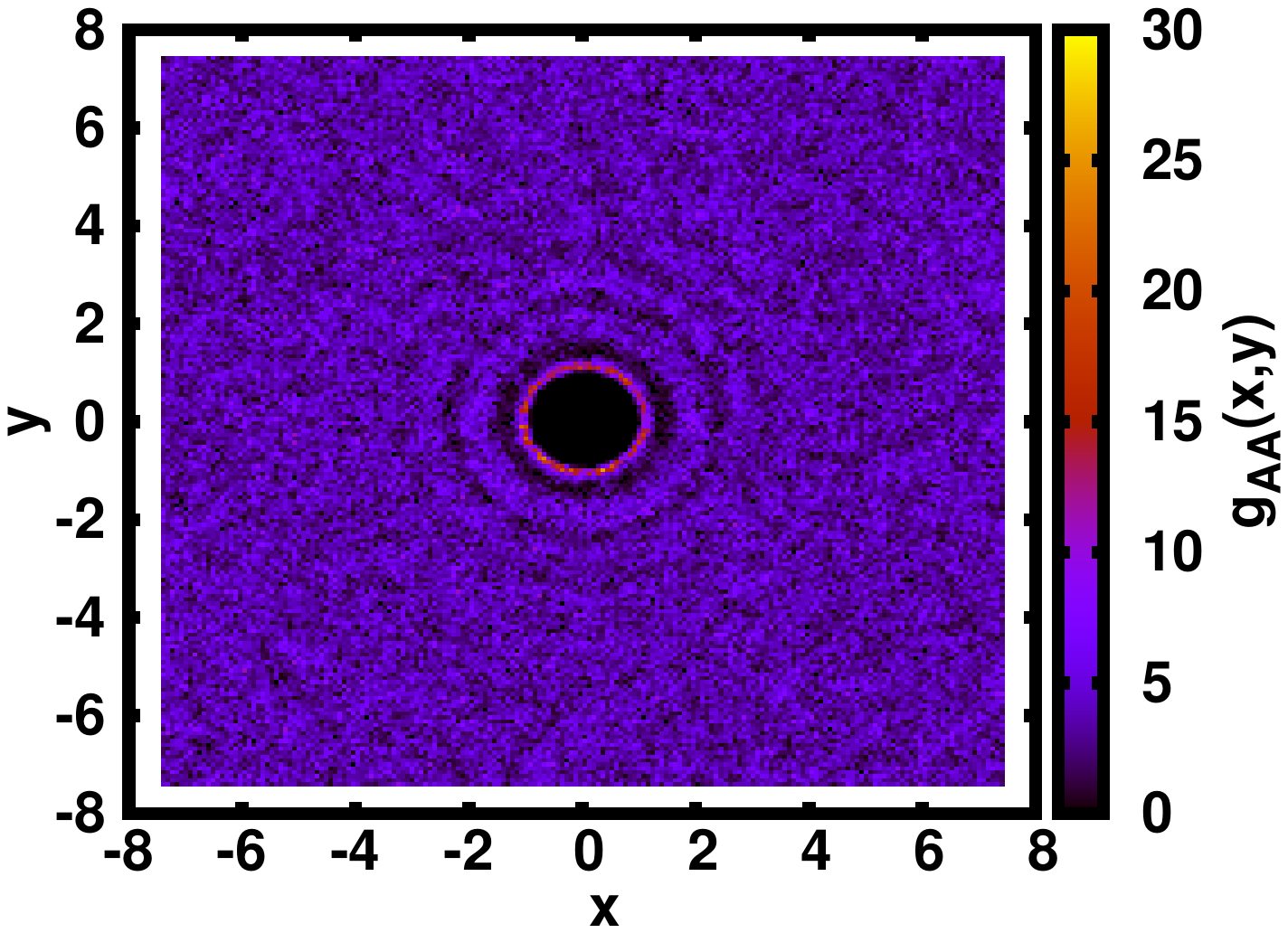}
\includegraphics[scale=0.25]{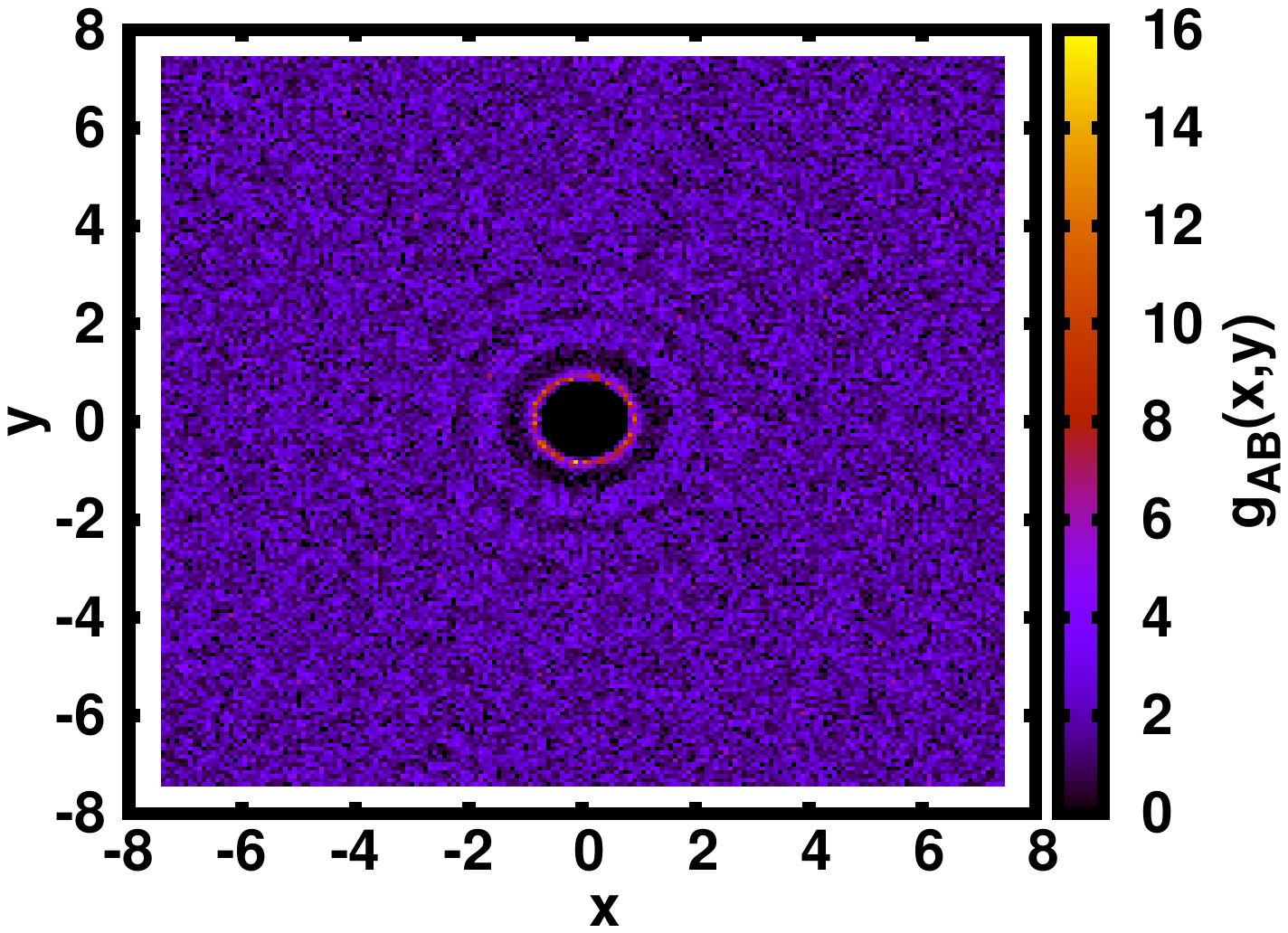}
\includegraphics[scale=0.25]{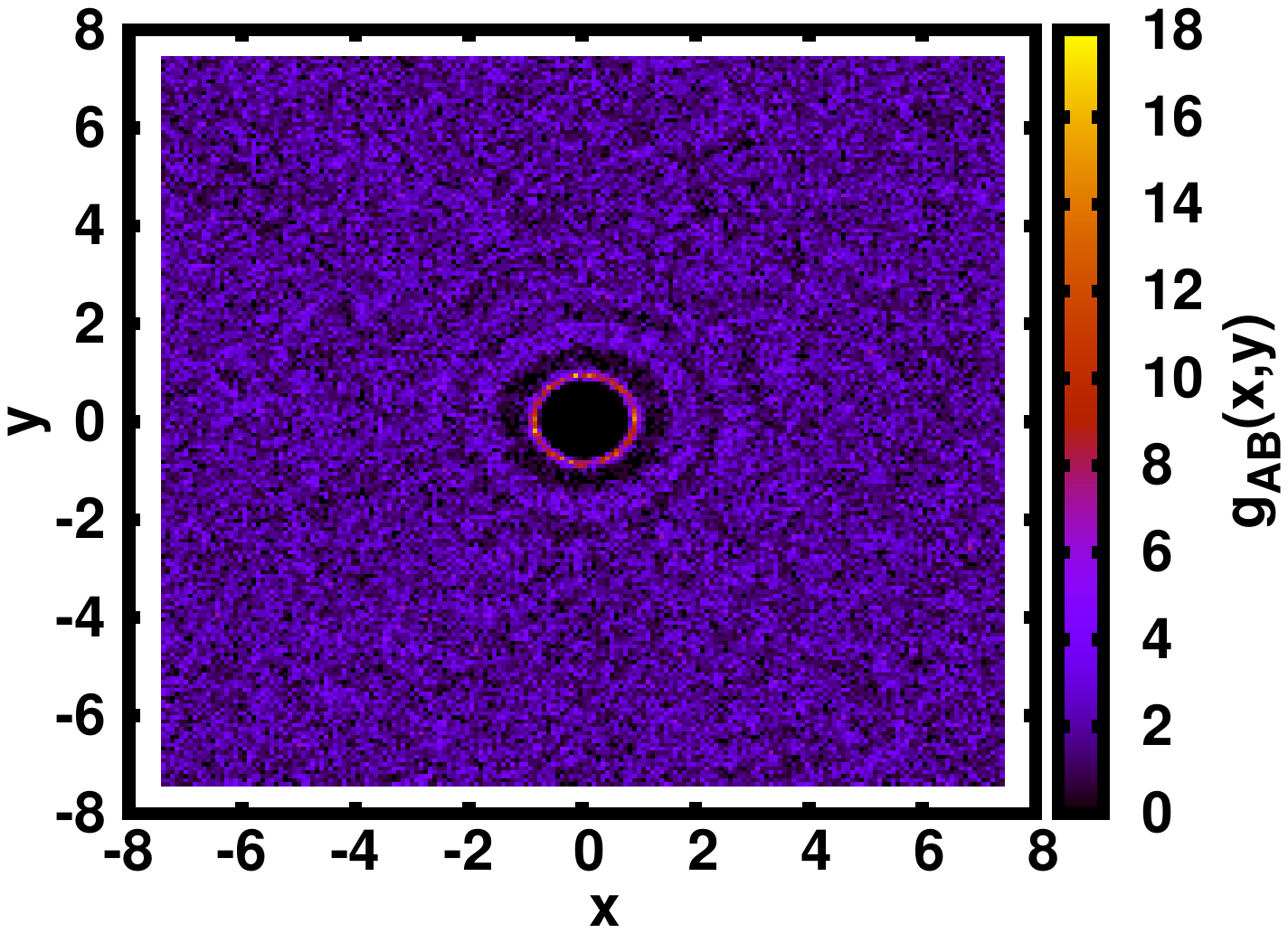}
\includegraphics[scale=0.25]{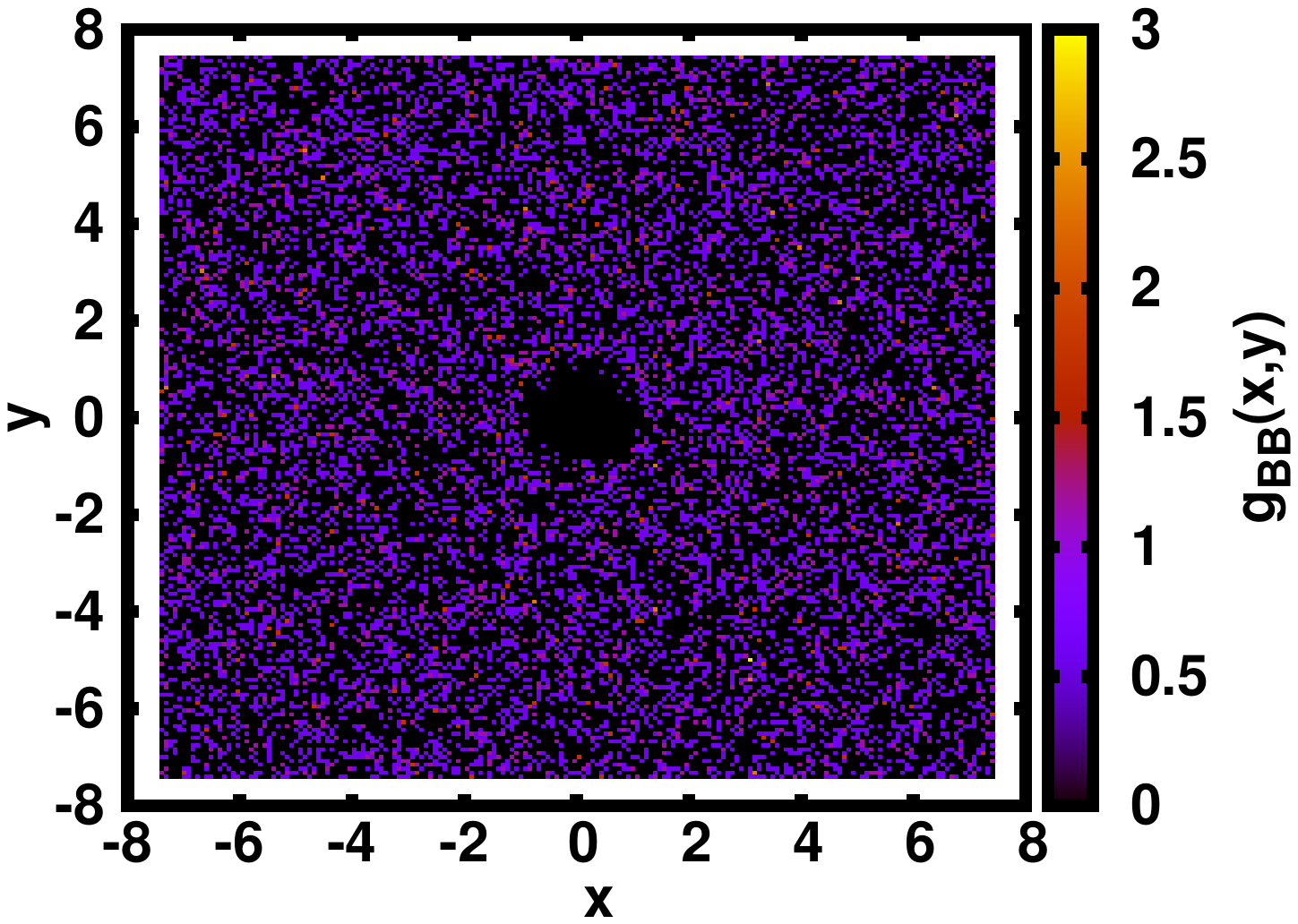}
\includegraphics[scale=0.25]{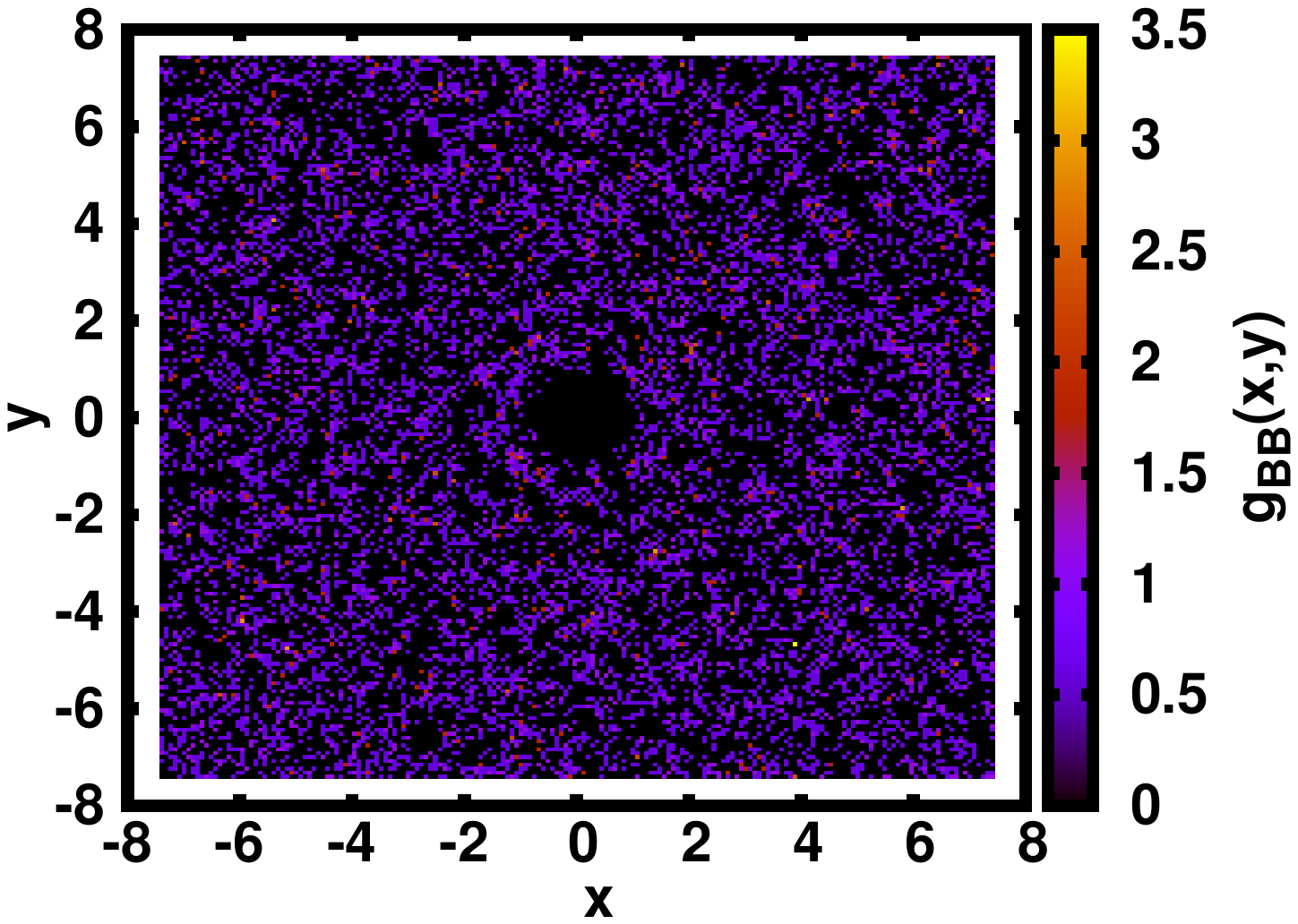}
\centering
\caption{Two dimensional, partial  pair correlation functions in the shear plane for (left panel) the initial configurations at $T = 0.466$, and (right panel) after 2000 shear cycles ($t ~ 4x10^7$). No indications of anisotropy is observed. }
\label{gofrdir}
\end{figure*}

As shown in Fig. \ref{gofrdir}, these correlation functions do not reveal any indications of anisotropy. 
%\textcolor{black}{We have also looked at if any structural anisotropy is manifested through the measure of fabric anisotropy (Fig.\ref{FAall}) which is defined as $FA=\sum_{N(i\neq j)}\frac{\vec{r_{ij}}}{|r_{ij}|}\otimes \frac{\vec{r_{ij}}}{|r_{ij}|} $ ($i,j$ are the particle pairs are considered within the interaction cutoff.)  }
%\clearpage

%\clearpage
\clearpage

\subsection{Energy barriers} 

\begin{figure}[h]
\centering
\includegraphics[scale=0.45]{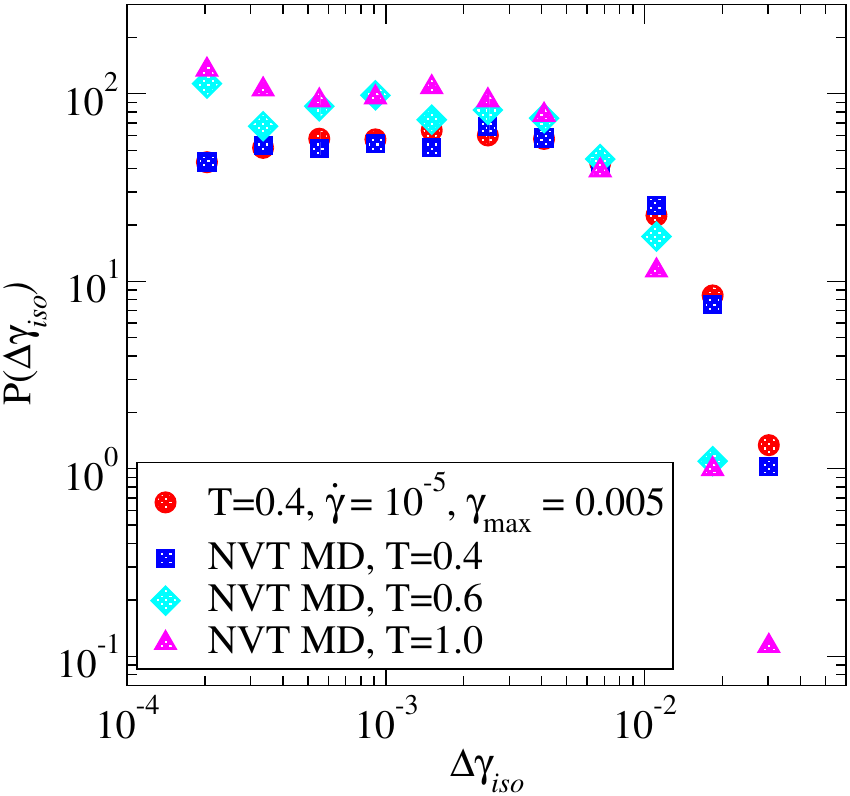} 
\caption{The distribution of the strain intervals ($\Delta \gamma_{iso}$) at which the first plastic deformation event occurs for inherent structures obtained through cyclic shear and from NVT molecular dynamics. The strain interval values are a measure of the energy barriers surrounding the undeformed inherent structures. For $T = 0.4$, the distributions from MD and cyclic shear are quantitatively the same, whereas they are easily distinguishable from higher temperature MD results.}
\label{aval}
\end{figure}

To obtain an estimate of the energy barriers surrounding the low energy minima we investigate the  distribution of strain required for the first plastic rearrangement when these configurations are sheared. These strain intervals provide a measure of the potential energy barriers of the corresponding minima \cite{karmakar2010statistical}. Here we have taken the inherent structures obtained from equilibrium MD and cyclic shearing at $T=0.4$. These configurations are subjected to a uniform athermal quasistatic deformation with a strain step of $10^{-5}$. With the application of uniform shear, the elastic energy of the system increases and this increase is punctuated by a drop in stress corresponding to plastic events. The distribution of the strain value ($\Delta \gamma_{iso}$) at which the inherent structure energy/stress decreases discontinuously for the first time has been for configurations obtained by equilibrium MD and cyclic shear. These distributions are qualitatively and quantitatively comparable (Fig. \ref{aval}), indicatiing that  the energy barriers surrounding the minima obtained through cyclic shearing is similar to the 
barriers surrounding minima obtained from MD simulations.  In order to ensure that the apparent agreement is meaningful, we compare these distributions to the strain interval ($\Delta \gamma_{iso}$) distributions for higher temperatures (obtained by molecular dynamics), and note that indeed, these higher temperature distributions are distinguishably different.

These observations together suggest that, in the cases where equilibrated configurations can be generated through equilibrium MD,  their properties  are indistinguishable from those obtained from cyclic shear deformation at a properly chosen shear strain amplitude. 

\subsection{Shearing in variable planes}
\begin{figure}[h]
\centering
\includegraphics[scale=0.25]{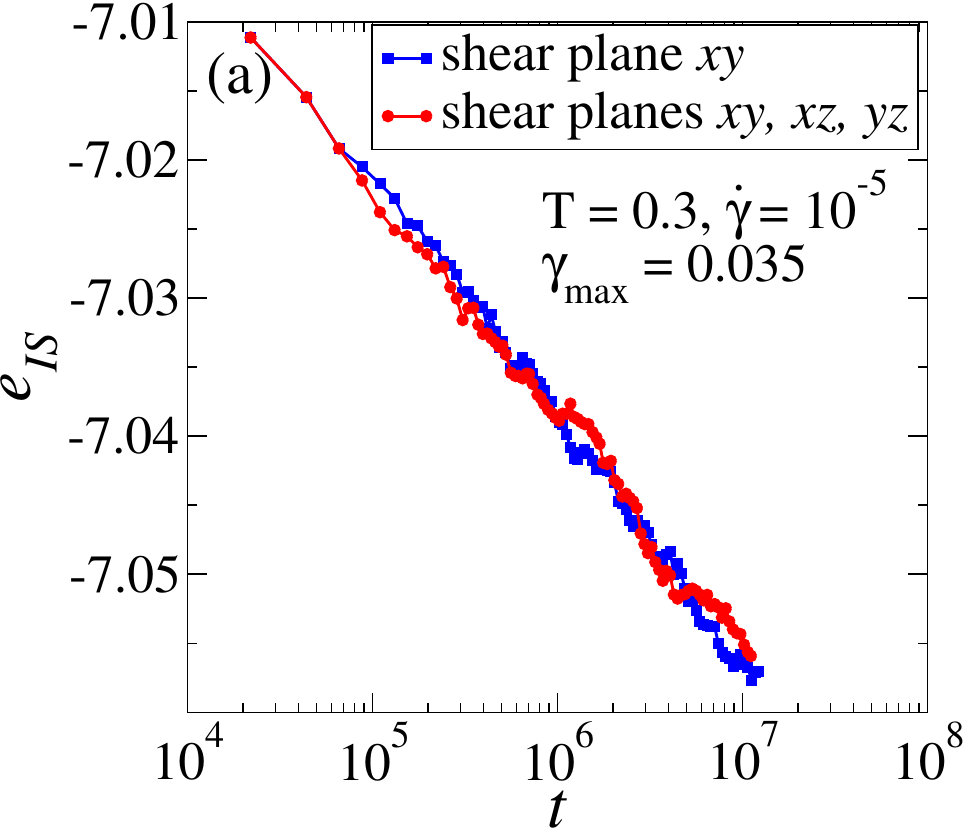}
\includegraphics[scale=0.25]{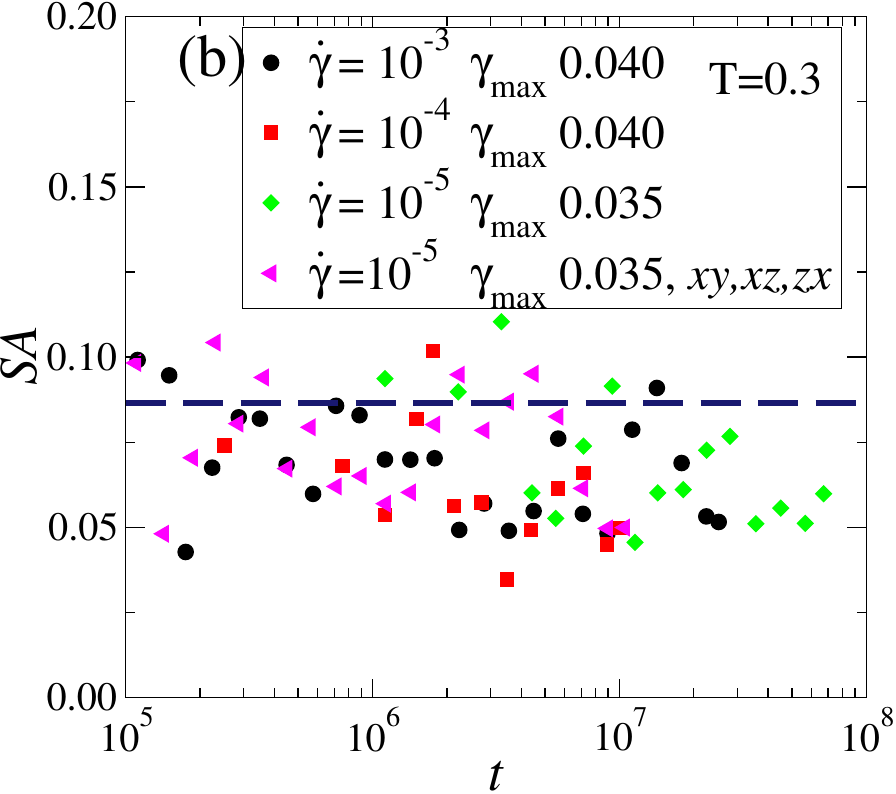}
\caption{(\emph{a}) Inherent structure (IS) energies (when shear is applied in alternating planes) vary in the same way with time as when only $xy$ shear is applied.
(\emph{c}) For strain amplitudes lower than or close to the yielding amplitude, stress anisotropies compare with those of isotropic inherent structures (indicated by the horizontal line) for different strain rates, and for the case when shear in applied in alternating shear planes. }

 \label{Fig:anisosm}
\end{figure}

We considered the effect of shearing in alternating shear planes ($xy$, $xz$, and $yz$, repeated after 3 cycles).  As shown in Fig. \ref{Fig:anisosm} (a) the generated energies are essentially the same (however, it has been reported that for larger systems at lower temperatures and shear amplitudes, alternating shear planes leads to better annealing \cite{PRIEZJEV2019,krishnan2021annealing}. In this case also we find the stress anisotropies to be small, and comparable to those of inherent structures quenched directly from liquid configurations, as shown in Fig. \ref{Fig:anisosm} (b). Fig. \ref{Fig:anisosm} (b)
also shows stress anisotropies for shear in a single plane obtained at different shear rates and strain amplitudes, for comparison. 

\subsection{Details of Algorithm for Equilibrium Sampling}

We provide below details concerning the equilibrium sampling algorithm. Much of these details are standard, and found, {\it e. g.} in \cite{ToddDaivisBook}, but they are included here for completeness. 

\subsubsection{Equations of Motion} 

The dynamics we will consider is the SLLOD dynamics, where the shear is applied through a sinusoidally varying $xy$ strain 

\beq 
\gamma_{xy}(t) = \gamma_{max} sin(\omega t)
\eeq 
which corresponds to to a strain rate 

\beq
\dot{\gamma}_{xy} = \gamma_{max} \omega cos(\omega t).
\eeq 

Nominally we can associate this with a strain rate $\dot{\gamma}_{xy} = \gamma_{max} \omega$ or more precisely, with $\gamma_{accum} = 4 \gamma_{max}$, $T = 2 \pi/ \omega$, with  $\dot{\gamma}_{xy} = (2/\pi) \gamma_{max} \omega$.  In the following, we will express the dynamics in terms of a general strain rate tensor ${\bf \nabla} {\bf v}$ but simple shear in the $xy$ plane will correspond to 

\begin{eqnarray} 
		{\bf \nabla} {\bf v} &=\begin{pmatrix}
				\nonumber  0 & 0 & 0 \\ 
				\nonumber   \dot{\gamma}_{xy} & 0 & 0 \\
				\nonumber 0 & 0 & 0
		\end{pmatrix}
\end{eqnarray}

The SLLOD equations of motion are written as 

\beq
\dot{{\bf r}}_i = { {\bf p}_i \over m} + {\bf r}_i. {\bf \nabla} {\bf v}
\eeq

\beq
\dot{{\bf p}}_i = { {\bf F}_i \over m} - {\bf p}_i. {\bf \nabla} {\bf v} - \alpha ~ {\bf p}_i
\eeq

With simple $xy$ shear, we have 

$$
 {\bf r}_i. {\bf \nabla} {\bf v} =  \dot{\gamma}_{xy} y_i {\bf i} 
 $$
 
 $$
 {\bf p}_i. {\bf \nabla} {\bf v} =  \dot{\gamma}_{xy} p_{y~i} {\bf i} 
 $$

and the equations of motion in component form are 

\be 
\nonumber \dot{x}_i & = & { p_{x~i} \over m} +  \dot{\gamma}_{xy} y_i \\
\nonumber  \dot{y}_i & = & { p_{y~i} \over m}\\
\nonumber  \dot{z}_i & = & { p_{z~i} \over m}\\
\nonumber \dot{p}_{x~i} & = & { F_{x~i} \over m} -  \dot{\gamma}_{xy} p_{y ~i} - \alpha ~ p_{x~i} \\
\nonumber \dot{p}_{y~i} & = & { F_{y~i} \over m}  - \alpha ~ p_{y~i} \\
\nonumber \dot{p}_{z~i} & = & { F_{z~i} \over m}  - \alpha ~ p_{z~i} 
\ee

We consider two thermostats, the  Gaussian isokinetic thermostat and the Nos\'e-Hoover thermostat. For the  Gaussian isokinetic thermostat, the damping coefficient is given by: 

\be
\alpha &=& \frac{\sum_i^N \mathbf{p}_i\cdot(\mathbf{F}_i - \mathbf{p}_i\cdot\nabla\mathbf{v})/m_i}{\sum_i^N \mathbf{p}_i^2/m_i}
\ee 

The corresponding expression for the Nos\'e-Hoover thermostat is:

\beq
\alpha = p_{\eta}/Q
\eeq 
where $p_{\eta}$ is the momentum associated with the thermostat coordinate $\eta$ and $Q$ is the corresponding mass. $p_{\eta}$ is governed by
\beq
\dot{p_{\eta}} = {\sum_i^N \mathbf{p}_i^2/m} - 3 N k_B T 
\eeq

\subsubsection{Modified Hybrid Monte Carlo Method} 

We first describe the steps in the standard Monte Carlo scheme, but in the case that the thermostat temperature is different from the sampling temperature. We consider a system that is characterized by a Hamiltonian 

\begin{equation} 
H = \sum_i {p_i^2 \over 2 m}   + \Phi(x); ~~  \Phi(x) = \sum_{i < j} \phi_{ij}(x_{ij})
\end{equation} 
where $x_i, i = 1, \dots 3 N$ refer to the coordinates of $N$ particles, and $p_i$ are the corresponding momenta. We wish to generate an equilibrium sampling of coordinates $x$,

\begin{equation}
p_{eq} (x) dx = {1 \over Z} \exp{ (-\beta_s  \Phi(x))} dx 
\end{equation} 
where $\beta_s = 1/k_B T_s$, where $T_s$ is the "sampling" temperature, and 

\begin{equation}
Z = \int dx  \exp{(-\beta_s  \Phi(x))}.
\end{equation} 

The Markov transition probability densities $p_M (x^{i} \rightarrow x^{i+1})$ need to satisfy detailed balance condition 

\begin{equation}
p(x) p_M(x \rightarrow x^{'}) dx ~dx^{'} = p(x^{'}) p_M(x^{'} \rightarrow x) dx ~dx^{'}.
\end{equation} 
Configurations $x^{'}$ are generated through a molecular dynamics run, starting with initial coordinates $x$ and momenta generated according to the Maxwell distribution, with a "thermostat" temperature $T_t = (k_B \beta_t)^{-1}$ that need not be the same as $T_s$. The procedure listed below can be verified to obey the  detailed balance condition mentioned above \cite{Palmer2017} for time reversible and phase space volume preserving dynamics:

 \begin{itemize}

\item[1.] Starting with $x$, generate momenta with the Maxwell distribution at temperature $T_t$ 

\begin{equation}
p_G (p) \propto \exp{(-\beta_t p^2 / 2 m)} 
\end{equation}  
 
 \item[2.] propagate $(x,p)$ using a time reversible, phase space volume preserving dynamics. For the simplest case, the Hamiltonian dynamics with $H_0$, we have for the probability of generating $(x^{'},p^{'})$ from $(x,p)$
 
 \begin{equation}
 p_H ((x,p) \rightarrow (x^{'},p^{'})) =  p_H ((x^{'},-p^{'}) \rightarrow (x,-p)) 
 \end{equation} 
 
\item[3.] Accept $(x^{'},p^{'}))$ with probability 
 
 \begin{equation}
 p_A ((x,p) \rightarrow (x^{'},p^{'})) =  min\{ 1, \exp{(-\beta_s \Delta \Phi )  \exp(-\beta_t \Delta K) }\} 
 \end{equation}  
 where $\Delta \Phi  = \Phi (x^{'},p^{'}) - \Phi (x,p)$ and  $\Delta K = \sum_i {p_i^{'~2} \over 2m} - \sum_i {p_i^2 \over 2m} $.  
 
 \item[4.] Return to step 1 and generate new momenta regardless of acceptance. 
 \end{itemize} 
 
 We next consider the case where the dynamics, while time reversible, does not conserve phase space volume, such as the thermostatted SLLOD dynamics we consider here. We consider for concreteness that the dynamics compresses the phase space volume. Thus, a volume we denote by $dx~dp$ around $(x,p)$ is mapped to a smaller volume $dx^{'}~dp^{'}$ around $(x^{'},p^{'})$ with a corresponding increase in the phase space density (see below). As a result, considering a distribution of points proportional to the equilibrium probability density, and the standard Metropolis acceptance probability, the net transition probability in the "forward" direction, proportional to 
 \begin{widetext}
 \begin{equation} 
 {\exp(-\beta_s H(x,p))  \over Q}  min \{ 1, exp(-\beta_s \Delta H) \} dx dp, ~~~~ \Delta H = H(x^{'},p^{'}) - H(x,p) 
\end{equation} 

does not balance that in the "reverse" direction, 

\begin{equation} 
 {\exp(-\beta_s H(x^{'},-p^{'}))  \over Q} dx^{'} dp^{'} min \{ 1, exp(-\beta_s   [ H(x,-p) - H(x^{'},-p^{'}) ]   \}
\end{equation} 

which can be manipulated to write as 

\begin{equation} 
 {\exp(-\beta_s H(x,-p))  \over Q}  min \{ 1, exp(-\beta_s \Delta H) \} dx^{'} dp^{'}.
\end{equation} 
\end{widetext}
The ratio of the forward to revere transition probability is the ratio of phase space volumes which we write as ${dx ~ dp \over dx^{'} dp^{'}}$. In order to obtain detailed balance, the acceptance probability must be corrected by a compensation factor $C = { dx^{'} dp^{'} \over dx dp  }$, leading to 

 \begin{equation} 
 p_A ((x,p) \rightarrow (x^{'},p^{'}) ) = min\{1,  exp(-\beta_s \Delta H) ~~ C \}, 
 \label{eq:Met1}
 \end{equation} 
which can be verified to obey detailed balance. 

\subsubsection{Compression of phase space volume}

The calculation of the correction term $C$ is done here in two ways. First, we consider how the volume element around the initial state point $(x,p)$ is modified by the dynamics. Second, we can alternately consider how the phase space density is modified by the dynamics. 

We consider the phase space point $\Gamma = (x,p)$, and write the dynamics governing it as 

\begin{equation} 
\dot{\Gamma} = G(\Gamma, t)
\end{equation} 

If we now consider a displacement from  $\Gamma$, $\delta  \Gamma$, the dynamics of the displacement (which we assume is always small) is given by 

\begin{equation} 
\dot{\delta \Gamma} = T.\delta \Gamma
\end{equation} 
where 

\begin{equation} 
T  = {\partial G \over \partial \Gamma} = {\partial \dot{\Gamma} \over \partial \Gamma} 
\end{equation}  

We can define a phase space volume by considering a set of independent $\delta \Gamma_i$, $ i = 1, 2 \dots 6 N$, and obtaining the determinant of the matrix $M = \{ \delta \Gamma_1, \delta \Gamma_2, \dots  \delta \Gamma_{6N} \}$, as discussed in the context of computing Lyapunov exponents. We have 

\begin{equation} 
\dot{M} = T. M 
\end{equation} 
Formally, we can write the solution as 

\begin{equation} 
M(t) = \exp\left[\int T(s) ds\right] M(0)
\end{equation}

To evaluate this, we consider dividing the time interval $[0,t]$ into infinitesimal segments $d t$ within which we treat the matrix $T$ as being constant (we can take the limit of $dt \rightarrow 0$ eventually). With this, we can write 

\begin{equation} 
M(t) = \Pi_i \exp[T_i ~dt] ~~M(0)
\end{equation} 

We now write the determinant of $M$ as 

\begin{eqnarray} 
det \left( M(t) \right) &=& det \left( \Pi_i \exp[T_i ~dt] ~~M(0) \right) \\ \nonumber 
	&=&  \left( \Pi_i det \left( \exp[T_i ~dt] \right) \right) \times det \left( M(0) \right) \\ \nonumber 
	&=&  (\Pi_i  \exp[ Tr. T_i ~ dt] ) \times det \left( M(0) \right)
\end{eqnarray} 
by the corollary to Jacobi's formula, $det \left( exp(A) \right) = exp(Tr.A)$. Thus,  

\begin{eqnarray} 
det \left( M(t) \right) &=& \exp\left( \sum_i Tr. T_i ~dt \right) \times det \left( M(0) \right) \\ \nonumber 
		&=& \exp\left(\int Tr.T(s) ~dt \right)  \times det \left( M(0) \right); ~~ dt \rightarrow 0
\end{eqnarray} 

We can write $det \left(M\right) = V_{\Gamma}$, the phase space volume. Further, writing $Tr.T = {\partial \over \partial \Gamma} . \dot{\Gamma} $, we have 

\begin{equation} 
V_{\Gamma}(t) = \exp\left[ \int  {\partial \over \partial \Gamma} . \dot{\Gamma}(s) ds \right] \times  V_{\Gamma}(0) . 
\end{equation}
The trace of $T$ integrated over time $t$ also equals the sum of Lyapunov exponents times $t$. We will consider later how this quantity can be calculated for specific dynamics. 

The compensation factor $C$ we are after is therefore 
\begin{equation} 
C(t) = \exp\left[ \int^{t} \Lambda(\Gamma) (s) ds \right] 
\end{equation}

where $\Lambda(\Gamma) =  {\partial \over \partial \Gamma} . \dot{\Gamma}$ is called the phase space compression factor. 

We can alternately consider the time evolution of the phase space density $f(\Gamma,t)$, and the phase space compression is given by $f(\Gamma,0)/f(\Gamma,t)$. Considering an ensemble of phase space points, and writing 
\begin{equation}
   f(\Gamma,t) = \sum_i\delta(\Gamma - \Gamma_i(t)), 
\end{equation}
we have 

\begin{eqnarray}
   \frac{df(\Gamma,t)}{dt}  &=& \dot{\Gamma}\cdot\frac{\partial f}{\partial \Gamma} + \frac{\partial f}{\partial t} \\
   &=& \dot{\Gamma}\cdot\frac{\partial f}{\partial \Gamma} -\frac{\partial}{\partial \Gamma}\cdot(\dot{\Gamma} f) \\
   &=& -f\frac{\partial}{\partial \Gamma}\cdot\dot{\Gamma} \\
   &=& -f\Lambda(\Gamma)
\end{eqnarray}

which we see leads to the same result as before. $\Lambda(\Gamma)$ can be written in more explicit notation as

\begin{equation}
\Lambda(\Gamma) = \sum_{i=1}^{3N}(\frac{\partial}{\partial q_i},\frac{\partial}{\partial p_i})\cdot(\dot{q_i},\dot{p_i}).
\end{equation}

Considering the SLLOD equations of motion with Gaussian thermostat, we can compute $\Lambda(\Gamma)$ by using expressions for $\dot{q_i}$ and $\dot{p_i}$, and obtain 

\begin{equation}
    \Lambda(\Gamma) = N(\nabla.\mathbf{v} - \nabla.\mathbf{v} - 3 \alpha) = -3N\alpha
\end{equation}

ignoring terms of ${\cal{O}}(1)$ that arise from differentiating $\alpha$, which are proportional to $\alpha$ and the strain rate. For SLLOD dynamics without thermostatting, the phase space compression factor is zero, and therefore phase space volume is conserved, even though the dynamics is non-Hamiltonian. Thus, the standard HMC can be used if we don't apply a thermostat. With the thermostat, the damping coefficient $\alpha$ allows computation of the correction factor $C$ needed for the modified HMC. 

\subsubsection{Work done and heat dissipation} 

We finally consider the time derivative of the internal energy of the system in order to obtain expressions for the rate of work done and the rate of heat change (or dissipation), which enables us to relate the compensation factor $C$ to heat. 

To understand how the internal energy changes can be related to $\Lambda$ consider the following:
\begin{equation}
    H = K + \Phi
\end{equation}
where $H$ is the total internal energy, K is the thermal (kinetic) energy and $\Phi$ is the internal (pairwise) potential energy.

\begin{eqnarray}
   K &=& \sum_{i=1}^N \frac{\mathbf{p}_i\cdot\mathbf{p}_i}{2m_i} \\
   \Phi &=& \frac{1}{2}\sum_{i=1}^N\sum_{j\neq i}^N \phi_{ij}(r_{ij})
\end{eqnarray}

The rate of change of energy is then
\begin{equation}
    \frac{dH}{dt} = \frac{dK}{dt} + \frac{d\Phi}{dt}
\end{equation}

We find

\begin{eqnarray} 
   \frac{dK}{dt} &=&  \sum_{i=1}^N \frac{ \mathbf{p}_i \cdot \dot{\mathbf{p}}_i}{m} \\ \nonumber 
		      & = & \sum_{i=1}^N { \mathbf{p}_i \over m} \cdot (\mathbf{F}_i - \mathbf{p}_i \cdot \nabla\mathbf{v} - \alpha \mathbf{p}_i )  \\ \nonumber
		      & = &  \sum_{i=1}^N {\mathbf{p}_i \over m} \cdot \mathbf{F}_i -  \sum_{i=1}^N { \mathbf{p}_i \mathbf{p}_i  \over m} :\nabla\mathbf{v} - \alpha \sum_{i=1}^N {p_i^2 \over m}
\end{eqnarray}

\begin{eqnarray}
   \frac{d\Phi}{dt} & = & \sum_i \nabla_i \Phi . \dot{\bf r}_i \\
   & = & - \sum_i \mathbf{F}_i . ( {\mathbf{p}_i \over m} + \mathbf{r}_i.\nabla\mathbf{v} ) \\  
       &=& -\sum_{i=1}^N \frac{\mathbf{p}_i}{m_i}\cdot\mathbf{F}_i - \sum_{i=1}^N (\mathbf{r}_{i}\mathbf{F}_{i})^T:\nabla\mathbf{v} \\       
    &=& -\sum_{i=1}^N \frac{\mathbf{p}_i}{m_i}\cdot\mathbf{F}_i - \frac{1}{2}\sum_{i=1}^N\sum_{j=1}^N(\mathbf{r}_{ij}\mathbf{F}_{ij})^T:\nabla\mathbf{v}
\end{eqnarray}

(NB: The time dependence of  $\nabla\mathbf{v}$ does not enter and make a difference to these equations.) Combining the two terms, we have 
\begin{widetext}
\begin{eqnarray}
    \frac{dH}{dt} &=& - \sum_{i=1}^N \frac{\mathbf{p}_i\mathbf{p}_i}{m_i}:\nabla\mathbf{v} - \frac{1}{2}\sum_{i=1}^N\sum_{j=1}^N(\mathbf{r}_{ij}\mathbf{F}_{ij})^T:\nabla\mathbf{v} - \alpha\sum_{i=1}^N \frac{\mathbf{p}_i\cdot\mathbf{p}_i}{m_i} \\
    &=& -V\mathbf{P}^T:\nabla\mathbf{v} - \alpha\sum_{i=1}^N \frac{\mathbf{p}_i\cdot\mathbf{p}_i}{m_i}
\end{eqnarray}
\end{widetext}

We have the pressure tensor defined as 

\beq 
\mathbf{P} = {1 \over V}  \left[  \sum_{i=1}^N \frac{\mathbf{p}_i\mathbf{p}_i}{m} + \sum_{i=1}^N \mathbf{r}_i\mathbf{F}_i \right]
\eeq 

The stress tensor $\mathbf{\sigma}$ is defined as the negative of the pressure tensor $ \mathbf{\sigma} = - \mathbf{P} $. For strain rate in the $xy$ plane with only ${\partial v_x \over \partial y} \ne 0$, we have 

\begin{eqnarray}
    \frac{dH}{dt} &=& - \sum_{i=1}^N \left[{ p_{xi} p_{yi} \over m} + F_{xi} y_i \right] \dot{\gamma}_{xy} - \alpha \sum_{i=1}^N {p_i^2 \over m} \\
    &=& -  \mathbf{P}^{T}_{xy} \dot{\gamma}_{xy} - \alpha \sum_{i=1}^N {p_i^2 \over m} 
\end{eqnarray}

Using $\sum_{i=1}^N {p_i^2 \over m} = 3 N k_B T$, we have 

 \begin{eqnarray}
    \frac{dH}{dt} &=& - \sum_{i=1}^N \left[{ p_{xi} p_{yi} \over m} + F_{xi} y_i \right] \dot{\gamma}_{xy} - 3 N \alpha  k_B T \\
    			& = & -V \mathbf{P}^{T}_{xy}  \dot{\gamma}_{xy} - 3 N \alpha  k_B T \\
    			& = & V \mathbf{\sigma}^{T}_{xy}  \dot{\gamma}_{xy} - 3 N \alpha  k_B T \\			
    			& = & V \mathbf{\sigma}^{T}_{xy}  \dot{\gamma}_{xy} + \Lambda  k_B T 
\end{eqnarray}

A change in energy over some interval $t$ is given by 
\begin{widetext}
\be
\Delta H(t) & = & \int^t  V \mathbf{\sigma}^{T}(s)_{xy}(s) \dot{\gamma}_{xy}(s) ds + k_B T\int^t \Lambda(s) ds  \\
		&=& \Delta W + \Delta Q 
\label{eq:WQ}		
\ee
\end{widetext}
Based on the above, the multiplicative factor in the Monte Carlo step for thermostatted systems discussed above will be 

\beq 
C = \exp\left( \int \Lambda(s) ds \right) = e^{\Delta Q \over k_B T}
\eeq 

Note that $\Delta W$ and $\Delta Q$ are work done on the system and heat input into the system as they both contribute to an increase in the internal energy. Heat dissipated, which we denote by $Q_d = -\Delta Q$. 

Finally, returning to Eq. \ref{eq:Met1}, we can write the acceptance probability as 
\begin{widetext} 
 \begin{eqnarray} 
 p_A ((x,p) \rightarrow (x^{'},p^{'}) ) &=& min\{1,  exp(-\beta_s \Delta H) exp(\beta_s \Delta Q)  \}\\
 &=& min\{1,  exp(-\beta_s (\Delta H -\Delta Q)  \}\\	
 &=& min\{1,  exp(-\beta_s \Delta W)   \}.
 \label{eq:Met2}
 \end{eqnarray} 

When the system and thermostat temperatures are not the same, we have 

  \begin{equation}
 p_A ((x,p) \rightarrow (x^{'},p^{'})) =  min\{ 1, \exp{(-\beta_s \Delta \Phi )  \exp(-\beta_t \Delta K)  \exp(\beta_t \Delta Q)   }\} 
  \label{eq:Met3}
 \end{equation}  
\end{widetext}
With the expressions above for $\Delta W$, $\Delta Q)$ given above (Eq. \ref{eq:WQ}), Eq.s \ref{eq:Met2} and \ref{eq:Met3} provide the acceptance probabilities for an equilibrium sampling algorithm.

\subsection{Additional Results from Equilibrium Sampling Monte Carlo Simulations}

%\clearpage

\begin{figure}[h]
\centering
\includegraphics[scale=0.26]{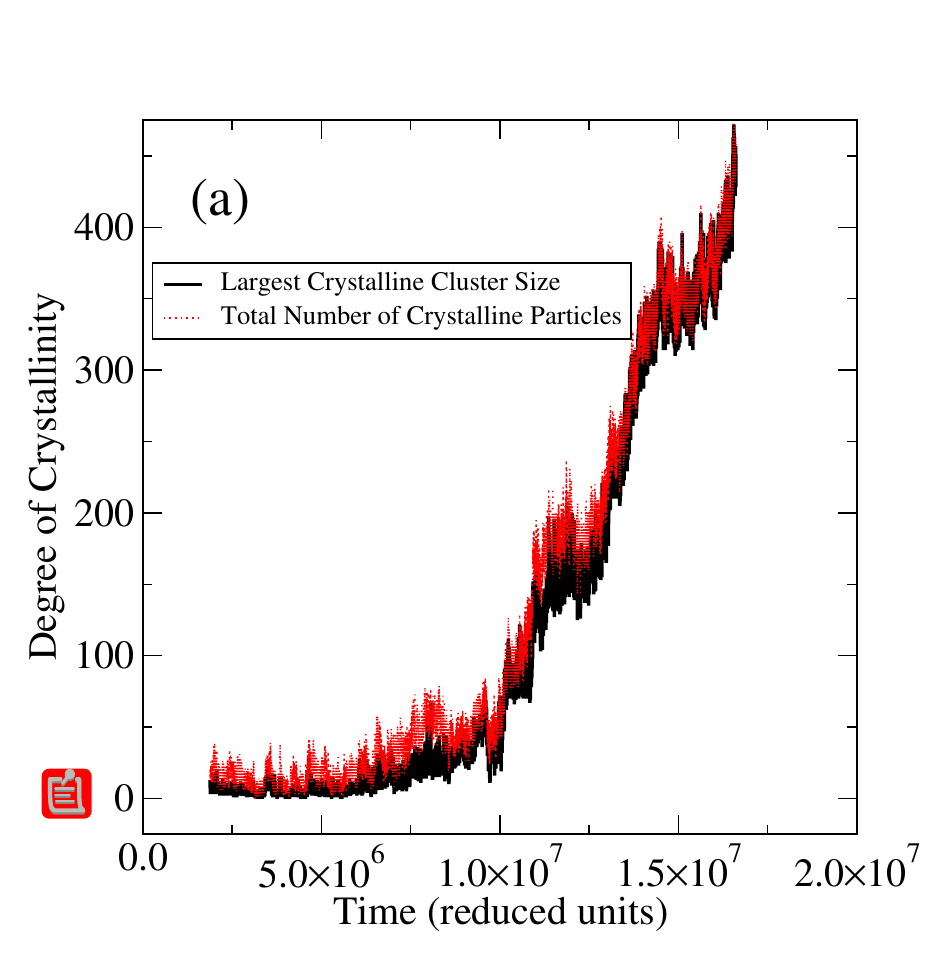}
\includegraphics[scale=0.26]{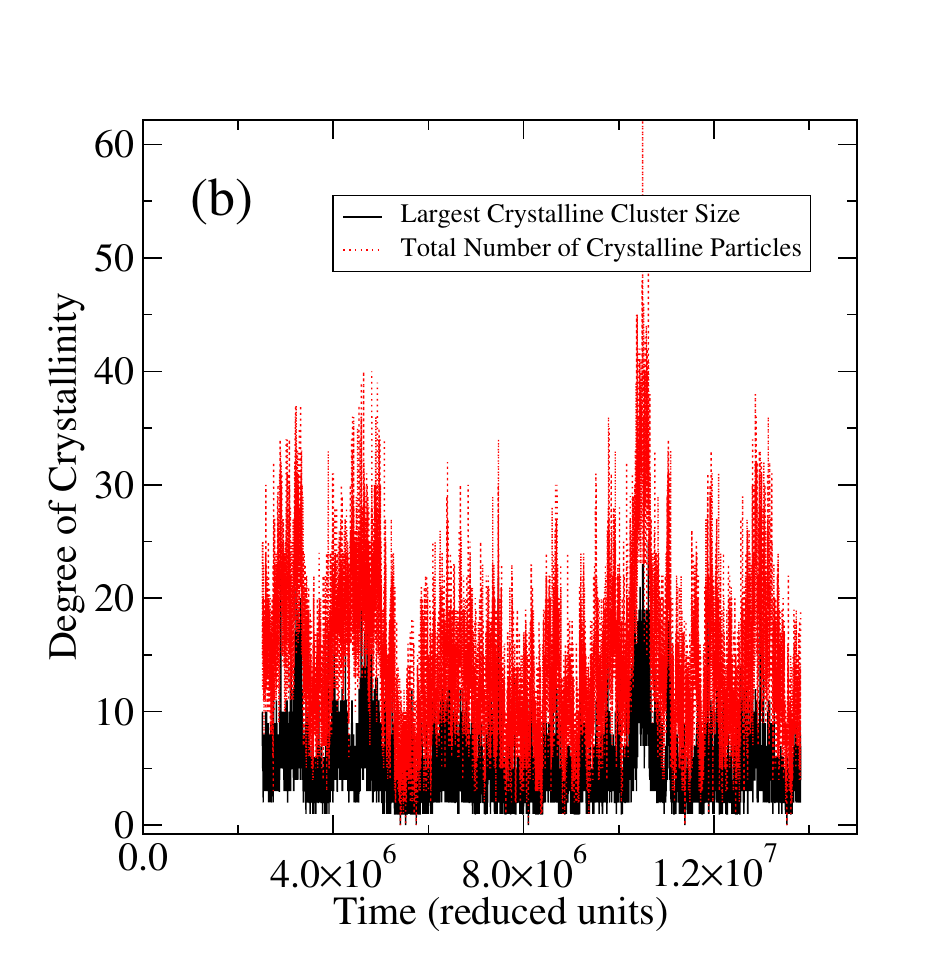}
\includegraphics[scale=0.26]{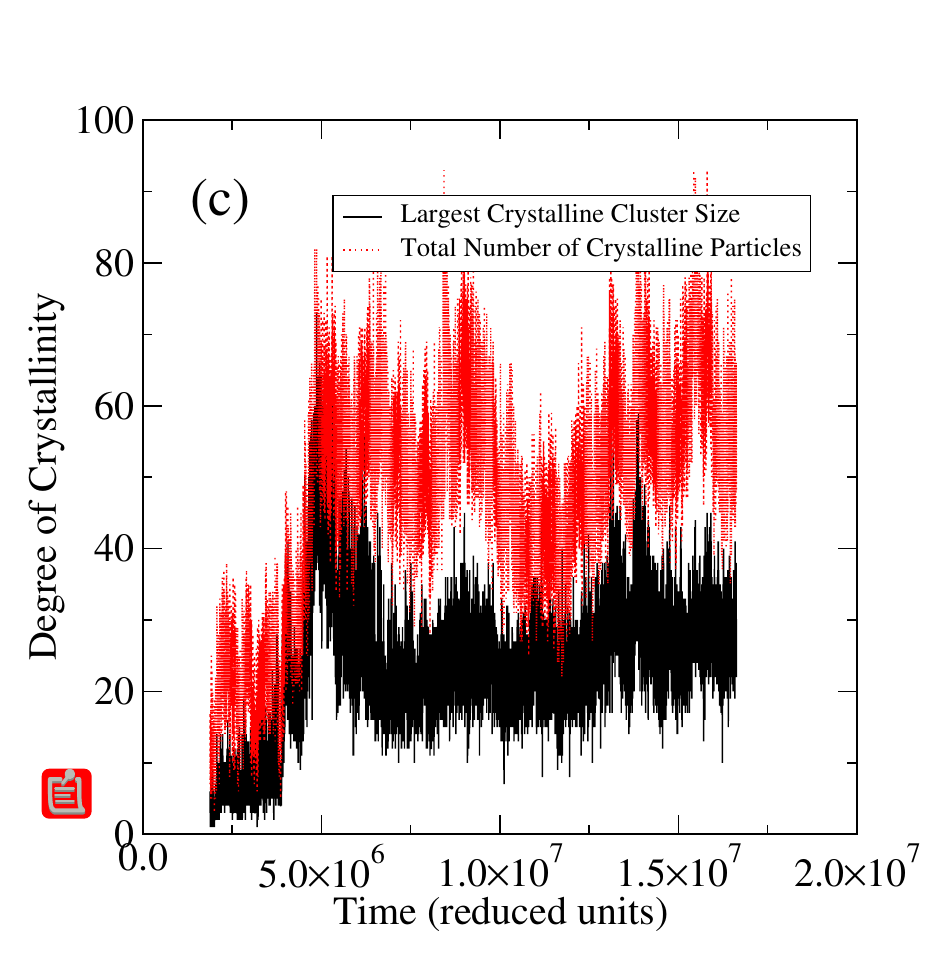}
\includegraphics[scale=0.26]{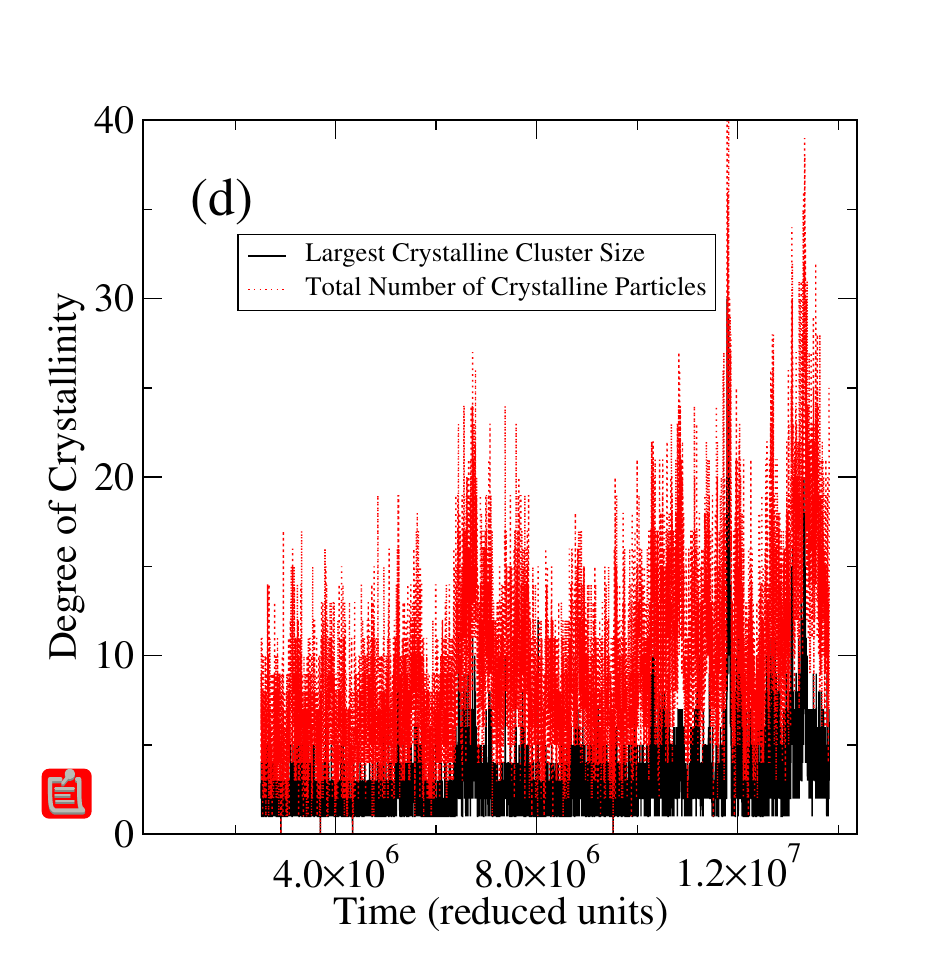}
\caption{Degree of crystallinity of sheared samples shown by considering the size of the largest crystalline cluster, as well as the total number of crystalline particles, for cyclically sheared trajectories without and with Monte Carlo sampling, for $T = 0.4$ and $T = 0.37$. (a) Cyclic shear, $T = 0.4$, (b) Cyclic shear, with Monte Carlo, $T = 0.4$, (c) Cyclic shear, $T = 0.37$, and (d) Cyclic shear, with Monte Carlo, $T = 0.37$.
}
\label{fig:crys}
\end{figure}
\subsubsection{Crystallization:} As noticed in previous work \cite{PhysRevXCryst2019}, the model studied here, the Kob-Andersen binary mixture, crystallizes on time scales that are accessible to long molecular dynamics simulations. In particular, at $T = 0.4$, for $N = 4000$, results in \cite{PhysRevXCryst2019} indicate that the crystallization time $\tau_{X} \sim 100 \tau_{\alpha}$. With $\tau_{\alpha} \sim 1.5 \times 10^5$ at this temperature, we do indeed see crystallization on the expected time scale, but on the lower side by a factor of $2$. We analyse the crystallization kinetics employing standard methods \cite{AuerFrenkelReview2004,PhysRevXCryst2019} that are based on defining bond orientational order parameters for each particle $i$, $q_{lm}(i) = {1\over n_b(i)} \sum_{j = 1}^{n_b(i)} Y_{lm} (\theta_{ij}, \phi_{ij})$, where $\theta_{ij}$ and $\phi_{ij}$ are angles formed by separation vectors ${\bf r}_{ij}$ between particles $i$ and neighbors $j$ in the laboratory frame. The $l$ value used is $6$. We employ cutoffs for defining neighbors from the location of the first minimum of the pair correlation functions $g_{\alpha \beta} (r)$, which are, $r^{cut}_{AA} = 1.4$,  $r^{cut}_{AB} = 1.25$ and  $r^{cut}_{BB} = 1.1$. Following \cite{PhysRevXCryst2019}, we define two neighbors to be bonded if the normalised dot product ${\bf q}_{6}(i).{\bf q}_{6}(j)/|{\bf q}_{6}(i)| |{\bf q}_{6}(j)|$ is bigger than $0.7$, and a particle is labeled as crystalline if it has at least $7$ such bonds. Crystalline particles are then connected if they are within a cutoff distance, for which we use slightly larger values 
$r^{clust}_{AA} = 1.5$,  $r^{clust}_{AB} = 1.4$ and  $r^{clust}_{BB} = 1.2$. We then perform a cluster finding procedure to identify all clusters of connected crystalline particles and report the largest cluster size of crystalline particles. Because the largest cluster size may not fully reflect the degree of crystallinity in cases where the cluster sizes are small, and the clustering definition may not capture physical proximity, we also report the total number of crystalline particles. These are shown in Fig. \ref{fig:crys} for one trajectory each with $T = 0.4$ and $T = 0.37$,  both without Monte Carlo sampling between cycles and with sampling. For $T = 0.4$ (Fig. \ref{fig:crys} (a)), the largest cluster grows significantly beyond time $t \sim 7.5 \times 10^6$ reaching large values. For the other cases, the degree of crystallinity is never very large, but for $T = 0.37$ (Fig. \ref{fig:crys} (c)), beyond time $t \sim 4.5 \times 10^6$, there is a rise in the number of crystalline particles. In \cite{PhysRevXCryst2019}, the critical nucleus size is estimated to be $50 - 100$ particles. In our analysis, when a trajectory exhibits persistently a largest crystalline cluster size of $30$ or above, or the total number of crystalline particles is above $60$ in a persistent fashion, or even otherwise, when the growth of crystallinity suggests the onset of crystallization, we exclude the corresponding trajectory segments ({\it i. e.} parts of the trajectory thereafter) from the computation of averages and statistics. Thus, for example, for the trajectory shown in Fig. \ref{fig:crys} (a), we do not consider the trajectory after 
$t \sim 7.5 \times 10^6$, and for the trajectory shown in Fig. \ref{fig:crys} (c), we do not consider the trajectory after 
$t \sim 4.5 \times 10^6$. All the data in Fig. \ref{fig:crys} (b) and (d) are included in the analysis.

%Even though it appears that the crystallites that form do not grow much over the simulation time scales, we do not analyse configurations beyond $t = 5 \times 10^6$ in this case. For the cases where Monte Carlo sampling is performed, there is no significant degree of crystallinity for all the times reported, but some indications of crystallization at the longest times shown. We thus restrict analysis to configurations before time $t = 10^7$. We restrict anlysis to times less than $t \sim 6.5 \times 10^6$. 

\begin{figure}[ht]
\centering
\includegraphics[scale=0.45]{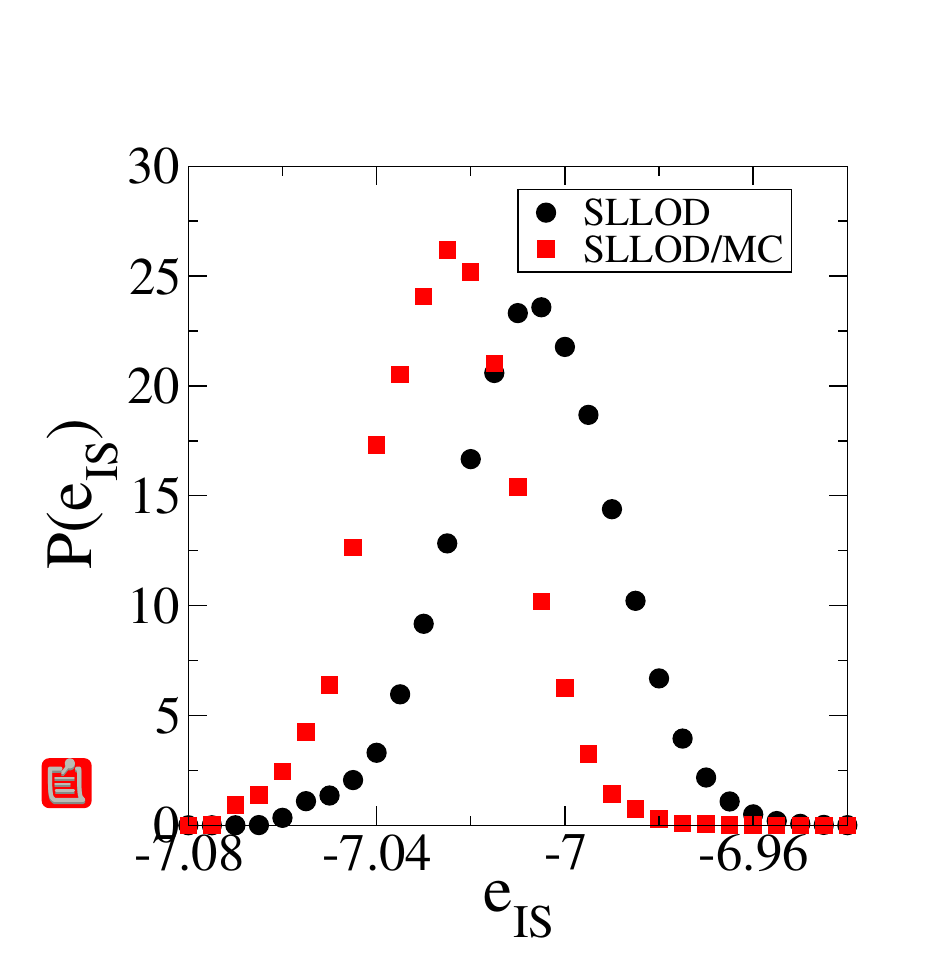}
\caption{Histograms of inherent structure energies obtained without (SLLOD) and with (SLLOD/MC) equilibrium Monte Carlo sampling, demonstrating that without Monte Carlo sampling, the energies are shifted to higher values compared to the expected equilibrium values, whereas with Monte Carlo sampling, they are close to the expected equilibrium values.}
\label{fig:MCtest}
\end{figure}
\subsubsection{Proper sampling under large amplitude shear:} Although results from cyclic shear without Monte Carlo sampling shown in the main text are close to those obtained from equilibrium Monte Carlo sampling, this is not guaranteed and is not always the case. We have shown above that with an increase in strain amplitude, the inherent structure energies attained increase beyond the optimal amplitude. In such a case, we should expect that the energy distribution obtained without equilibrium Monte Carlo sampling will deviate from that obtained with such sampling, which should generate an equilibrium ensemble. We demonstrate here that such is indeed the case. However, with an increase in strain amplitude (and/or system size; see below), we expect that the acceptance rate of the Monte Carlo procedure may also decrease, and sampling will be less efficient. Nevertheless, we should expect to converge to the correct distribution of energies. In order to demonstrate this, we consider a system of $N = 500$ particles, simulated at $T = 0.4$, at a large strain amplitude of $\gamma_{max} = 0.03$. We generate $5$ independent trajectories of length $t = 9.42 \times 10^6$ without Monte Carlo sampling, and $10$ independent trajectories of length $t = 1.88 \times 10^7$ with Monte Carlo sampling and obtain the histogram of energies in each case by averaging over the independent runs. Results shown in Fig. \ref{fig:MCtest} do indeed confirm that the Monte Carlo procedure generates a distribution of energies with a mean that is close to the expected equilibrium value (in this case, by extrapolation, $e_{IS} = -7.028$, based on results for $N = 1000$ described above) whereas without Monte Carlo sampling, the energies are clearly higher. We note in passing that there appears to be a small shift in energies to higher values with a decrease in system size but do not analyse such system size dependence further. 

\begin{figure}[h]
\centering
\includegraphics[scale=0.45]{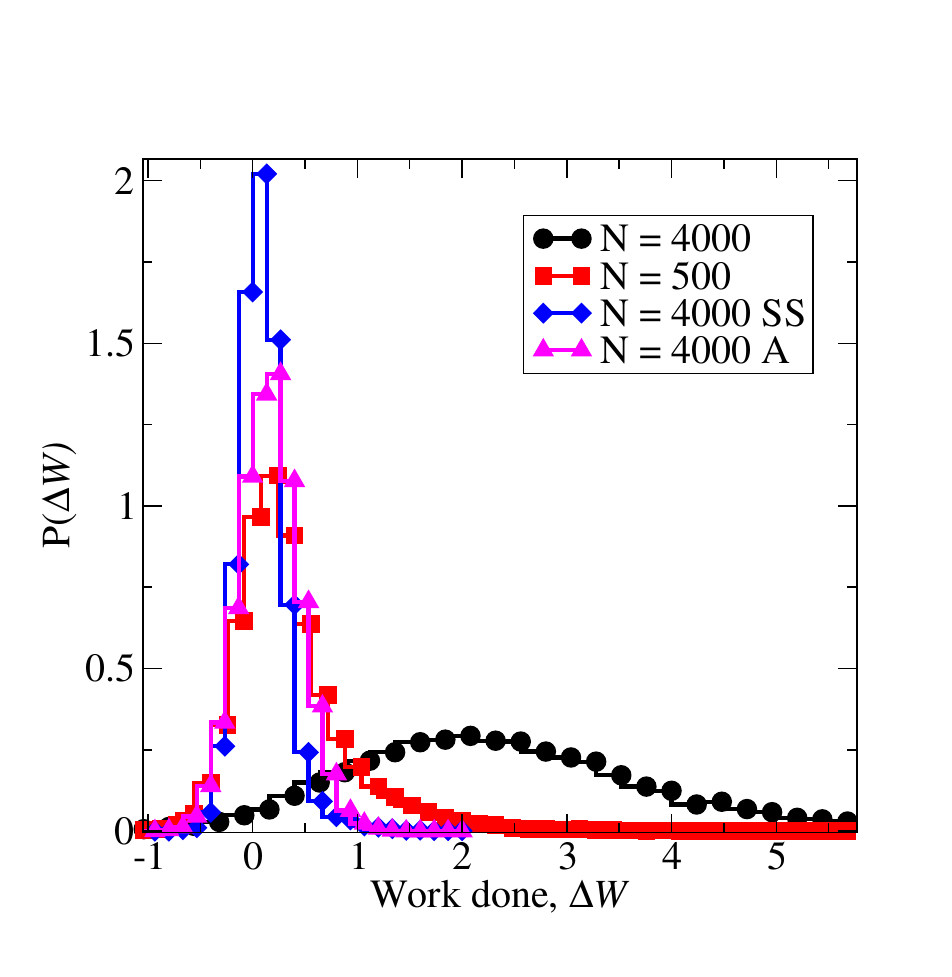}
\caption{The histogram of work done during a cycle for cyclic shear runs with $T = 0.3$, for (a) $N = 4000$, $\gamma_{max} = 0.03$, (b) $N = 500$, $\gamma_{max} = 0.03$, (c) $N = 4000$, $\gamma_{max} = 0.03$, where shear is applied to a subvolume of $\approx 500$ particles (SS), and (d) $N = 4000$, with adaptive tuning of $\gamma_{max}$ (here, on average (A), $\gamma_{max} = 0.004$).}
\label{fig:dw}
\end{figure}

\subsubsection{Work Done:} The work done over a cycle depends on the system size $N$ and the strain amplitude $\gamma_{max}$, among other variables. For $N = 4000$, with $\gamma_{max} = 0.03$, the histogram work done is shown in Fig. \ref{fig:dw}. The average value is about $2$, with makes the acceptane probabilities of Monte Carlo moves negligible. For $N = 500$, with $\gamma_{max} = 0.03$, the distribution of values is nearly centered around zero. Also shown are the histograms of work done for $N = 4000$, $\gamma_{max} = 0.03$, when only a subvolume of $\approx 500$ partices is sheared (labeled SS), and for  $N = 4000$, when the  $\gamma_{max}$ is varied adaptively to ensure a specified acceptance probability (labeled A). In the case shown, the the average values of $\gamma_{max} \approx 0.004$. In all cases, cyclic shear is applied without Monte Carlo rejection, with $T_s = T_t = 0.3$.

%\clearpage 

%\bibliography{annealSIbib}
% Bibliography
\bibliography{cyclicshear}
\end{document}